\documentclass[10pt]{article}

\usepackage{geometry}
\usepackage[superscript,biblabel]{cite}
\usepackage{hyperref}
\usepackage{authblk}
\usepackage{amssymb,amsmath,amsthm,amsfonts}
\usepackage[bb=pazo]{mathalpha}
\usepackage{bm}
\usepackage{graphicx,color,enumitem,xspace}
\usepackage[textsize=tiny]{todonotes}
\usepackage{lineno,etoolbox}
\usepackage{units}
\usepackage{xcolor}
\usepackage{placeins}
\newcommand{\linenomathpatch}[1]{%
  \cspreto{#1}{\linenomath}%
  \cspreto{#1*}{\linenomath}%
  \csappto{end#1}{\endlinenomath}%
  \csappto{end#1*}{\endlinenomath}%
}
\linenomathpatch{equation}
\linenomathpatch{gather}
\linenomathpatch{multline}
\linenomathpatch{align}
\linenomathpatch{alignat}
\linenomathpatch{flalign}

\begin{document}

\newcommand{\bu}{{\boldsymbol u}} 
\newcommand{\bv}{{\boldsymbol v}}
\newcommand{\bw}{{\boldsymbol w}}
\newcommand{\br}{{\boldsymbol r}}
\newcommand{\bff}{{\boldsymbol f}}
\newcommand{\bF}{{\boldsymbol F}}
\newcommand{\bg}{{\boldsymbol g}}
\newcommand{\bG}{{\boldsymbol G}}
\newcommand{\bn}{{\boldsymbol n}} 
\newcommand{\bx}{{\boldsymbol x}} 
\newcommand{\bX}{{\boldsymbol V}}
\newcommand{\bsig}{{\boldsymbol \sigma}}
\hyphenation{bio-mark-er bio-mark-ers}
\newcommand{\RSV}{R_{\mathrm{SV}}}
\newtheorem{remark}{Remark}

\newcommand{\turbleg}{{The legend is given in Figure~\ref{fig:turb_legend}.}}
\newcommand{\impsvr}{{Impact of SVR variation: }}
\newcommand{\impturb}{{Impact of turbulence model: }}

\renewcommand\Affilfont{\itshape\small}

\title{Impact of Turbulence Modeling on the Simulation of Blood Flow in Aortic Coarctation} 

\author[1]{Sarah Katz}
\author[1]{Alfonso Caiazzo}
\author[1]{Baptiste Moreau}
\author[1]{Ulrich Wilbrandt}
\author[2]{Jan Br\"uning}
\author[2,4]{Leonid Goubergrits}
\author[1,3]{Volker John}

%

\affil[1]{Weierstrass Institute for Applied Analysis and Stochastics (WIAS), Mohrenstr.~39, 10117~Berlin, Germany}
\affil[2]{Institute of Computer-assisted Cardiovascular Medicine, Charité -- Universit\"atsmedizin Berlin, Augustenburger~Platz~1, 13353~Berlin, Germany}
\affil[3]{Department of Mathematics and Computer Science, Freie Universit\"at Berlin, Arnimallee~6, 14195~Berlin, Germany}
\affil[4]{Einstein Center Digital Future, Wilhelmstraße~67, 10117~Berlin, Germany}
%
%
%

%
%

\maketitle

\begin{abstract}
  Numerical simulations of pulsatile blood flow in an aortic coarctation require
  the use of turbulence modeling. This paper considers three models from the
  class of large eddy simulation (LES) models (Smagorinsky, Vreman,
  $\bsig$-model) and one model   from the class of variational multiscale models
  (residual-based) within a finite element framework. The influence of these
  models on the estimation of clinically relevant biomarkers used to assess the
  degree of severity of the pathological condition (pressure difference,
  secondary flow degree, normalized flow displacement, wall shear stress) is
  investigated in detail.
  The simulations show that most methods are consistent in terms of severity
  indicators such as pressure difference and stenotic velocity. The numerical
  results indicate that second order velocity elements outperform first order
  elements in terms of accuracy.
  Moreover, using second order velocity finite elements, different turbulence
  models might lead to considerably different results concerning other
  clinically relevant quantities such as wall shear stresses.
  These differences may be attributed to differences in numerical dissipation
  introduced by the turbulence models.
\end{abstract}

\section{Introduction}

Coarctation of the aorta (CoA) is a congenital heart defect consisting in a local narrowing
in a portion of the aorta, resulting in hypertension of the upper body and with
potentially severe complications. The most relevant diagnostic parameter for
this pathology is the trans-stenotic pressure gradient/difference, which can
only be measured directly via invasive catheterization.
Non-invasive imaging-based techniques for assessing the severity of CoA
rely on measuring patient anatomy, blood velocities and flow rates in the area
by cardiac MRI or (Doppler) ultrasound echocardiography. Estimating pressure
gradients from velocity information using a simplified Bernoulli equation has
remained common practice well into the present century, despite its
well-documented
limitations\cite{de_mey_limitations_2001}\cite{circulation-2005}.
Clinical guidelines\cite{circulation-2019-guidelines}\cite{esc_aorta_2014}
provide diagnostic criteria in terms of these biomarkers.

The severity of the disease does not only depend on the anatomical condition and the
pressure gradient/difference, but can be assessed via different biomarkers that
are related to abnormal flow conditions, such as increased flow asymmetries and
abnormal oscillatory behaviors of the wall shear stresses (WSS).
However, due to the relatively low spatial resolution of MRI, these biomarkers
can only be quantified directly from medical imaging with reduced accuracy.
Furthermore, these methods are time-consuming and costly. Numerical blood flow
simulations can therefore play an important role in supporting available medical
data, such as anatomical images and flow fields, for the estimation of these
quantities of interest\cite{goubergits-etal-2015,schubert-etal-2020}.

The pulsatile blood flow in the ascending aorta reaches moderate to high
Reynolds number (larger than $2000$\cite{goubergrits-2013}) and the flow
disturbances caused by aortic narrowing can yield to a transition to turbulence.
Understanding the behavior of a turbulent flow is therefore relevant from the
clinical point of view, since turbulence might have implications for the
pathophysiology of vascular diseases and for the design of cardiovascular
devices such as stents or artificial valves \cite{pietrasanta-2022,rigatelli-2019}.

The dynamics of turbulent flows spans a wide range of spatial scales, from
Kolmogorov lengths of the order of $\unit[10-70]{\mu m}$ up to the diameter of
the blood vessel. Direct numerical simulations (DNS) of the whole scale spectrum
are beyond computationally affordable resolution of numerical discretizations.
However, the smallest scales cannot be neglected, since otherwise a laminar flow
would be simulated with the corresponding high inaccuracy of the computational
results. The purpose of turbulence modeling consists in modeling the impact of
the unresolved scales onto the resolved ones so that important properties of
turbulent flows, like boundary layers, are present in the simulated flow fields.

A popular approach for modeling the effect of turbulence are so-called Reynolds
Averaged Navier--Stokes (RANS) methods. These approaches focus only on the
largest scales of motion and model all turbulent scales via additional terms in
the momentum equations called Reynolds stress terms. Although their ability to
predict transitional and relaminarizing types of flows has been
criticized\cite{mittal-etal-2001}, RANS methods are still popular in the context
of cardiovascular simulations.
Recent studies focused, e.g., on evaluation of aortic WSS in a phantom model
of aortic coarctation\cite{perinajova-etal-2021}, as well as on the anisotropy
of turbulent blood flow in patient-specific
settings\cite{andersson-kalrsson-2021}.

As an alternative to RANS, Large Eddy Simulation (LES) methods attempt to model
the large turbulent scales, applying a convolutional low-pass filter to the
Navier--Stokes equations and surrogating the effect of the small scales into
explicit models for the stress tensor.
Widely used models in this class are the original Smagorinsky model\cite{Sma63},
as well as the Vreman\cite{vreman_eddy-viscosity_2004} and
Nicoud\cite{nicoud_using_2011} $\bsig$-viscosity models.
In recent relevant works, a LES $\bsig$-model was used to investigate the impact
of turbulence in the context of abdominal aortic
aneurysms\cite{vergara-etal-2017}, while a LES Leroy decay model was used to
study the sensitivity of simulated WSS in the
aorta\cite{xu-baroli-veneziani-2021}.

A conceptually different turbulence model considered in this paper is the
residual-based variational multiscale approach\cite{bazilevs_variational_2007}
(RB-VMS). This method is based on a two-scale decomposition of the analytic
function spaces for velocity and pressure, where the fine space represents the
scales which cannot be represented by the considered finite element
discretization. The influence of these scales is then surrogated in the coarse
dynamics using additional terms in the variational formulation. RB-VMS methods
have shown promising results in recent studies of turbulent channel
flows\cite{AJ20}.
To the best of our knowledge, variational multiscale models have not yet been
investigated in the context of blood flow simulation, nor are detailed studies
comparing different choices available in the literature.

The question studied in this paper can be formulated as follows: using a
reasonably fine computational mesh which is still affordable from the point of
view of computing times, how much do results differ for several clinically
relevant quantities of interest if different turbulence models are used in the
discretization? This question addresses the common practical situation where a
mesh of the domain is given, which is chosen fine in order to obtain accurate
results, but any (uniform) refinement of the mesh is prohibitive due to the
increasing computational costs.

The purpose of this work is to investigate in detail the impact of turbulence
modeling on the simulation of blood flow in an aortic coarctation. In
particular, LES models and and RB-VMS models are considered. Exemplarily for
the RB-VMS model, the impact of the order of the finite element velocity
space is investigated.
For the comparison we focus on selected quantities of interest which are
commonly used to characterize abnormal or pathological flow conditions such as
the variation of pressure along the aorta, the secondary flow degree, and the
normalized flow displacement\cite{vanderpalen-etal-2020}. Furthermore, the
sensitivity of the wall shear stress (WSS) and of the related oscillatory shear
index (OSI) are analyzed, as these biomarkers have been linked with the
deposition of atheromatous plaque in blood vessels\cite{feng-2005}.

The numerical simulations are based on an aortic geometry obtained from medical
imaging, with patient-specific boundary conditions defined using available data.
In particular, measurements are limited to a space-dependent cardiac outflow
profile, which is prescribed as Dirichlet inlet boundary condition, and peak
outflow rates on the brachiocephalic artery, the left carotid, the left
subclavian, and the descending aorta, which are used to tune lumped parameter
models.
To this purpose, purely resistive outflow boundary conditions are used, and a
sequential approach for the estimation of boundary condition parameters based on
the available flow rates is proposed.

The remainder of the article is structured as follows.
Section~\ref{sec:materials} describes the available data which were used to
build the computational model and the simulation setup. Section~\ref{sec:model}
introduces the blood flow model, the numerical methods, the proposed approach
for estimating boundary parameters, and the considered turbulence models.
The results are presented in Section~\ref{sec:numres}, while
Section~\ref{sec:conclusions} summarizes the conclusions.

\section{Materials}\label{sec:materials}

Available data were acquired on a $\unit[1.5]{T}$ clinical MR system (Achieva;
Philips Healthcare, Best, Netherlands) with a five-element cardiac phased-array
coil\cite{schubert-etal-2020}. Within the cardiac MRI protocol velocity-encoded
MRI (4D VEC MRI) was acquired in planes perpendicular to the ascending aorta
distally to the valve and in the descending aorta at the level of the diaphragm
to assess inflow conditions in three flow encoding orientations and outflow
towards the abdominal aorta.

\begin{figure}[t!]
  \centerline
  {
    \includegraphics[height=0.35\textwidth]{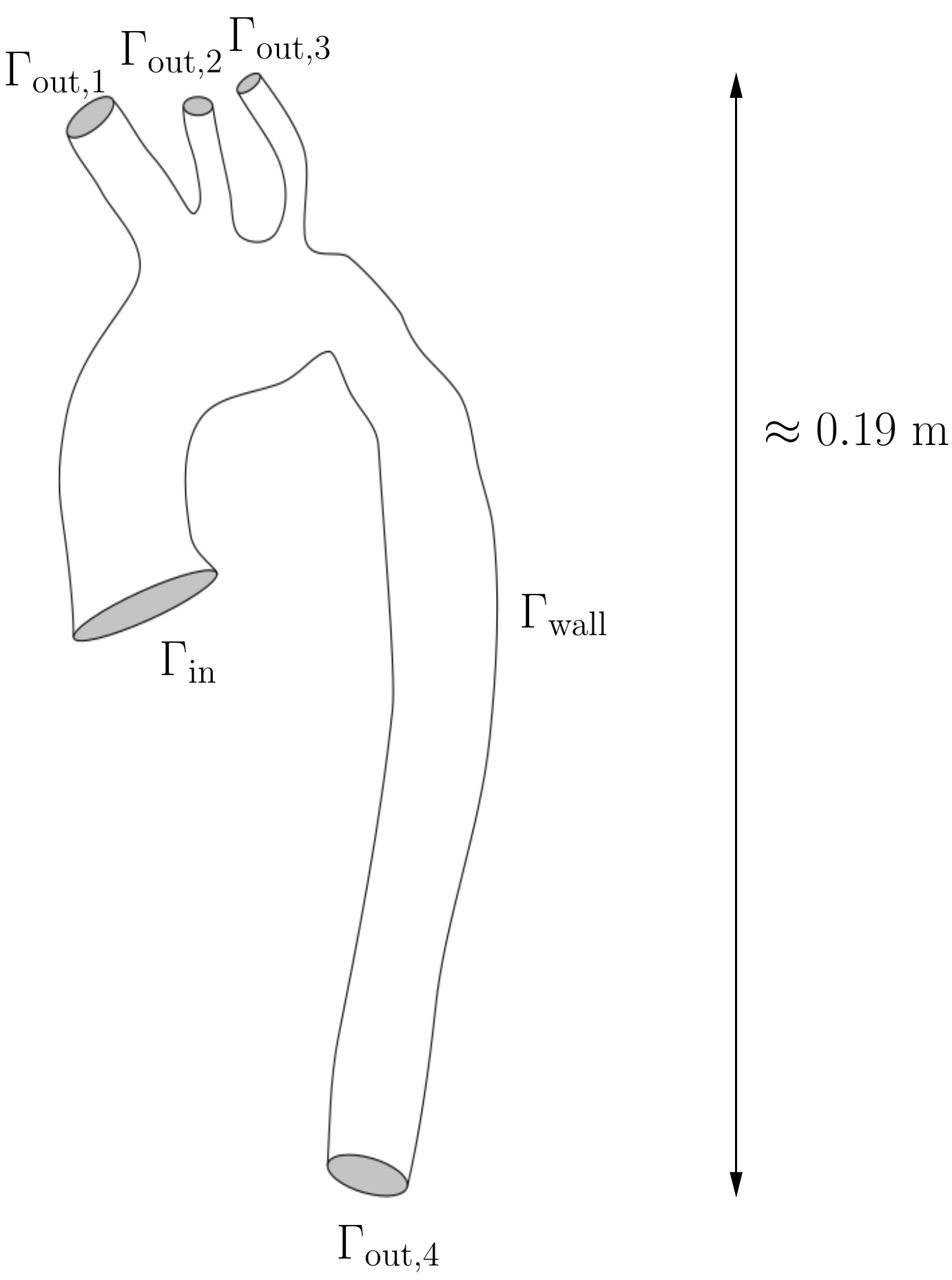}\hspace*{3em}
    \includegraphics[height=0.35\textwidth]{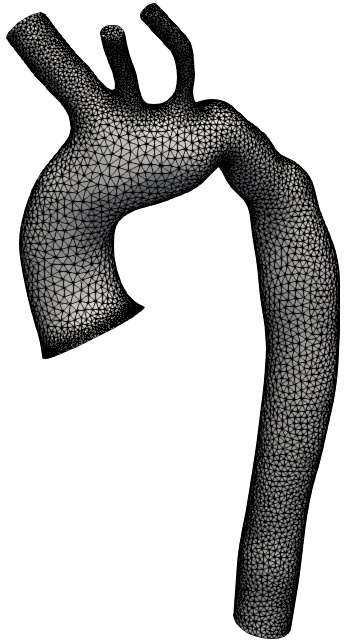}\hspace*{2em}
    \includegraphics[height=0.15\textwidth]{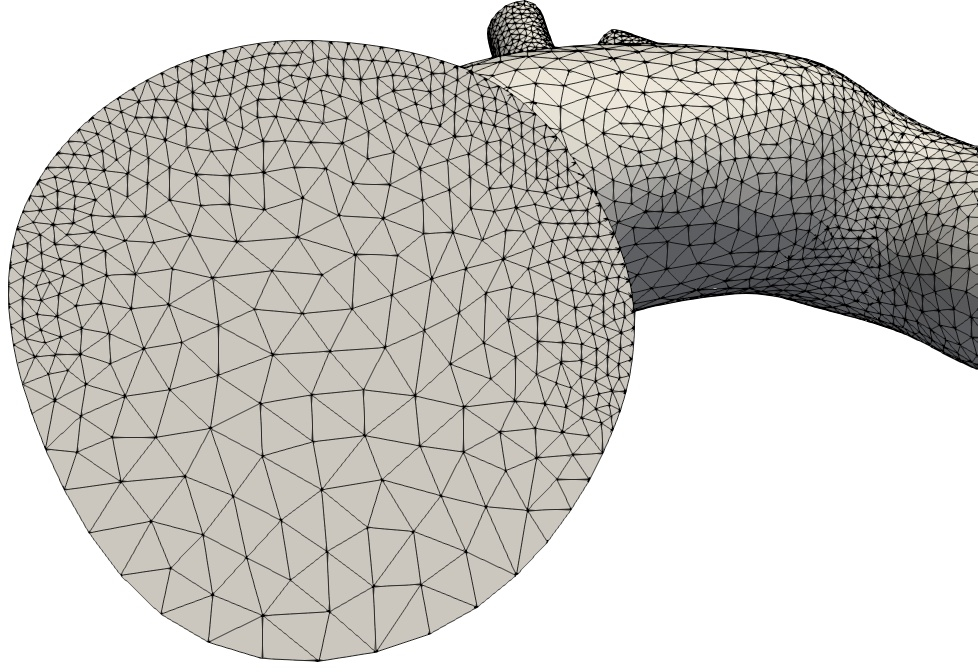}
  }
  \caption[Computational domain]
  {
    Left: Sketch of the computational domain, representing a segment of the
      aorta, and the corresponding decomposition of the boundary.
    Center and right: Surface mesh (coarser version), and zoom on the inlet
      boundary $\Gamma_{\mathrm{in}}$.
  }
  \label{fig:domain_sketch}
\end{figure}

The considered domain (Figure \ref{fig:domain_sketch}, left) for the numerical
simulation consists of a portion of the aorta from the sino-tubular junction to
the descending aorta at the level of the left ventricular apex (about
$\unit[20]{cm}$ length).

The computational mesh was obtained by segmenting the anatomy of the aorta based
on the diastolic 3D SSFP cine images using ZIBAmira (v. 2015.28, Zuse Institute
Berlin, Germany), as in a previous work\cite{goubergrits-2013}.
From the obtained surface triangular mesh, Figure \ref{fig:domain_sketch}, center
and right, a tetrahedral volume mesh was generated using
{\sf TetGen}\cite{tetgen} prescribing the maximal allowed volume of the
tetrahedra.
The resulting mesh $\mathcal T$ consisted of $106,983$ volume elements. For the
computation, also a uniform refinement $\mathcal T^\prime$ was utilized,
composed of $855,864$ tetrahedra. Table~\ref{tab:grid_stat} shows further mesh
statistics, notably the maximum boundary layer height $y_\mathrm{max}$ and the
area-weighted average boundary layer height $\bar y$, as defined by the height
above each boundary face of the single adjacent tetrahedron.

On the one hand, the meshes were chosen to be
fine in order to perform simulations with small spatial errors. But on the other
hand, they were chosen to be sufficiently coarse in order to perform simulations
in affordable computing times. Altogether, one encounters here a situation
typical in practice, namely that any further (uniform) refinement of the meshes,
which increases the computational costs at least by a factor of eight, is
prohibitive from the point of view of computing times. It should be also noted
that on the coarser mesh most turbulence models are applied with second order
finite elements for the velocity, so that the resolution in this respect
corresponds to the mesh width of the fine mesh. Since most quantities of
interest studied in this paper are based on the bulk flow, at different
locations of the aorta segment, we decided to use rather uniform meshes for the
whole domain, without special local adaptions.

In order to setup the numerical simulations, the patient-specific peak systolic
velocity vector profiles measured using planar 4D VEC MRI were mapped onto the
nodes of the meshed inlet boundary $\Gamma_{\rm in}$ using a linear
interpolation scheme.
The volume flow in the descending aorta was prescribed according to the MRI
measurements. The flow difference between ascending and descending aorta was
then distributed in the remaining outlets using the following assumptions:
(i) the volume flow within the brachiocephalic artery (right arm and head)
equals the volume flowing in both the left common carotid artery and left
subclavian artery, and
(ii) the flow distribution in the left common carotid artery and left subclavian
artery depends on the cross sectional areas of the outlets according to Murray's
law, i.e., $Q_i = \frac{D_i^3}{D_2^3 + D_3^3}$, where $i=2,3$, and $Q_i$ stands
for the flow rate in outlet $\Gamma_{{\rm out},i}$. The resulting flow rates
$Q_i^*$, $i =1,\hdots,4$, on the four outlet open boundaries (see
Table~\ref{tab:flowsplit}) were used to define the lumped parameter models used
as boundary conditions for the CFD simulation.

\section{Methods}\label{sec:model}

\subsection{Blood Flow Modeling}\label{sec:flow_problem}
Let $\Omega \subset \mathbb R^3$ denote the computational domain and decompose
its boundary $\partial \Omega$ as
\[
  \partial \Omega = \Gamma_{\mathrm{in}} \cup \Gamma_{\mathrm{wall}}
  \cup \Gamma_{\mathrm{out},1} \cup \ldots \cup \Gamma_{\mathrm{out},4},
\]
(following the notation introduced in Figure~\ref{fig:domain_sketch}, left).
The inlet boundary $\Gamma_{\mathrm{in}}$ is situated close to the left
ventricle, the arterial wall is denoted by $\Gamma_{\mathrm{wall}}$, and
$\Gamma_{\mathrm{out},1}, \ldots, \Gamma_{\mathrm{out},4}$ denote the artificial
outlet boundaries created by cutting the physical domain and neglecting the
downstream circulation.

In the considered physiological regime, the blood flow in $\Omega$ is modeled as
an incompressible, Newtonian fluid, whose dynamics is described by the
incompressible Navier--Stokes equations in terms of a velocity field
$\bu\ [\unitfrac{m}s]: \Omega \to \mathbb R^3$ and a pressure field
$p \ [\unit{Pa}]: \Omega \to \mathbb R$ satisfying the system of equations
\begin{equation}\label{eq:nse}
\begin{array}{rcll}
\rho \partial_t \bu - 2\mu \nabla\cdot \mathbb D(\bu) + \rho (\bu \cdot \nabla) \bu + \nabla p & = & \boldsymbol 0 &
\mbox{in } (0,T] \times \Omega, \\
\nabla \cdot \bu & = & 0 & \mbox{in } (0,T] \times \Omega.
\end{array}
\end{equation}
In \eqref{eq:nse}, $T\ [\unit{s}]$ is the final simulation time,
$\rho = \unitfrac[1060]{kg}{m^3}$ stands for the blood density,
$\mu = \unit[3.5\cdot 10^{-3}]{Pa\cdot s}$ is the dynamic viscosity, and
$\mathbb D(\bu) = \big(\nabla \bu + (\nabla \bu)^T\big) / 2$ denotes the
velocity deformation tensor (i.e., the symmetric part of the velocity gradient).

The characteristic peak velocity scale of the blood velocity in the ascending
aorta is of the order of $U = \unitfrac[\mathcal O(1) \,]{m}s$. Using the
diameter of the aorta $L = \unit[0.03]{m}$ as characteristic length scale, the
Reynolds number of the flow is
\[
\mathrm{Re} = \frac{\rho LU}{\mu} \approx 9086,
\]
which indicates a turbulent flow.

For deriving the non-dimensional equations used in the numerical simulations, a
characteristic length scale of $\tilde L = \unit[1]{m}$ was utilized, leading to
the dimensionless viscosity coefficient
\begin{equation*}
\nu = \frac{\mu}{\rho \tilde LU} \approx 3.3\cdot10^{-6}.
\end{equation*}
Dividing \eqref{eq:nse} by $\rho$ and using the dimensionless viscosity
coefficient, the time-dependent incompressible Navier--Stokes equations can be
written in fully dimensionless form:
\begin{equation}\label{eq:dimless_nse}
\begin{array}{rcll}
\partial_t \bu - 2\nu \nabla\cdot \mathbb D(\bu) + (\bu \cdot \nabla) \bu + \nabla p & = & \boldsymbol 0 &
\mbox{in } (0,T] \times \Omega, \\
\nabla \cdot \bu & = & 0 & \mbox{in } (0,T] \times \Omega.
\end{array}
\end{equation}
In what follows, with a slight abuse of notation, $\bu$ and $p$ will be used
also to denote the (dimensionless) velocity and pressure fields.

\subsubsection{Initial and Boundary Conditions}

\begin{figure}[t!]
	\centering
	\includegraphics[width=0.4\textwidth]{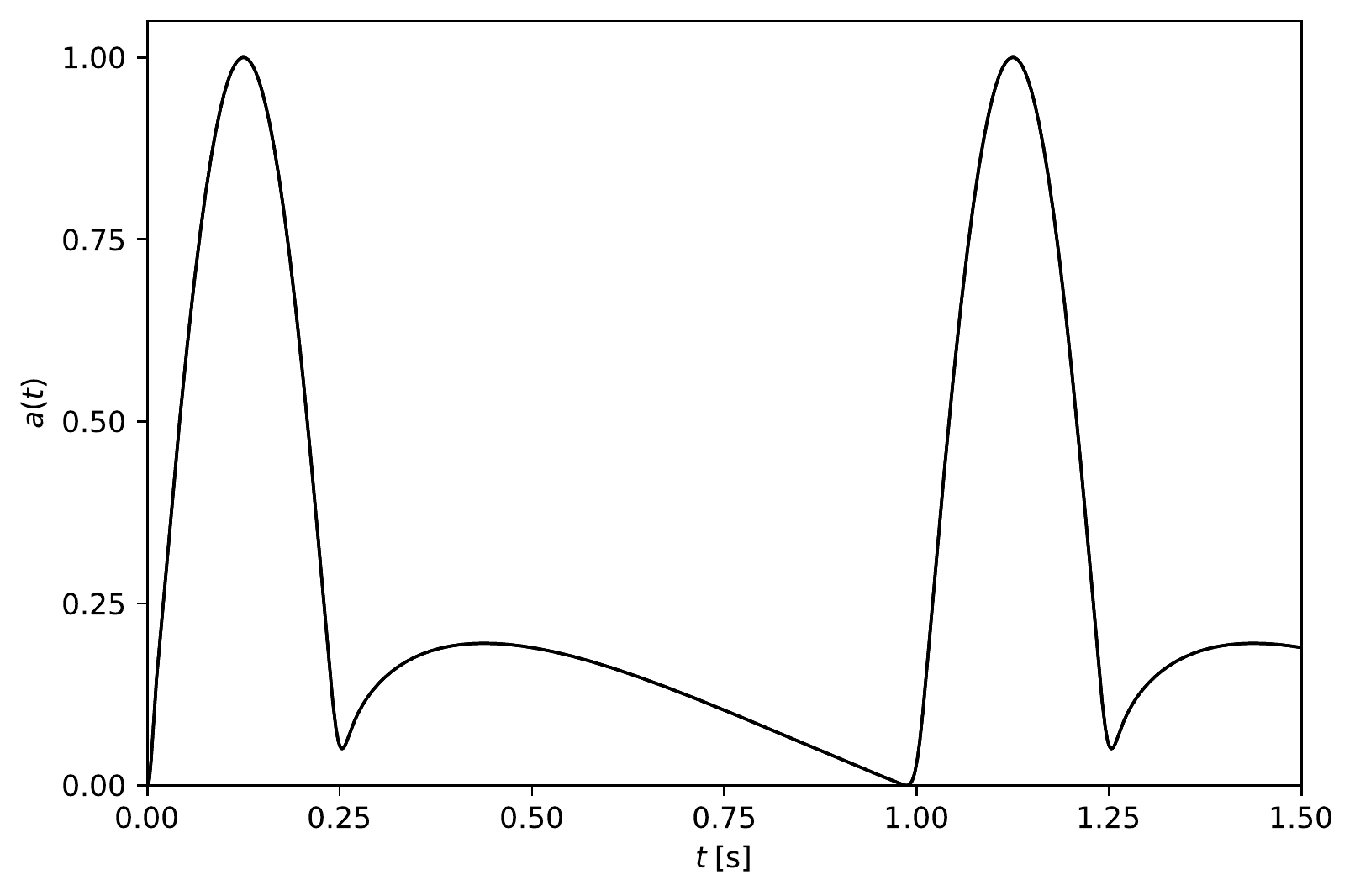}
	\caption{The inflow pulse profile $a(t)$ plotted over the first $1.5$ seconds.}
	\label{fig:inflow_amplitude}
\end{figure}

The system of equations \eqref{eq:nse} is completed by the following initial and
boundary conditions, whose definition is motivated by the availability of data:
\begin{equation}\label{eq:nse_bc}
\begin{array}{rcll}
\bu(0,\bx) &=& \boldsymbol{0} &\mbox{in } \Omega,\\
\bu(t,\bx) &=& \bu_\mathrm{in}(t, \bx) := a(t) \bu^*_\mathrm{in}(\bx) &\mbox{on } [0,T] \times \Gamma_{\mathrm{in}},\\
\bu(t,\bx) & = & \boldsymbol 0 &\mbox{on } [0,T] \times \Gamma_{\mathrm{wall}}, \\
2\mu \mathbb D(\bu) \bn - p \bn & = & -P_i(\bu) \bn & \mbox{on } [0,T] \times \Gamma_{\mathrm{out},i}, \ i=1,\ldots,4.
\end{array}
\end{equation}
In \eqref{eq:nse_bc}$_2$, $a: \mathbb R \to \mathbb R$ is a smooth function such
that $a(0) = 0$ and which becomes periodic with period $T_0 = \unit[1]{s}$ after
a given time $t_0 = \unit[0.01]{s}$, i.e., $a(t + T_0) = a(t)$, for all
$t > t_0$. (see Figure~\ref{fig:inflow_amplitude}).
Equation \eqref{eq:nse_bc}$_3$ models the arterial wall as rigid, while the
Neumann boundary conditions \eqref{eq:nse_bc}$_4$ are imposed via lumped
parameter models $P_i(\bu)$, $i=1,\hdots,4$, which shall be defined in order to
obtain a simulation setup in agreement with the measured outlet flow rates.

The selection of the boundary conditions at the outlets has to take into account
the amount of available data. Since only outflow measurements at systole are at
hand, one has to choose a model whose parameters can be determined with these
data. The boundary conditions on the outlets are thus defined by the purely
resistive model
\begin{equation} \label{eq:bc_resistive}
P_i(t) = R_i Q_i(t), \quad i = 1, \ldots, 4,
\end{equation}
where
\begin{equation*} 
Q_i := \int_{\Gamma_{\mathrm{out},i}} \bu \cdot \bn\ d\boldsymbol s,
\end{equation*}
for $i=1,\hdots,4$, denotes the outgoing flow through the outlet
$\Gamma_{\mathrm{out},i}$. The iterative approach for defining the boundary
resistances will be discussed in more detail in Section~\ref{ssec:resistive}.

\begin{remark}
Model \eqref{eq:bc_resistive} does not take into account more complex
interactions with the downstream circulation; an obvious option would be a
Windkessel model with 3 or 4 elements. However, increasing the model complexity
requires additional parameters and assumptions, which cannot realistically be
adjusted to a patient-specific context without additional data.
\end{remark}

\subsubsection{Outflow stabilization}

It is well known that blood flow simulations of aortic flow might be affected by
backflow instabilities, i.e., spurious oscillations at the open boundaries, when
the flow is directed ``back'' into the computational
domain\cite{caiazzo-bertoglio-benchmark-2015}. To overcome this issue, a
directional do-nothing
condition\cite{bruneau_new_1996,braack_directional_2014} is considered, which
can be seen as a modification of the Neumann boundary conditions
\eqref{eq:nse_bc}$_3$ of the form
\begin{equation*} 
\big(2\mu \mathbb D(\bu) - p I\big) \bn = -P_i(t) \bn + \frac{\beta} 2 (\bu \cdot \bn)_{-} \bu
\mbox{ on } [0,T] \times \Gamma_{\mathrm{out},i},
\end{equation*}
$i = 1,\ldots,4$, where $(\bu \cdot \bn)_{-} = \min \{\bu \cdot \bn, 0\}$ is the
negative part of the boundary velocity's normal component.
This approach has been extensively used in computational
hemodynamics\cite{bazilevs-2009,moghadam-2011} and has been shown to be a
natural means to control a priori energy
estimates\cite{braack_directional_2014} for $\beta = 1$.
Alternative backflow stabilization approaches have been recently proposed,
considering, among others, tangential regularization of the boundary
flow\cite{bertoglio-caiazzo-2014}, stabilization based on the residue of a
surrogate Stokes problem \cite{bertoglio-caiazzo-2016}, or on a rotational
velocity correction\cite{dong_convective-like_2015}. The interested reader is
referred to a benchmark study\cite{caiazzo-bertoglio-benchmark-2015} and the
references therein.

\subsection{Spatial and Temporal Discretizations}\label{ssec:disc}

The system of equations \eqref{eq:dimless_nse} is discretized in space using a
finite element method. In order to introduce the formulation, let $\mathcal T_h$
denote the considered regular tetrahedral mesh, let $h$ be its characteristic
size, and let $\bX_h$ and $Q_h$ denote continuous piecewise polynomial spaces
defined on $\mathcal T_h$.
Furthermore, let
\[
\bX_{h,0} = \left\{ \bv_h \in \bX_h : \bv_h \equiv \boldsymbol 0 
\text{ on } \Gamma_{\mathrm{in}} \cup \Gamma_{\mathrm{wall}} \right\}
\]
be the subspace of $\bX_h$ including the essential boundary conditions on the
inlet boundary and the arterial wall.

Let us introduce the nonlinear form
\begin{equation*} 
 \mathrm{Gal}\Big((\bu_h, p_h), (\bv_h, q_h)\Big) :=
 \nu a(\bu_h, \bv_h) + b(\bu_h; \bu_h, \bv_h) - (\nabla \cdot \bv_h,p_h) + (\nabla \cdot \bu_h, q_h) - f(\bu_h, \bv_h)
\end{equation*}
with
\begin{equation*} 
a(\bu_h, \bv_h) := 2 \left(\mathbb{D}(\bu_h), \mathbb{D}(\bv_h)\right), \quad 
b(\bu_h; \bv_h, \bw_h) := \big( (\bu_h \cdot \nabla) \bv_h, \bw_h \big),
\end{equation*}
and
\begin{equation*} 
f(\bu_h, \bv_h) := \sum_{i=1}^4 \left(P_i(\bu_h) \bn + \frac 1 2 (\bu_h \cdot \bn)_{-} \bu_h, \bv_h \right)_{\Gamma_{\mathrm{out},i}}\,.
\end{equation*}

The standard Galerkin discrete formulation of \eqref{eq:dimless_nse} reads: Find
$(\bu_h, p_h): [0, T] \to \bX_h \times Q_h$ such that $\bu_h \equiv \boldsymbol 0$ on
$[0,T] \times \Gamma_{\mathrm{wall}}$, $\bu_h$ satisfies \eqref{eq:nse_bc}$_1$,
and
\begin{equation} \label{eq:nse_gal}
\left(\partial_t \bu_h, \bv_h\right) = -\mathrm{Gal} \Big((\bu_h, p_h), (\bv_h, q_h)\Big)
\end{equation}
for all $t \in (0, T)$ and for all $(\bv_h, q_h) \in \bX_{h,0} \times Q_h$.

Equation \eqref{eq:nse_gal} is discretized in time using a BDF-2 scheme, an
A-stable second-order method that has the advantage of requiring residuals of
only one time step. Since the scheme requires the solution at two previous time
instants, the first time iteration is performed using a backward Euler method.

At each time step, a nonlinear problem in the velocity and pressure has to be
solved. This is handled via a Picard method. Namely, the convective term and the
boundary condition term in \eqref{eq:nse_gal} are linearized using the velocity
field $\hat{\bu}_h$ computed at the last iteration, thus resulting in a linear
system with the following linearization of the Galerkin term at each Picard
iteration
\begin{equation} \label{eq:nse_gal_linear}
\widehat{\mathrm{Gal}}\left((\bu_h, p_h), (\bv_h, q_h)\right) =
\nu a(\bu_h, \bv_h) + b(\hat{\bu}_h; \bu_h, \bv_h) - (\nabla \cdot \bv_h,p_h) + (\nabla \cdot \bu_h, q_h) - f(\hat{\bu}_h, \bv_h),
\end{equation}
where the solution $(\bu_h,p_h)$ denotes the next iterate.

\begin{remark}
The linear problems to be solved in each Picard iteration have a saddle point
structure. As will be shown in Section~\ref{ssec:turbmod}, the same structure is
retained for each considered turbulence model, with the exception of the RB-VMS
model discussed in Section~\ref{sssec:rbvms}.
\end{remark}

\subsection{Estimation of boundary resistances} \label{ssec:resistive}
The outlet resistances $R_i, \; i = 1, \hdots ,4$, were tuned in order to obtain
simulated outflow rates $Q_i$ at systole close to the measured reference values
$Q_i^*$ given in Table~\ref{tab:flowsplit}. To this purpose, a sequential
estimation approach was implemented, in which the parameter values are optimized
during the time iteration depending on the difference between simulated values
and available data.

The approach is motivated by two observations. Firstly, considering a surrogate
0D model of the downstream circulation, the systemic vascular resistance (SVR),
the resistance to blood flow offered by all of the systemic vasculature
excluding the pulmonary tract and the small resistance of the upper aorta
itself, can be defined by the relation
\begin{equation}\label{eq:rsv}
 \RSV = \left( \sum_{i=1}^4 R_i^{-1} \right)^{-1} = \left( \sum_{i=1}^4 G_i \right)^{-1},
\end{equation}
where $G_i := R_i^{-1}$ denote the outlet conductances. Secondly, mass
conservation guarantees
\begin{equation}\label{eq:q_i_sum}
\sum_{i=1}^4 Q_i(t) = \sum_{i=1}^4 Q_i^* = Q_{\rm in} \, .
\end{equation}
On the one hand, mass conservation implies therefore that the four available
measurements are not independent. On the other hand, if the inlet flow is
constant in time, one obtains
\begin{equation}\label{eq:dQ_i_sum}
\sum_{i=1}^4 \dot{Q}_i(t) = 0.
\end{equation}

For a given value of the systemic vascular resistance $\RSV$\footnote
{
  See Section~\ref{ssec:svr_impact} for our choices.
},
the parameter estimation method is based on the solution of a Navier--Stokes problem
\begin{equation}\label{eq:nse-estim}
\begin{array}{rcll}
\rho \partial_t \bu - 2\mu \nabla\cdot \mathbb D(\bu) + \rho (\bu \cdot \nabla) \bu + \nabla p & = & \boldsymbol 0 &
\mbox{in } (0,T] \times \Omega, \\
\nabla \cdot \bu & = & 0 & \mbox{in } (0,T] \times \Omega \\
\bu(t, \bx) = \bu_\mathrm{in}(t, \bx) &= &\hat a(t) \bu^*_\mathrm{in}(\bx), & \mbox{for}\;(t, \bx) \in [0, T] \times \Gamma_{\mathrm{in}}, \\[0.2em]
2\mu \mathbb D(\bu) \bn - p \bn & = & -G_i(t)^{-1}Q_i(t) \bn & \mbox{on } [0,T] \times \Gamma_{\mathrm{out},i}, \ i=1,\ldots,4, \\
\bu(0, \bx) & = & \bu_0(\bx) & \mbox{in } \Omega\,,
\end{array}
\end{equation}
coupled to an additional ODE
for the conductances:
\begin{equation}\label{eq:G-estim}
\dot{G}_i(t)  = \frac{\gamma_0}{Q_{\rm in}} \left( Q_i^* - Q_i(t) \right), \quad
G_i(0) = G_i^0,\; i=1,\hdots,4,
\end{equation}
and with the additional condition \eqref{eq:rsv}, i.e.,
\begin{equation}\label{eq:rsv-estim}
\sum_{i=1}^4 G_i^0 = \RSV^{-1}\,.
\end{equation}
Equation \eqref{eq:rsv-estim} is imposed in order to overcome the dependency of
the outlet measurements stated in \eqref{eq:q_i_sum}.
In \eqref{eq:nse-estim}, the function $\hat a(t)$ defines a smooth transition to
a constant profile, i.e., it is such that $\hat a(0) = 0$,
$\mathrm d_t \hat a(0) = 0$, and $\hat a(t) = 1$ for all
$t > t_1 = \unit[0.05]{s}$\,.
In particular, it follows from \eqref{eq:dQ_i_sum} that, when the inflow is
constant (for $t > t_1$), the sum of conductances also remains constant over
time. In \eqref{eq:G-estim}, $\gamma_0$ is a positive parameter and the initial
values of $G_i^0$ can be obtained, e.g., by successive simulations with
decreasing viscosity.

If \eqref{eq:nse-estim}, \eqref{eq:G-estim} reaches a steady state, then
$Q_i^* = Q_i(t)$ for $i=1,\hdots,4$, and the corresponding stationary values of
$R_i = G_i^{-1},\,$ $i = 1,\hdots,4$, can be used for the blood flow simulation.
In practice, due to the presence of turbulence, the values for the resistances
are defined taking a suitable long-term average of the solution once a
quasi-periodic state has been reached.

Once the outlet resistances $R_i = R_i^*(\RSV)$ have been determined for a
certain $\RSV$ solving \eqref{eq:nse-estim}-\eqref{eq:rsv-estim}, the values
$R_i^\prime = R_i^*\left(\RSV^\prime\right)$ for a different systemic vascular
resistance $\RSV^{\prime}$ have been computed considering that the difference
between $\RSV$ and $\RSV^{\prime}$ induces a shift in the overall blood pressure
in $\Omega$. Namely, let us denote by $\left(\bu,p\right)$ and $G_1,\hdots,G_4$
the solutions to \eqref{eq:nse-estim}, \eqref{eq:G-estim}, and
\eqref{eq:rsv-estim}.
Then, there exists a constant $\Delta P$ such that
$\left(\bu,p + \Delta P\right)$ satisfies $\eqref{eq:nse-estim}$ at the steady
state with $G_i^{\prime} = (R_i^{\prime})^{-1}$, and
\begin{equation*}
R_i^{\prime} = R_i + \frac{\Delta P}{Q_i^*}, \; i=1,\hdots,4,
\end{equation*}
i.e., such that the pressure at each outlet increases by $\Delta P$ when
$Q_i = Q_i^*$.
Hence, the value of $\Delta P$ can be computed from equation
\eqref{eq:rsv-estim} as a function of $\RSV^{\prime}$ solving
\begin{equation}\label{eq:RSV-dP}
\dfrac{1}{\RSV^{\prime}} = \sum_{i=1}^4 \frac{1}{R_i^\prime}
= \sum_{i=1}^4 \frac{1}{R_i + \frac{\Delta P}{Q_i^*}} \,.
\end{equation}
It can be shown that $\RSV^{\prime}$ in \eqref{eq:RSV-dP} is a smooth and
monotonous function of $\Delta P$, for $\Delta P > -\min_i Q_i^* R_i$.
Moreover, since
\[
\RSV^\prime (\Delta P) \to \infty \text{ for } \Delta P \to \infty
\quad \mbox{ and } \quad
\RSV^\prime (\Delta P) \to 0 \text{ for } \Delta P \to -\min_i Q_i^* R_i,
\]
one can conclude that there exists a unique $\Delta P$ that satisfies
\eqref{eq:RSV-dP} for a given $\RSV^{\prime}$, or, equivalently, such that
$R_i^\prime Q_i^* =Q_i^* R_i + \Delta P$ for all outlets.

In practice, the approach delivered satisfactory results for moderate values of
$\gamma_0$ (see Section~\ref{ssec:svr_impact} for more details) and with a
negligible influence of the discretization used. However, rigorous convergence
estimates are out of the scope of this work.%

\begin{remark}[Average blood pressure]
Prescribing the systemic vascular resistance is equivalent to fixing the
pressure constant for the solution of the Navier--Stokes problem
\eqref{eq:nse-estim}. In fact, since $\left(\bu,p + \Delta P\right)$ satisfies
\eqref{eq:nse-estim} for the new value of the systemic vascular resistance,
$\Delta P$ in equation \eqref{eq:RSV-dP} determines also the shift in the
average blood pressure in the considered aortic segment.
\end{remark}

\subsection{Turbulence Modeling} \label{ssec:turbmod}

This section provides a brief presentation of the turbulence models that were investigated in the
numerical studies.

\subsubsection{The Smagorinsky model}\label{sssec:smago}

The Smagorinsky model\cite{Sma63} is certainly one of the most popular LES
models, but also one of the simplest. The model relies on the Boussinesq
hypothesis that the effect of small-scale fluctuations on large-scale flow
behavior is mostly dissipative. Motivated by this assumption, the deviatoric
part of the subgrid-scale (SGS) stress tensor $\boldsymbol \tau$,
\begin{equation*} 
\tau_{ij} = \overline{u_i u_j} - \overline{u}_i \overline{u}_j,
\end{equation*}
is modeled by a nonlinear scalar multiple of the velocity deformation tensor
$\mathbb D(\bu)$, i.e.,
\begin{equation} \label{eq:les_eddy_viscosity}
\tau_{ij} \approx -2 \nu_t \mathbb D(\bu)_{ij} + \delta_{ij} \frac {\mathrm{trace}(\boldsymbol \tau)} 3,
\end{equation}
with a suitable turbulent or eddy viscosity $\nu_t = \nu_t(\bu)$.
Equation \eqref{eq:les_eddy_viscosity} results in an additional nonlinear
viscous term in the momentum balance of the Navier--Stokes equations
\eqref{eq:nse}, which has the form
\begin{equation*} 
-2 \nu_t \nabla \cdot \mathbb D(\bu),
\end{equation*}
while the trace part of the SGS stress tensor is hidden in the modified filtered
pressure
\[
\tilde p = p + \frac {\mathrm{trace}(\boldsymbol \tau)} 3,
\]
requiring no further modification of \eqref{eq:nse}.

In the semidiscrete formulation \eqref{eq:nse_gal}, the Smagorinsky model
amounts to adding the term $-(\nu_t \mathbb D(\bu_h), \mathbb D(\bv_h))$ to the
right-hand side. Linearization for the Picard iteration is achieved by computing
$\nu_t$ from the current iterate $\hat{\bu}_h$.

The Smagorinsky model uses dimensional arguments at Kolmogorov scales to
arrive at the Smagorinsky eddy viscosity
\begin{equation} \label{eq:smago}
\nu_t = C_{\mathrm{Sma}} \delta^2 \|\mathbb D(\bu)\|_{\mathrm{F}}.
\end{equation}
In \eqref{eq:smago}, $\|\mathbb D(\bu)\|_{\mathrm{F}}$ is the Frobenius norm of
the velocity deformation tensor, $\delta$ is a local length scale, and
$C_{\mathrm{Sma}}$ is a user-chosen constant. The advantages and drawbacks of
the Smagorinsky model for practical simulations are well
known\cite[Chapter~5.3]{Sag06}. From the mathematical point of view, concerning
questions such as existence and uniqueness of a weak solution and finite element
error analysis, the Smagorinsky model belongs to the best understood turbulence
models\cite[Chapter~8.3]{Joh16}.

For the simulations presented in this paper, the local length scale was chosen
to be piecewise constant on each mesh element, i.e.,
$\delta = 2h_{K,\mathrm{sh}}$, where $h_{K,\mathrm{sh}}$ is the length of the
shortest edge of mesh cell $K$.
This choice of $\delta$ proportional to the shortest edge has been proven to be
better than other approaches, e.g. using the diameter of the cell\cite{JR07}.
%
The constant $C_{\mathrm{Sma}}$ is a free parameter of the model. Different values
$C_{\mathrm{Sma}} \in \{0.01, 0.005\}$ were investigated in our simulations.

\subsubsection{The Vreman model}\label{sssec:vreman}

The Vreman model\cite{vreman_eddy-viscosity_2004} proposes an alternate eddy
viscosity. It is motivated by the Smagorinsky model's excessively dissipative
behavior in laminar and transitional flows, including shear flows near walls.
Using algebraic arguments, involving the classification of local flow behaviors
for which the subgrid dissipation
\[
D_\tau = -\boldsymbol \tau : \nabla \bu
\]
vanishes compared to various functionals of the velocity gradient field, the
following form of eddy viscosity is considered:
\begin{equation} \label{eq:vreman}
\nu_t = \frac {C_{\mathrm{Vre}}} {\|\nabla \bu\|_F} \sqrt{B_\beta}.
\end{equation}
In \eqref{eq:vreman}, $\|\nabla \bu\|_F$ is the Frobenius norm of the velocity
gradient, $C_{\mathrm{Vre}}$ is a free parameter, and
\[
B_\beta = \frac 1 2 \sum_{i,j = 1}^3 \left(\beta_{ii} \beta_{jj} - \beta_{ij}^2\right)
= \sum_{1 \leq i < j \leq 3} \left(\beta_{ii} \beta_{jj} - \beta_{ij}^2\right),
\]
where
\[
\beta_{ij} = \sum_{k=1}^3 \delta_k^2 \partial_k u_i \partial_k u_j\,,
\]
is a rotational invariant of the symmetric positive definite tensor
$\boldsymbol \beta = \nabla \bu \boldsymbol \delta^2 (\nabla \bu)^T$ with
anisotropic filter widths
\[
\boldsymbol \delta = \begin{pmatrix}
\delta_1 &        0 &        0 \\
       0 & \delta_2 &        0 \\
       0 &        0 & \delta_3
\end{pmatrix}.
\]
Note that if $\boldsymbol \beta$ 
has eigenvalues $\lambda_1, \lambda_2, \lambda_3$, then
\[
B_\beta = \lambda_1 \lambda_2 + \lambda_1 \lambda_3 + \lambda_2 \lambda_3.
\]
The $k$-th length scale $\delta_k$ is again chosen piecewise constant. On each
mesh cell $K$, $\delta_k = \delta_k(K)$ is the width of $K$ in the $k$-th
coordinate direction:
\[
\delta_k(K) = \max_{\bx, \bx^\prime \in K} \left|x_k - x_k^\prime\right|.
\]
In regions where $\nabla \bu$ is (nearly) zero, the eddy viscosity is taken to
be zero. This choice is consistent:
$|\boldsymbol \beta| \lesssim \|\nabla \bu\|^2$, so
$\sqrt{B_\beta} \lesssim \sqrt{|\boldsymbol \beta|^2} \lesssim \|\nabla \bu\|^2$
and $\nu_t \lesssim \|\nabla \bu\|$, as in the Smagorinsky model.

Using the Vreman model, the flow in the considered segment of the aorta was
simulated with $C_{\mathrm{Vre}} = 0.07$, i.e., the value suggested by
Vreman\cite{vreman_eddy-viscosity_2004} based on scaling arguments.

\subsubsection[The sigma-model]{The $\bsig$-model}\label{sssec:sigma}

The $\bsig$-model\cite{nicoud_using_2011}, developed by Nicoud et al.,
is an eddy viscosity model motivated by similar arguments as those used for the
Vreman model, namely the prevention of spurious artificial dissipation in
certain flow configurations. To this purpose, the model postulates an eddy
viscosity of the form
\begin{equation*}
\nu_t = (C_\sigma \delta)^2 \mathcal D_\sigma,
\end{equation*}
where $\delta$ is the filter width, $C_\sigma$ is a scaling parameter, and
$\mathcal D_\sigma = \mathcal D_\sigma(\nabla \bu) $ is a nonlinear differential
operator which satisfies the following properties:
\begin{itemize}
    \item[\textbf{P0}:]
      $\mathcal D_\sigma \geq 0$, i.e., no negative viscosity and no additional
      filtering steps,
    \item[\textbf{P1}:]
      cubic behavior near solid boundaries, i.e.,
      $\mathcal D_\sigma \sim y^3$ near $y = 0$ for shear flows above the
      $xz$-plane,
    \item[\textbf{P2}:]
      $\mathcal D_\sigma = 0$ for less than three-dimensional flows, i.e., when
      $\mathrm{rank}(\nabla \bu) \leq 2$,
    \item [\textbf{P3}:]
      $\mathcal D_\sigma = 0$ for axisymmetric (and, in the case of compressible
      flows, isotropic) expansion or contraction,
    \item [*\textbf{P4}:]
      $\mathcal D_\sigma$ should scale with frequency, i.e.,
      $\mathcal D_\sigma(\lambda \boldsymbol g) = |\lambda| \mathcal D_\sigma(\boldsymbol g)$.
\end{itemize}
These requirements are justified largely by arguments from experimental
observation and engineering constraints. The $\bsig$-model satisfies them by
taking
\begin{equation*} 
\mathcal D_\sigma (\nabla \bu) = \frac{\sigma_3(\sigma_1 - \sigma_2)(\sigma_2 - \sigma_3)}{\sigma_1^2},
\end{equation*}
where $\sigma_1 \geq \sigma_2 \geq \sigma_3 \geq 0$ are the singular values of
$\nabla \bu$, taking $\mathcal D_\sigma(\boldsymbol 0) = 0$. This choice
fulfills \textbf{P0} by the ordering of the singular values, \textbf{P2} and
\textbf{P3} by the product in the numerator, and *\textbf{P4} by the scale
factor $\sigma_1^{-2}$; \textbf{P1} is justified using Taylor expansion near
$y = 0$\cite[Section~II.B]{nicoud_using_2011}.

In the numerical simulation, the value $C_\sigma = 1.35$ was used, as obtained
by Nicoud et al.\cite[Section~III]{nicoud_using_2011} using both a simple
randomized procedure and a dynamic tuning approach applied to a high-fidelity
decaying isotropic turbulence simulation.

\subsubsection{The RB-VMS model} \label{sssec:rbvms}
The last considered turbulence model is
the residual-based variational multiscale (RB-VMS) approach proposed by Bazilevs
et al. in\cite{bazilevs_variational_2007}.
The major conceptual difference with respect to the eddy viscosity LES models
(Sections~\ref{sssec:smago}, \ref{sssec:vreman}, \ref{sssec:sigma}) lies in how
the scale separation is achieved. LES models typically proceed from the notion
of applying a convolutional low-pass filter to the Navier--Stokes equations
\eqref{eq:nse}, exchanging convolution and differentiation, and modelling the
remaining term involving the SGS stress tensor.
Variational multiscale models are instead based on a decomposition of both the
velocity and pressure spaces of the Navier--Stokes problem's variational form
into two or more ``coarse'' and ``fine'' spaces.

The RB-VMS model is a two-scale model \cite{ACJR17}. In the context of a finite
element method for discretizing the variational problem, the coarse scales are
defined as those resolved by the finite element discretization, whilst the fine
scales are the remaining (unresolved) ones.

Let $(\bu, p) = (\bu_h, p_h) + (\bu^\prime, p^\prime)$ denote the decomposition
into coarse and fine scales and let $\mathrm{Res}(\bu_h, p_h) = (\br_m, r_c)$
denote the (pointwise) residual of the coarse solution.
Following Bazilevs et al.\cite{bazilevs_variational_2007}, the major modelling
assumptions behind RB-VMS are (i) a representation of the fine-scale components
by a truncated perturbation series of
$\varepsilon = \|\mathrm{Res}(\bu_h, p_h)\|$, i.e.,
\[
\left(\bu^\prime, p^\prime\right) = \sum_{k \geq 1} \varepsilon^k \left(\bu_k^\prime, p_k^\prime\right) \approx \varepsilon \left(\bu_1^\prime, p_1^\prime\right),
\]
and (ii) an approximation of the fine-scale Green's operator relating
$(\bu_k^\prime, p_k^\prime)$ to
$(\bu_1^\prime, p_1^\prime)$, $\ldots$, $(\bu_{k - 1}^\prime, p_{k - 1}^\prime)$
and $\mathrm{Res}(\bu_h, p_h)$ by a $4 \times 4$ diagonal tensor
\[\boldsymbol{\tau} = \begin{pmatrix}
\tau_m \mathbb I_3 & 0 \\
0                  & \tau_c
\end{pmatrix}
\]
times a Dirac distribution, with momentum and continuity stabilization
parameters $\tau_m$ and $\tau_c$, which will be discussed in more detail below.
The model for the fine scales then reads
\begin{equation*} 
\begin{aligned}
\bu^\prime \approx -\tau_m \br_m(\bu_h, p_h) &= -\tau_m \left(\partial_t \bu_h + (\bu_h \cdot \nabla) \bu_h + \nabla p_h - \nu \Delta \bu_h \right) ,\\
p^\prime \approx -\tau_c r_c(\bu_h) &= -\tau_c (\nabla \cdot \bu_h).
\end{aligned}
\end{equation*}
Note that $\Delta \bu_h$ is typically not well-defined in terms of pointwise or
weak derivatives of $\bu_h$ in $\Omega$, as $\bu_h$ is only continuous piecewise
polynomial.
In our numerical simulations, pointwise second derivatives on the interior of
each tetrahedral cell are used, but projection-based methods of dealing with
this term may also be explored\cite[Page 181]{bazilevs_variational_2007}.

Finally, by considering interactions between fine and coarse scales (with a few
additional assumptions\footnote
{
    Stationary test functions, zero fine velocity on the boundary, velocity test
    function gradients orthogonal to fine velocity gradient.
}) and using integration by parts to avoid derivatives of
the residuals, the following modified semi-discrete problem is obtained:

Find $(\bu_h, p_h) : [0, T] \to \bX_h \times Q_h$ such that
\begin{equation} \label{eq:rbvms_variational}
\begin{aligned}
\left( \partial_t \bu_h, \bv_h \right) &=
- \mathrm{Gal}\left((\bu_h, p_h), (\bv_h, q_h)\right)
- \tau_m \left(\br_m(\bu_h, p_h), (\bu_h \cdot \nabla) \bv_h \right)
- \tau_m \left(\br_m(\bu_h, p_h), \nabla q_h \right) \\
& \quad
- \tau_c \left(r_c(\bu_h), \nabla \cdot \bv_h \right)
- \tau_m \left(\br_m(\bu_h, p_h), (\nabla \bv_h)^T \bu_h \right)
+ \tau_m^2 \left(\br_m(\bu_h, p_h) \otimes \br_m(\bu_h, p_h), \nabla \bv_h \right)
\end{aligned}
\end{equation}
at all times $t \in (0, T]$ and for all $\bv_h \in \bX_{h, 0}$, $q_h \in Q_h$.
In \eqref{eq:rbvms_variational}, $\mathrm{Gal}(\cdot, \cdot)$ denotes the terms
resulting from the Galerkin discretization of \eqref{eq:nse} as in
\eqref{eq:nse_gal}, and the remaining terms result from cross stresses (i.e, the
interactions between coarse and fine scales) and, in the case of the last term,
SGS stresses (i.e., fine-fine interactions).

Note that, except for the grad-div term
$\tau_c (r_c(\bu_h), \nabla \cdot \bv_h)$, all the additional terms introduced
by the RB-VMS model are at least quadratic in $\bu_h$, and the SGS term is
quadratic in $p_h$.
As a consequence, different approaches are possible when linearizing the problem
for a Picard iteration scheme.

Let $(\hat \bu_h, \hat p_h)$ denote the initial guess or last Picard iterate.
The stabilization parameters $\tau_m$, $\tau_c$ may depend on $\bu_h$. In this
case they are computed from $\hat \bu_h$. For legibility, this dependency will
not be marked in the notation.
Linearizing the term $\br_m$ as
\[
\hat \br_m(\bu_h, p_h) = \partial_t \bu_h + (\hat \bu_h \cdot \nabla) \bu_h + \nabla p_h - \nu \Delta \bu_h,
\]
one obtains the linear problem:
\begin{equation*} 
\begin{aligned}
\left( \partial_t \bu_h, \bv_h \right) &=
- \widehat{\mathrm{Gal}}\left((\bu_h, p_h), (\bv_h, q_h)\right)
- \tau_m \left(\hat \br_m(\bu_h, p_h), (\hat \bu_h \cdot \nabla) \bv_h \right)
- \tau_m \left(\hat \br_m(\bu_h, p_h), \nabla q_h \right) \\
& \quad
- \tau_c \left(r_c(\bu_h), \nabla \cdot \bv_h \right)
- \tau_m \left(\hat \br_m(\bu_h, p_h), (\nabla \bv_h)^T \hat \bu_h \right)
+ \tau_m^2 \left(\hat \br_m(\hat \bu_h, \hat p_h) \otimes \hat \br_m(\bu_h, p_h), \nabla \bv_h \right),
\end{aligned}
\end{equation*}
where $\widehat{\mathrm{Gal}}(\cdot, \cdot)$ denotes the linearization of the
Galerkin terms, as in \eqref{eq:nse_gal_linear}.
The momentum residual $\br_m$ (and the linearized version ${\hat \br}_m$) depend
on time derivatives of the velocity. This dependency is addressed by shifting
terms involving $\partial_t \bu_h$ to the time discretization's modified mass
matrix and discretizing ${\partial_t \hat \bu}_h$ as
\[
{\partial_t \hat \bu}_h \approx \frac 1 {\Delta t} \left({\hat \bu}_h -\bu_{h, \mathrm{prev}}\right),
\]
where $\bu_{h, \mathrm{prev}}$ is the previous time step's velocity.

Bazilevs et al.\cite[Equations~(63)~and~(64)]{bazilevs_variational_2007} suggest the
following formulas for $\tau_m$ and $\tau_c$ for equal-order pairs, based on
asymptotic scaling arguments for stabilized finite element methods:
\begin{equation} \label{eq:rbvms_tau_equalorder}
\tau_m(K,\bu_h) = \left( \frac 4 {\Delta t^2} + \bu_h \cdot \bG \bu_h + C_I \nu^2 (\bG : \bG) \right)^{-\frac 1 2}, \quad
\tau_c(K,\bu_h) = \frac 1 {\tau_m(K,\bu_h) |\bg|^2},
\end{equation}
In \eqref{eq:rbvms_tau_equalorder}, $K$ denotes a cell of the finite element
mesh, $\bG = (\nabla \bF^{-1})^T \nabla \bF^{-1}$ and
$\bg = \mathbb 1^T \nabla \bF^{-1}$ are derived from the local reference
transformation $\bF^{-1} : K \to \hat K$, and $C_I$ is the constant of an
element-wise inverse estimate. This inverse estimate is not clearly specified
and, in general, not trivial to obtain. However, as the $C_I$ term scales with
$\nu^2$, its influence can be assumed to be negligible in a highly turbulent
situation. For our computations with equal-order pairs, we used $C_I = 1$.

Using inf-sup stable pairs, the stabilization parameters were defined as
\begin{equation*} 
\tau_m = \max \left(\delta_0 h_{K,\mathrm{sh}}^2, \frac {\Delta t}{2} \right), \quad
\tau_c = \delta_1
\end{equation*}
with scaling parameters $\delta_0$, $\delta_1$ and the local cell's shortest
edge length $h_{K,\mathrm{sh}}$. In the numerical simulation, the values
$\delta_0 = 1$, $\delta_1 = 0.25$ were chosen.

\subsection{Simulation setup}
\label{sec:sim_setup}

Numerical simulations with the three eddy viscosity models described in
Sections~\ref{sssec:smago}, \ref{sssec:vreman}, and \ref{sssec:sigma} were run
using inf-sup stable Taylor--Hood finite elements, i.e., continuous piecewise
quadratic velocities ($\bX_h = \mathrm P_2(\mathcal T)^3$) and continuous
piecewise linear pressures ($Q_h = \mathrm P_1(\mathcal T)$). This pair of
spaces is probably the most popular inf-sup stable pair.
The RB-VMS model (Section~\ref{sssec:rbvms}) includes a stabilizing
pressure-pressure term and therefore does not necessarily require inf-sup stable
finite element spaces. In this case, results using
$\mathrm P_2^3 \times \mathrm P_1 =:\mathrm P_2 / \mathrm P_1$ elements on
$\mathcal T$ were also compared to those obtained with equal-order
$\mathrm P_1 / \mathrm P_1$ elements on $\mathcal T$ and on a refinement
$\mathcal T^\prime$.
Table~\ref{tab:grid_dof} provides information on the dimensions of the resulting
discrete problems depending on the different choices for the discretization for
the two computational meshes.

The time discretization was based on a BDF-2 scheme with a fixed time step
length of $\Delta t = \unit[\frac 1 8 \cdot 10^{-3}]{s}$. The resulting
nonlinear systems were solved using a Picard iteration, stopping the iteration
when the Euclidean norm of the residual vector was less than or equal to
$10^{-10}$. This was usually achieved in one or two iterations. The
corresponding linear systems were solved by a flexible GMRES iteration, using a
least-squares commutator preconditioner\footnote
{
    Here we used an iterative FGMRES/BiCGSTAB solver for the velocity problems
    (as for the whole system in the RB-VMS case) and a direct solver (MUMPS) for
    the pressure problems.
}\cite{elman_block_2006}
for the eddy viscosity models, which performed very efficiently in the numerical studies of \cite{ABJW18},
and a hybrid FGMRES/BiCGSTAB approach\footnote
{
    FGMRES preconditioned with a few iterations of BiCGSTAB at each step, itself
    with a basic Jacobi preconditioner.
}
for the RB-VMS models.

\begin{remark}[Preconditioning]
Since the RB-VMS method includes a pressure-pressure coupling term, the
system matrix always includes a nonzero pressure-pressure block, rendering
classical saddle point solvers or preconditioners inapplicable.
Although there exist methods for extending the LSC approach to stabilized
discretizations\cite{elman_least_2008}, in our experience they turned out to be
inefficient for the systems resulting from the RB-VMS method. Notice also that
due to the coupling of $\partial_t \bu_h$ with $\nabla q_h$, the modified mass
matrix appearing in the time-discretized system will have nonzero pressure rows.
However, a common mixed-method iterative solver provided an approach with
satisfactory efficiency.
\end{remark}

All computations were run using the finite element library
\textsc{ParMooN}\cite{wilbrandt_parmoonmodernized_2017} developed at WIAS
Berlin. The simulations were run with 60 parallel processes on an HPE Synergy
660 Gen10 compute server with four Intel Xeon Gold 6254 CPUs, each with 18 cores
clocked at $\unit[3.1]{GHz}$.

\subsection{Quantities of Interest} \label{ssec:qoi}
This section introduces the quantities used to assess the sensitivity of the
numerical results with respect to the utilized turbulence models.

\subsubsection{Pressure difference}
Pressure difference across the aorta is an important quantity used to
characterize the severity of the coarctation. In the numerical simulations,
the pressure difference between selected planar cross-sections roughly
orthogonal to the vessel centerline will be monitored.
Specifically, given two cross-sections $S$ and $S^\prime$, we will consider the
difference between averaged pressures:
\begin{equation*}
P_{S^\prime} - P_{S} = \frac 1 {|S^\prime|} \int_{S^\prime} p \: \mathrm{d}\mu_{S^\prime} - \frac 1 {|S|} \int_S p \: \mathrm{d}\mu_S.
\end{equation*}

\subsubsection{Maximum velocity}
The value of blood velocity in the stenotic region is also a relevant indicator used in
clinical practice to assess the severity of aortic stenoses.
In the upcoming studies the maximum velocity $\max_X |\bu| \; [\unitfrac{m}{s}]$
through certain regions will be monitored; $X$ may be a selected cross-section
or a portion of the domain enclosed between two cross-sections.

\subsubsection{Secondary flow degree (SFD)}
The secondary flow degree (SFD) is a dimensionless quantity defined over a given
planar cross-section as the ratio between the mean tangential (in-plane)
velocity magnitude and the mean orthogonal (through-plane) velocity.
Let $S$ be a cross-section, and let $\bn_{S}$ denote the unit normal vector on
$S$. Then the SFD on $S$ is defined as
\begin{equation*}
\mathrm{SFD}_{S} = \frac{\int_S |\bu - (\bu\cdot\bn_{S})\bn_{S} | \mathrm d \mu_S}{\int_S |\bu\cdot\bn_{S}| \mathrm d \mu_S}.
\end{equation*}

\subsubsection{Normalized flow displacement (NFD)}
The normalized flow displacement (NFD) is a dimensionless number that
quantifies, on a given planar cross-section, the distance of the moment of the
velocity normal to the plane from the cross-section's geometric center of mass,
normalized by the hydraulic radius of the cross-section. Let $S$ denote a
cross-section with geometric center of mass $\bx_S$, unit normal vector $\bn_S$,
area $A$ and perimeter $P$. Then $r_H:= \frac A P$ is its hydraulic
radius\footnote
{
    Note that in the case of a perfectly circular cross-section, the
    hydraulic radius is half the geometric radius.
}. Now the NFD is defined as
\begin{equation*}
\mathrm{NFD}_S = \frac {|\bx_\mathrm{n}(\bu, S) - \bx_S|} {r_H},
\end{equation*}
where
\[
\bx_\mathrm{n}(\bu, S) := \frac {\int_S |\bu \cdot \bn_S| \bx \mathrm d \mu_S} {\int_S |\bu \cdot \bn_S| \mathrm d \mu_S}.
\]

\subsubsection{Wall shear stress (WSS) and oscillatory shear index (OSI)}
The wall shear stress (WSS) quantifies the force per unit area exerted by the
blood flow on the vascular endothelium, directed on the local tangent plane. Let
$\bx \in \partial \Omega$ be a point on the boundary, and let $\bn$ be the outer
unit normal at $\bx$. Then the WSS at $\bx$ is given by the dynamic viscosity
times the normal derivative of the tangential component of the velocity, i.e.,
\[
\tau_\mathrm{w}(t,\bx) = \mu \frac {\partial} {\partial \bn} \Big(\bu (t,\bx) - \big(\bu (t,\bx) \cdot \bn\big) \bn\Big).
\]
In our studies, the WSS $\tau_\mathrm{w}$ was computed considering a piecewise
constant normal vector on each triangular face of the boundary $\partial \Omega$
and the gradient of the velocity at the face's centroid.

The WSS is a tangential pressure exerted on the boundary, and is essentially a
two-dimensional quantity. Taking a constant forward unit vector $\bv$
roughly aligned with the main direction of flow near the region of interest, one
can decompose the WSS $\tau_\mathrm{w}$ into a forward (or backward) component
$\tau_\mathrm{w} \cdot \bv$ and a lateral component
$\tau_\mathrm{w} \cdot \bw$, where $\bw = \bw(\bx)$ is a unit vector orthogonal
to both $\bv$ and the outer normal $\bn(\bx)$ at each point $\bx$ on the
reference patch.
In the upcoming studies, to avoid choosing an orientation of $\bw$ at each point and
possibly eliminating the lateral components of boundary-touching eddies,
the average of the magnitude of the lateral component
\[
  \vert \tau_\mathrm{w} \cdot \bw \vert = \left\vert \tau_\mathrm{w} - (\tau_\mathrm{w} \cdot \bv) \bv \right\vert
\]
rather than the lateral component itself will be considered.

The Oscillatory Shear Index\cite{ku-1985} (OSI) is an adimensional quantity that
measures the extent to which shear stress oscillates by the relative
difference between the temporal mean of the shear stress vector and the mean of
its magnitude, i.e.,
\begin{equation*}
\mathrm{OSI}_I(\bx) = \frac12 \left( 1 - \frac{ \left| \int_I \tau_\mathrm{w}(t,\bx) \mathrm{d}t \right|}{ \int_I |\tau_\mathrm{w}(t,\bx)| \mathrm{d}t}\right)
\end{equation*}
for a point $\bx \in \partial \Omega$ on the boundary and a time interval $I$.
The OSI varies between $0$ (shear stress always directed along the same
direction) and $\frac12$ (oscillating shear stress with zero average).

\subsubsection{Regions of interest}\label{ssec:S-planes}

\begin{figure}[h!]
\centering
\includegraphics[height=0.35\textwidth]{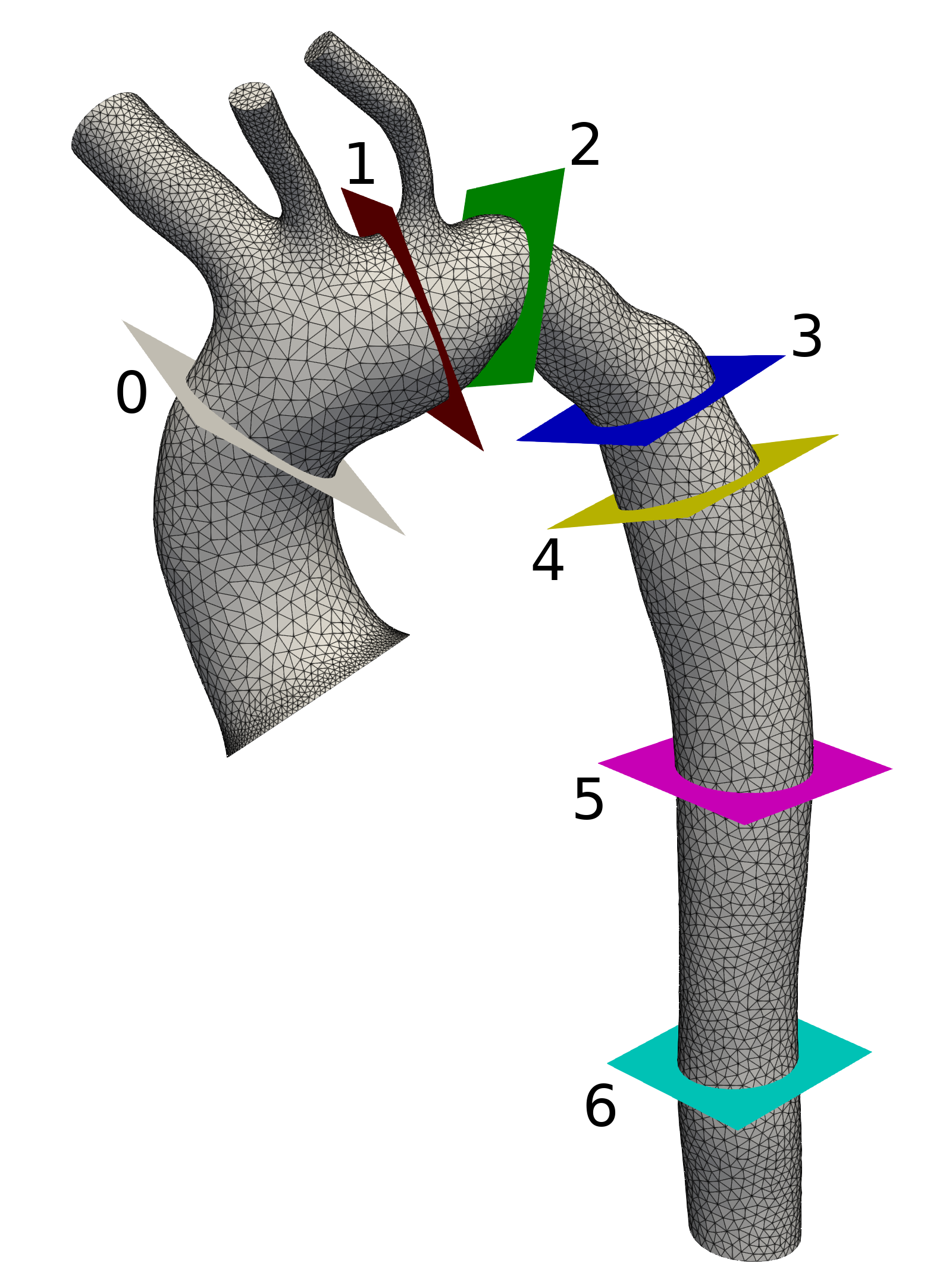}
\hspace{1cm}
\includegraphics[height=0.35\textwidth]{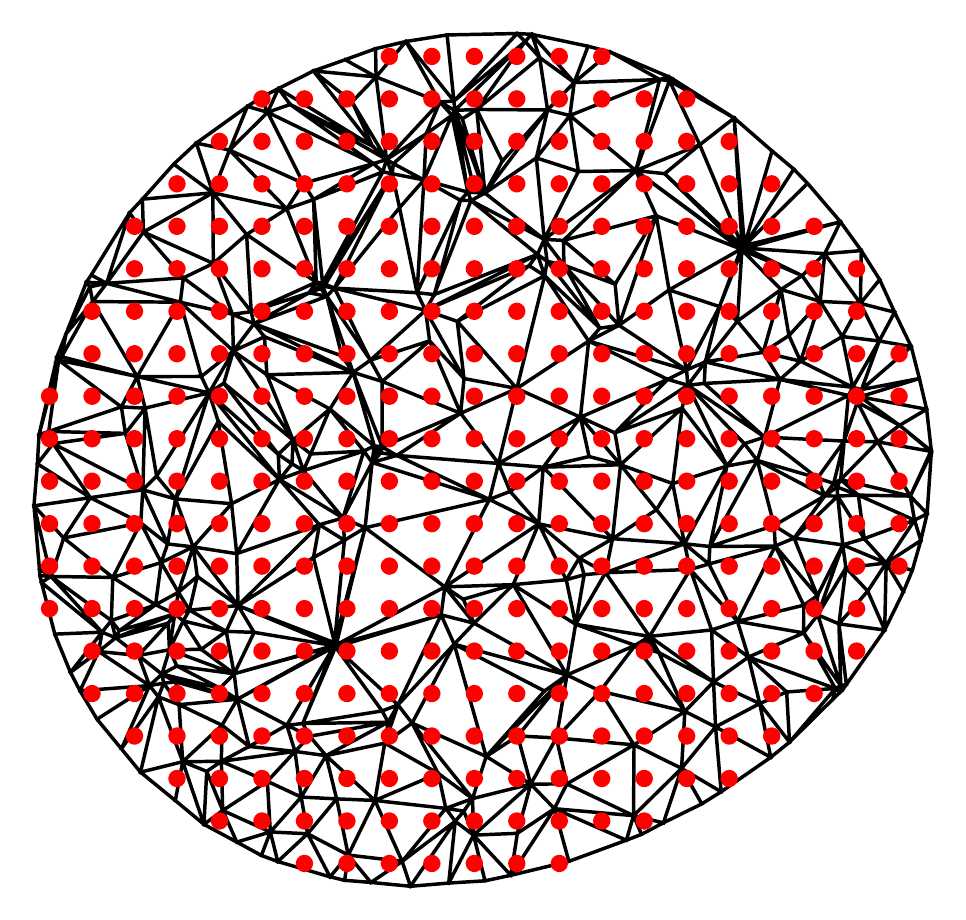}
\includegraphics[height=0.35\textwidth]{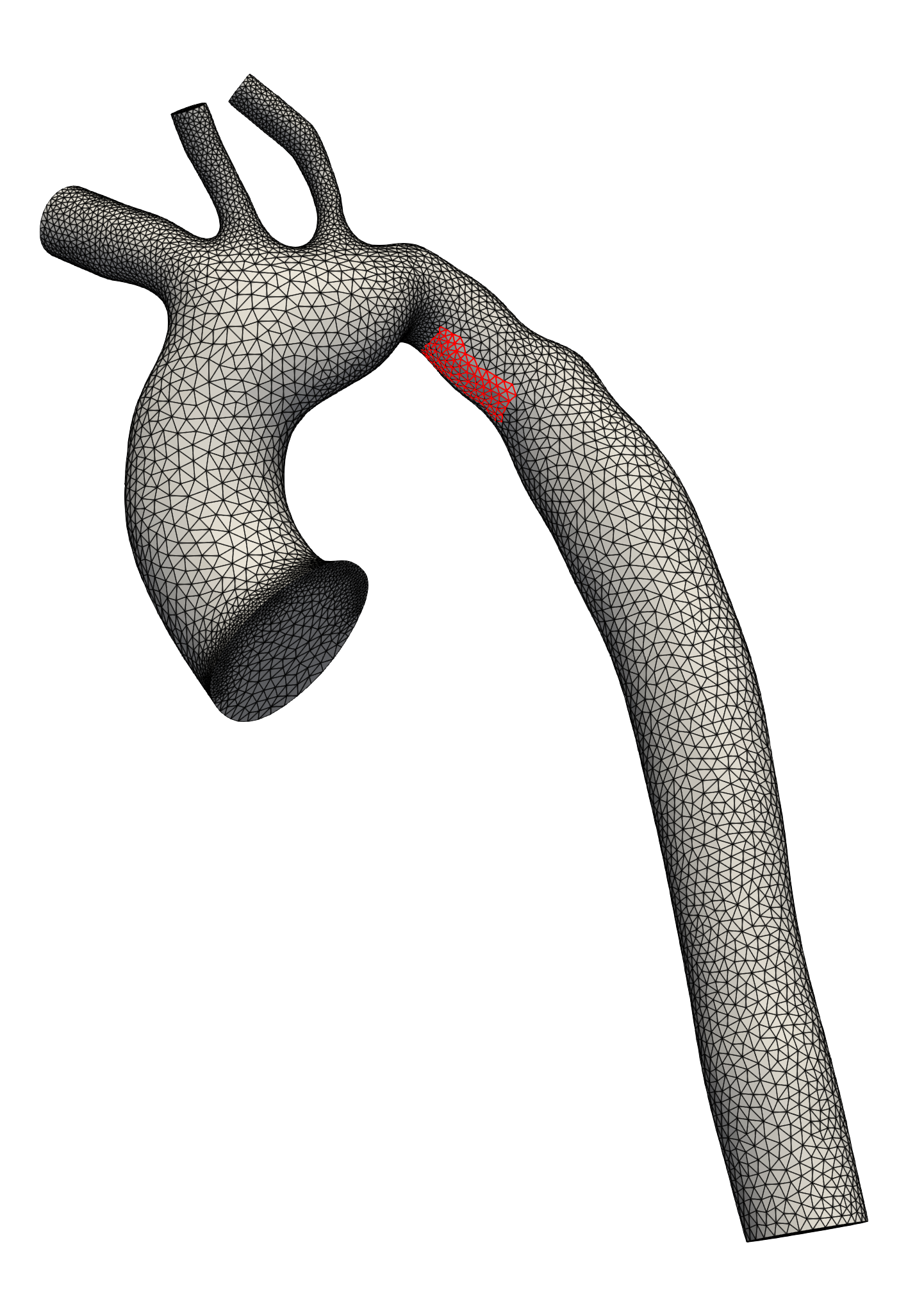}
  \caption[Positions of planar cross-sections and wall shear stress reference patch]
  {
    Left: positions of the planar cross-sections 0 through 6 of the aorta segment used
    for monitoring the quantities of interest.
    Middle: quadrature points on a $\unit[1]{mm}$-resolution
    grid on cross-section 4.
    Right: reference patch for wall shear stress computation.
  }
  \label{fig:qoi-planes}
\end{figure}

Average pressure, SFD, and NFD are evaluated on seven planar cross-sections
$S_i$ of the aorta segment under consideration, depicted in
Figure~\ref{fig:qoi-planes} (left), with $i$ taking the following values:
\begin{enumerate}[label = {\textbf{\arabic*}:}, start = 0]
    \item close to the inlet boundary, at the beginning of the aortic arch,
    \item between the left common carotid and left common subclavian arteries,
    \item immediately before the coarctation, where the flow narrows at the turn
    of the aortic arch and jet formation is expected,
    \item at the end of the aortic arch,
    \item at the beginning of the descending aorta, where the effects of the jet
    formed by inertia and the narrowing of the flow would be observable,
    \item half-way between the coarctation and the outlet boundary,
    where the flow should
    begin transitioning to a simpler form, and
    \item further down the descending aorta, close to the outlet boundary, where near-laminar
    flow is expected.
\end{enumerate}

Figure~\ref{fig:qoi-planes} (right) highlights a patch on the underside of the
transition from aortic arch to descending aorta; as a ``backward facing step''
effect with substantial vortex formation is to be expected here, this is an
interesting region on which to study the wall shear stress.

The evaluation of the other quantities of interest requires integration of the
numerical solution over arbitrary planar cross-sections of the computational
mesh. In our numerical studies, these integrals were approximated by defining,
on each considered cross section $S_i$, a Cartesian grid of quadrature points at a
resolution of $\unit[1]{mm}$ in each tangential direction; these points were
given equal weights corresponding to $\unit[1]{mm^2}$ each.
Additional computations with increased resolution of the grid used for numerical
quadrature showed negligible influence on the QOI estimates.
Table~\ref{tab:points} lists the number of quadrature points on each plane.

We also evaluate maximum velocities on the cross-sections $S_i$ and within the
``wedges'' $W_j$ between the $j$-th and $(j+1)$-th cross-sections,
$j = 0, \ldots, 5$.

\section{Results}\label{sec:numres}

\subsection{Impact of the variation of SVR} \label{ssec:svr_impact}

First, an appropriate estimate for the systemic vascular resistance, see
\eqref{eq:rsv}, should be identified and the impact of this choice
studied.
To this end, simulations were performed for three values
$\RSV \in \unitfrac[\{70,115,160\}]{MPa \cdot s}{m^3}$. The obtained results for
the quantities of interest are compared below. The chosen values correspond
roughly to the lower end, middle, and upper end of the adult human clinical
reference range\cite{washington-2008}.

Table~\ref{tab:resistances_coarse} shows the estimated outlet resistances
(Section~\ref{ssec:resistive}) depending on the selected turbulence model and
value of $\RSV$.
The estimated values were tuned based on the outflow fractions listed in
Table~\ref{tab:flowsplit}. As turbulent fluctuations produce small irregular
oscillations in the outflow rates, the quality of these estimates must be
evaluated over a longer time interval rather than at a single instant.
We performed constant-inflow simulations with the resistances listed in
Table~\ref{tab:resistances_coarse}; the resulting outflow errors averaged over
the time interval $[0.25, 0.5]$ satisfy
\[4 \left\vert \; \int_{0.25}^{0.5} \left( \frac {Q_i(t)}{Q_i^*} - 1 \right) \mathrm dt \right\vert < 10^{-3}\]
for each outlet $i=1,\hdots,4$.

It turned out that the impact of varying the SVR on the
quantities of interest is relatively small, as shall be discussed
in more detail throughout the remainder of this section.
Exemplarily, results of numerical simulations performed using the
Smagorinsky model with $C_\mathrm{Sma} = 0.01$ and with the values of the
systemic vascular resistance $\RSV$ listed in Table~\ref{tab:resistances_coarse}
will be presented.
All the results are based on a simulation time of one heartbeat, concretely in the time interval $[0.5,1.5]~\unit{s}$.

\subsubsection{Pressure difference} \label{sssec:svr_p}

\begin{figure}[t!]
    \begin{center}
        \includegraphics[width = 0.49 \textwidth]{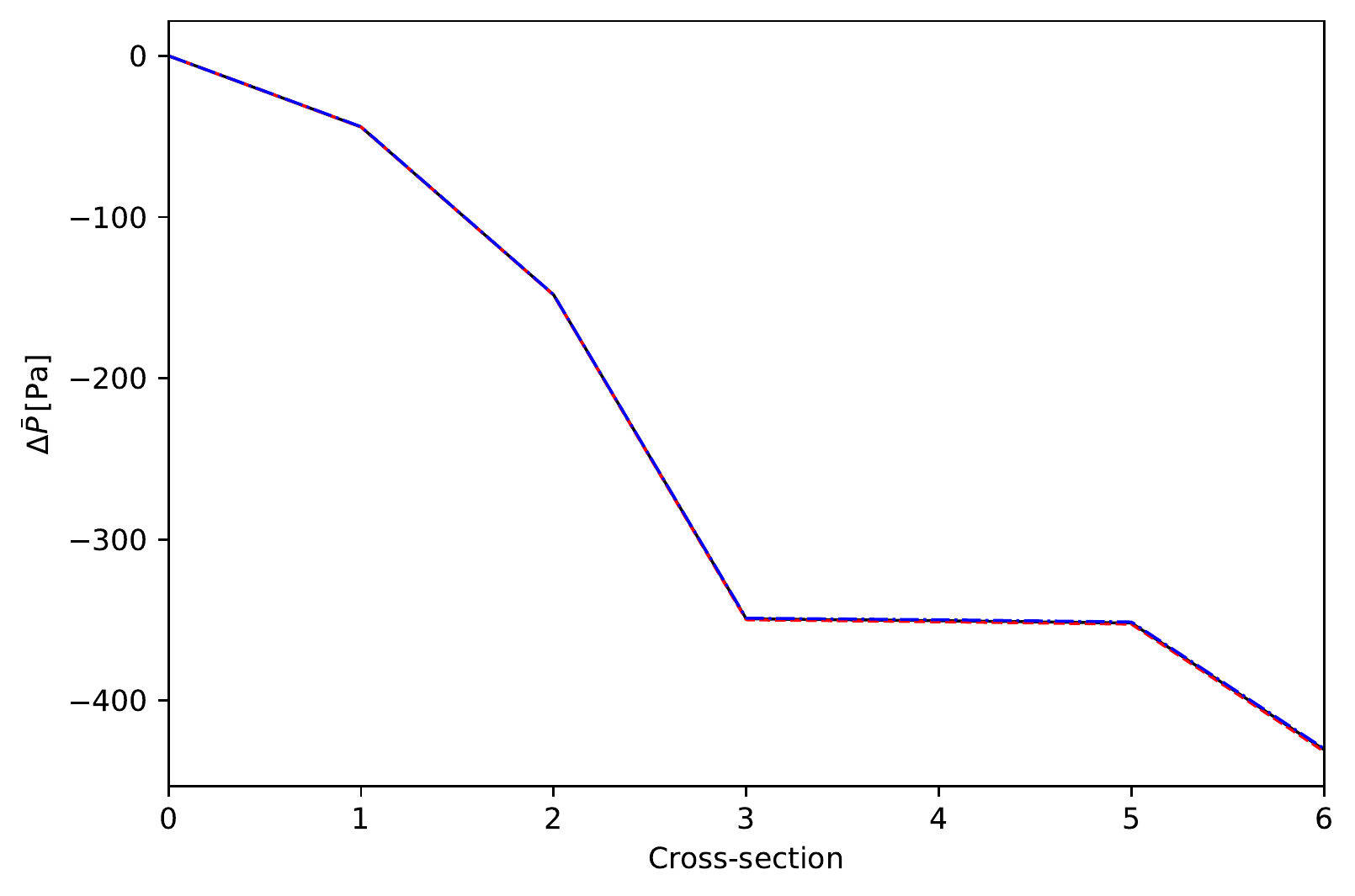}
        \includegraphics[width = 0.49 \textwidth]{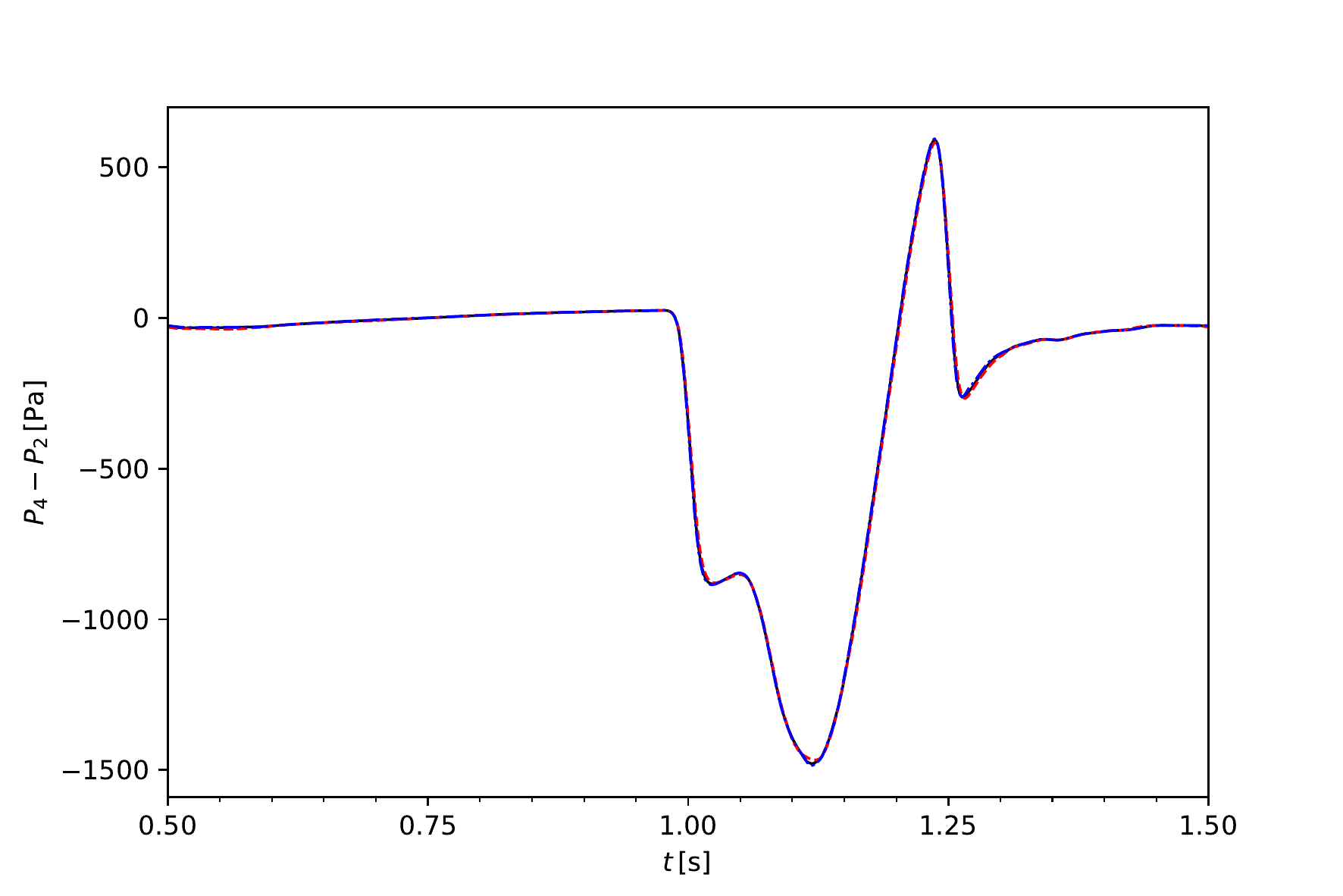}
        \caption[\impsvr pressure differences]
        {
          Impact of the variation of SVR. Left: time-averaged pressure difference on each cross-section.
          Right: pressure difference between cross-sections 4 and 2 over time.
          Simulations with Smagorinsky model, $C_\mathrm{Sma} = 0.01$,
          $\RSV  [\unitfrac{MPa \cdot s}{m^3}] = 160$ (dash-dot blue line), 115 (solid black line), 70 (dashed red line).
        }
        \label{fig:svr_p}
    \end{center}
\end{figure}

Figure~\ref{fig:svr_p} shows the pressure difference between each cross-section
and cross-section~0 averaged over one pulse period (left) as well as the
difference between cross-sections~4 and 2 (right), i.e., between the aortic arch
just past the left common subclavian artery and the upper descending aorta,
straddling the coarctation.

The time-averaged pressure difference varies by less than $\unit[2]{Pa}$
between the three values of $\RSV$. The largest variations in the difference
between cross-sections~4 and 2 over time occur just before systole and at the
end of the decelerating phase (around $\unit[1]{s}$ and $\unit[1.25]{s}$), with
maximum differences around $\unit[62]{Pa}$ and a mean of less than
$\unit[4]{Pa}$. The pressure differences between other pairs of cross-sections
behave comparably.

\subsubsection{Maximum velocity} \label{sssec:svr_max_u}

\begin{figure}[t!]
  \begin{center}
    \includegraphics[width = 0.49 \textwidth]{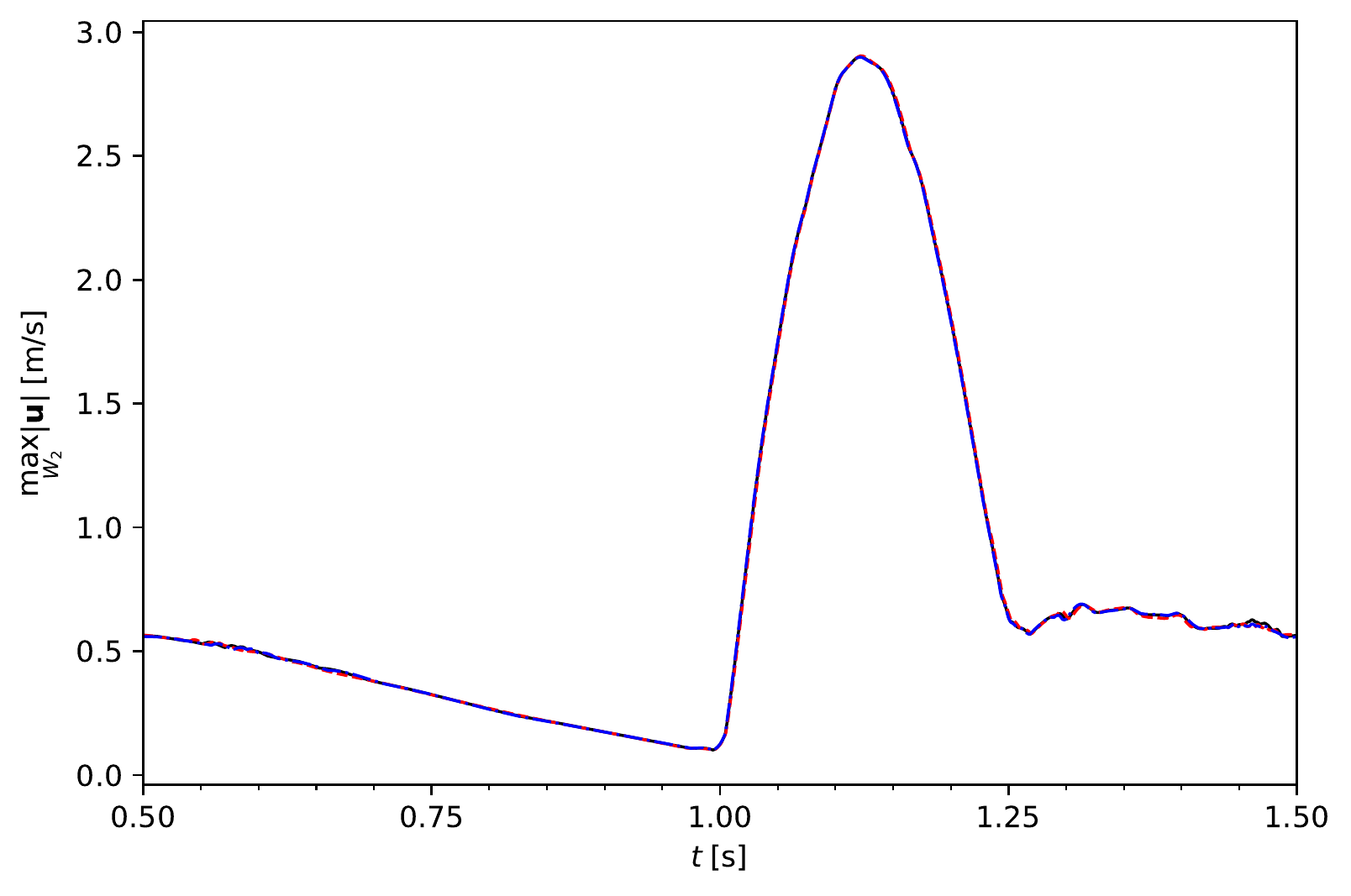}
    \includegraphics[width = 0.49 \textwidth]{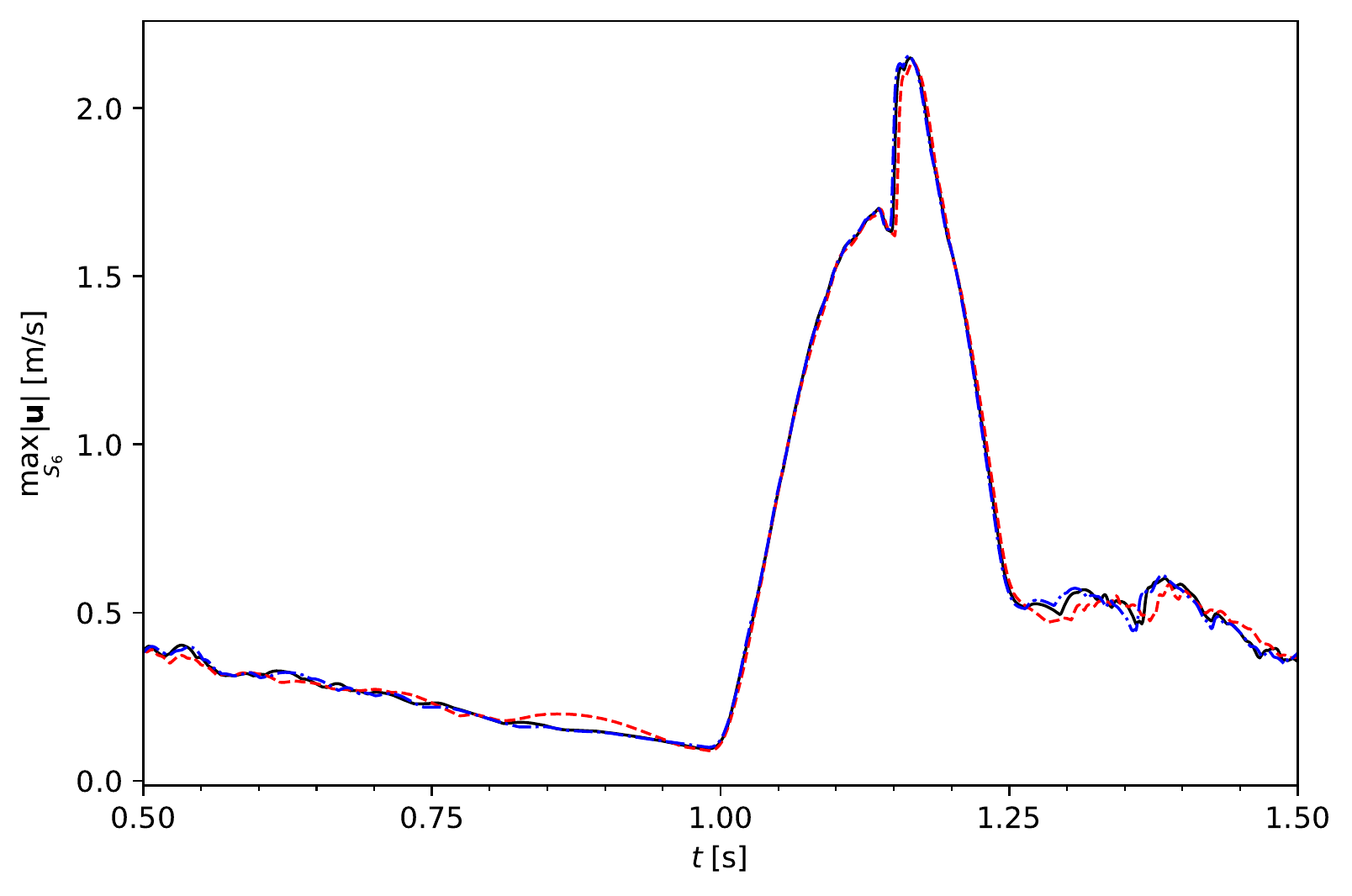}
    \caption[\impsvr maximum velocity]
    {
      Impact of the variation of SVR.
      Left: maximum velocity through the wedge between cross-sections~2 and 3
        over time.
      Right: maximum velocity through cross-section~6 over time.
      Simulations with Smagorinsky model, $C_\mathrm{Sma} = 0.01$,
      $\RSV  [\unitfrac{MPa \cdot s}{m^3}] = 160$ (dash-dot blue line), 115 (solid black line), 70 (dashed red line).
    }
    \label{fig:svr_max_u}
  \end{center}
\end{figure}

Figure~\ref{fig:svr_max_u} compares the maximum velocities through
the wedge between cross-sections~2 and 3 (left) and through cross-section~6
(right) over time. The results are again very close, though minor quantitative
differences appear particularly when the flow is less rapid overall.

\subsubsection{Secondary flow degree} \label{sssec:svr_sfd}

\begin{figure}[t!]
    \begin{center}
        \includegraphics[width = 0.49 \textwidth]{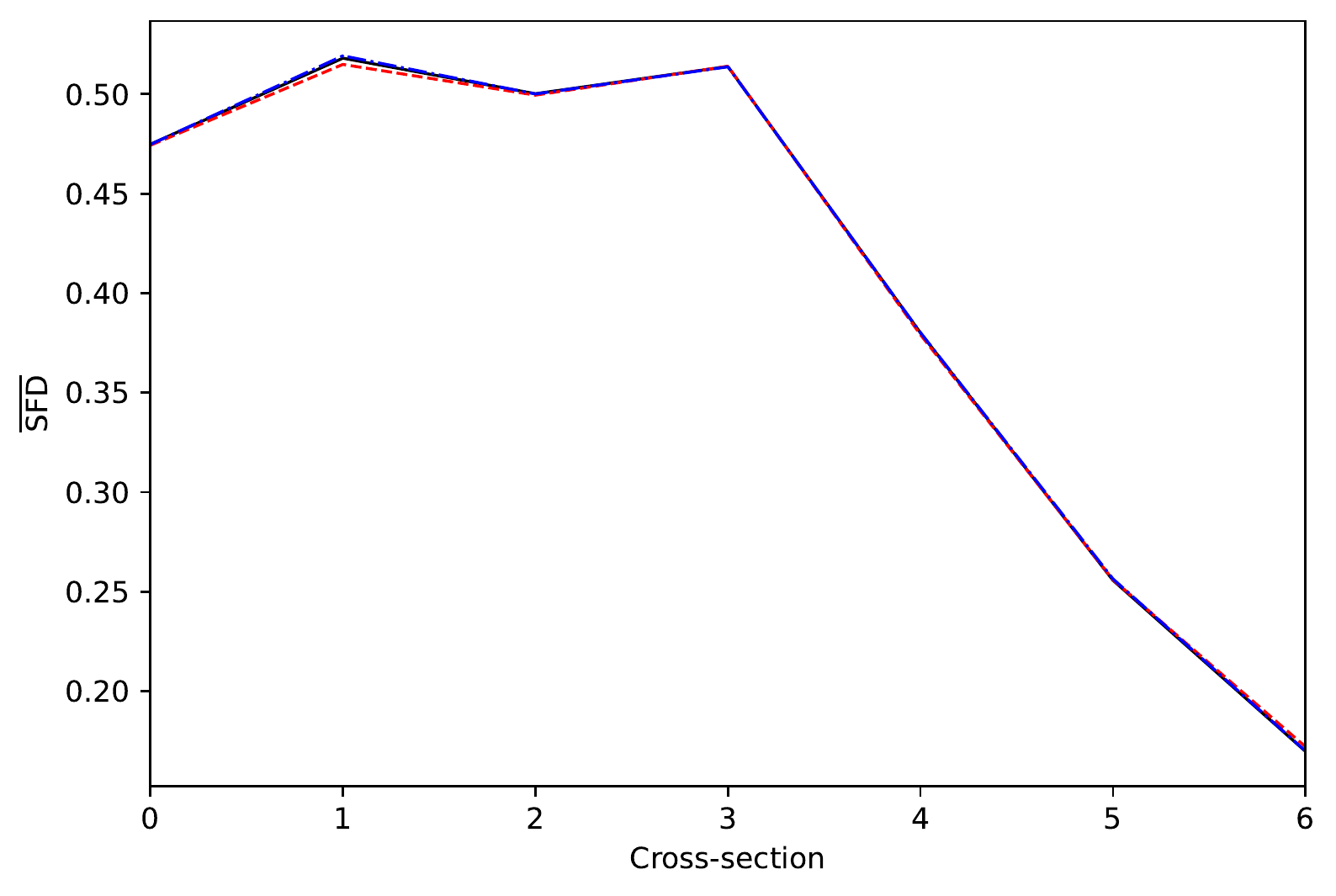}
        \includegraphics[width = 0.49 \textwidth]{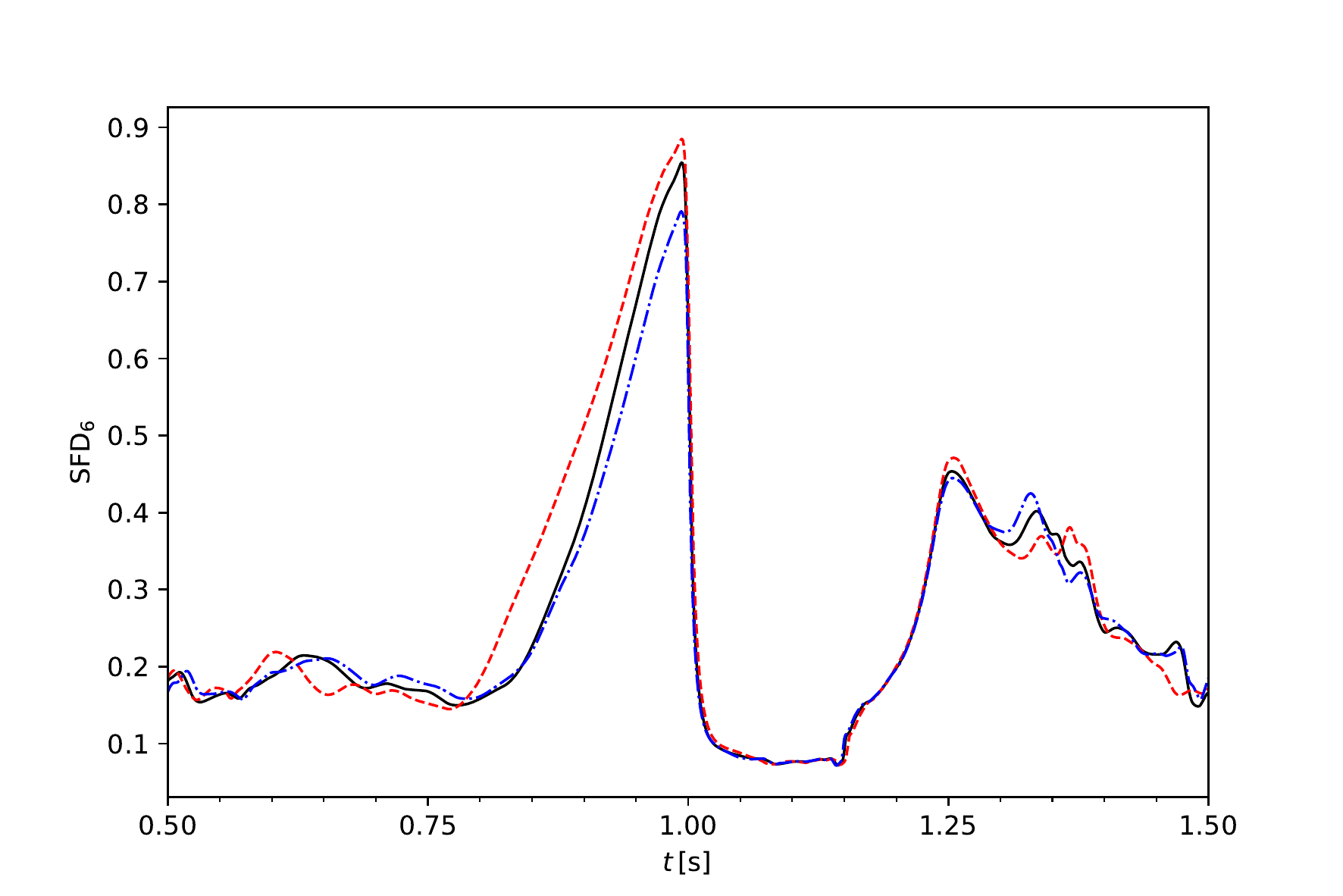}
        \caption[\impsvr secondary flow degree]
        {
          Impact of the variation of SVR. Left: time-averaged secondary flow degree per cross-section.
          Right: secondary flow degree across cross-section~6 over time.
          Simulations with Smagorinsky model, $C_\mathrm{Sma} = 0.01$,
          $\RSV [\unitfrac{MPa \cdot s}{m^3}] = 160$ (dash-dot blue line), 115 (solid black line), 70 (dashed red line).
        }
        \label{fig:svr_sfd}
    \end{center}
\end{figure}

The secondary flow degree across each cross-section averaged over one pulse
period (left) as well as across cross-section~6 over time (right), i.e., the
last cross-section before the lower end of the computational domain, are
depicted in Figure~\ref{fig:svr_sfd}. This cross-section was chosen because it
exhibits the most visible differences. Note that the time-averaged SFD was
computed not by time-averaging the instant SFD but by the ratio of
cumulative tangential flow to cumulative normal flow:
\[\overline{\mathrm{SFD}}_S = \frac
{\int_{0.5}^{1.5} \int_S \vert \bu - (\bu \cdot \bn) \bn \vert \mathrm{d}\mu_S \mathrm{d}t}
{\int_{0.5}^{1.5} \int_S \vert \bu \cdot \bn \vert \mathrm{d}\mu_S \mathrm{d}t}.
\]

As for the pressure difference, the values of time-averaged SFD vary negligibly
for different choices of $\RSV$. The largest absolute difference is found at
cross-section~1 (just before the brachiocephalic artery), where the values range
from $0.50439$ to $0.50743$.

Larger differences are visible plotting the SFD across cross-section~6 over
time. However, the qualitative behavior is largely unaltered, lower resistances
corresponding roughly to a higher peak just before systole and shifts in time
and amplitude of the irregular oscillation during diastole.

\subsubsection{Normalized flow displacement} \label{sssec:svr_nfd}

\begin{figure}[t!]
    \begin{center}
        \includegraphics[width = 0.49 \textwidth]{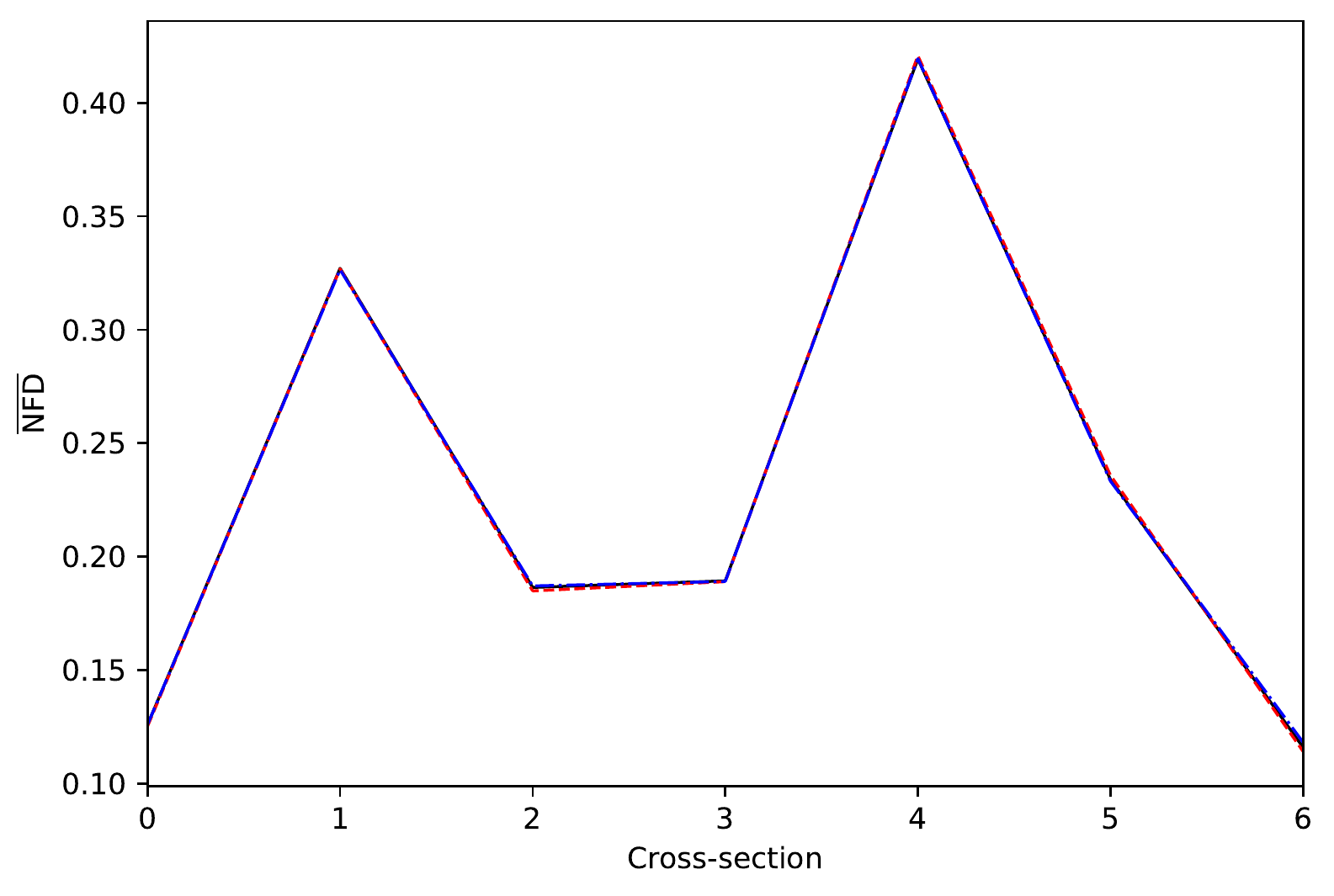}
        \includegraphics[width = 0.49 \textwidth]{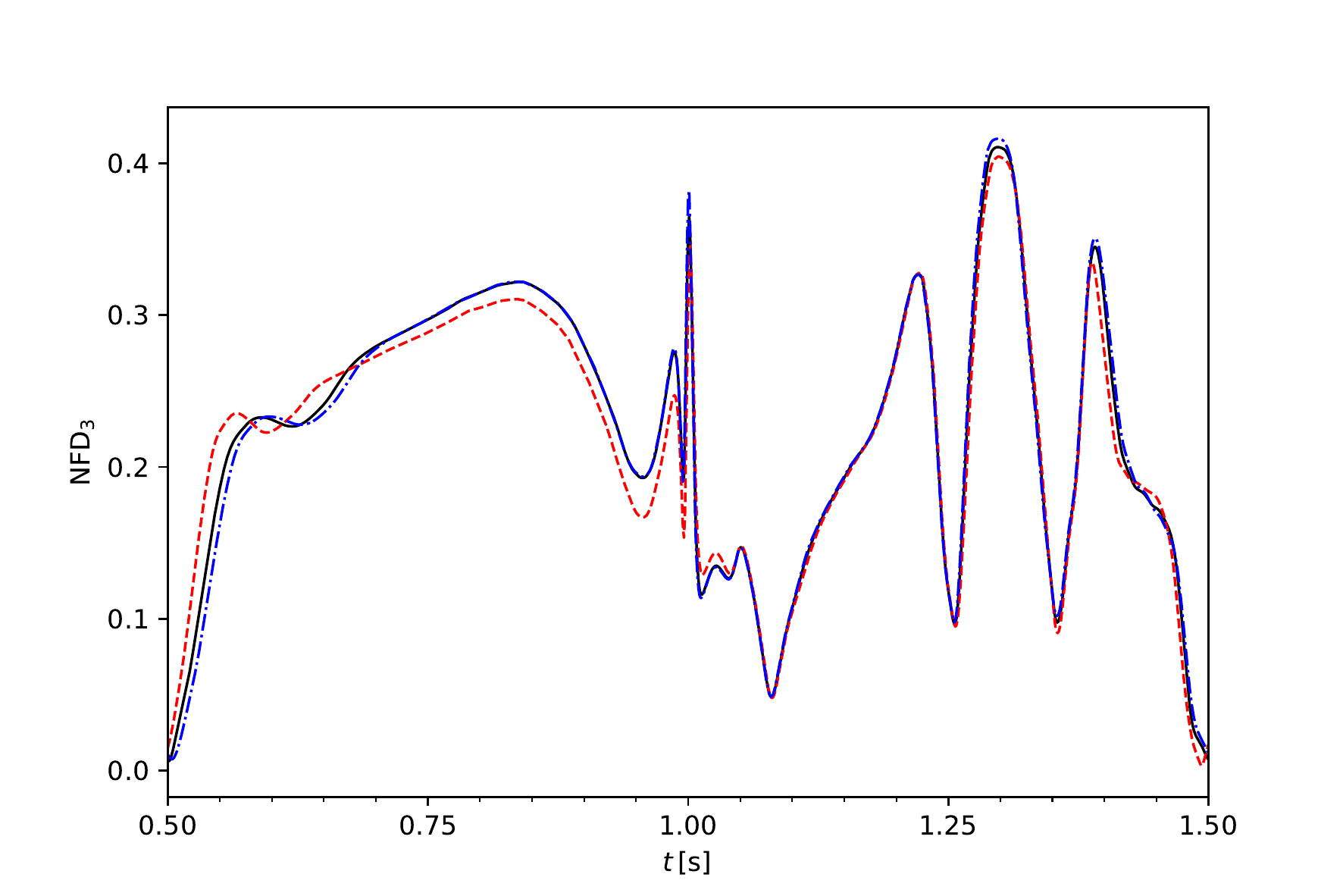}
        \caption[\impsvr normalized flow displacement]
        {
          Impact of the variation of SVR. Left: time-averaged normalized flow displacement per cross-section.
          Right: normalized flow displacement across cross-section~3 over time.
          Simulations with Smagorinsky model, $C_\mathrm{Sma} = 0.01$,
          $\RSV [\unitfrac{MPa \cdot s}{m^3}] = 160$ (dash-dot blue line), 115 (solid black line), 70 (dashed red line).
        }
        \label{fig:svr_nfd}
    \end{center}
\end{figure}

Figure~\ref{fig:svr_nfd} shows the normalized flow displacement across each
cross-section averaged over one pulse period (left) as well as across
cross-section~3 over time (right), i.e., the first cross-section past the
coarctation, chosen due to its position near the center of a prominent jet. Note
that the time-averaged NFD has been weighted by the normal flow rate:
\[\overline{\mathrm{NFD}}_S = \frac
{\int_{0.5}^{1.5} \int_S \vert \bu \cdot \bn \vert \mathrm{d}\mu_S \mathrm{NFD}_S(t) \mathrm{d}t}
{\int_{0.5}^{1.5} \int_S \vert \bu \cdot \bn \vert \mathrm{d}\mu_S \mathrm{d}t}.
\]

Also in this case, the time-averaged quantity shows only negligible differences:
the largest absolute differences are at cross-section~6, where the NFD ranges
from $0.05704$ to $0.05885$. As for the SFD, the effect of the different SVR on
the temporal variation amounts to slight shifts of the peaks and valleys in time
and amplitude.

\subsubsection{Wall shear stress} \label{sssec:svr_wss}

\begin{figure}[t!]
    \begin{center}
        \includegraphics[width = 0.49 \textwidth]{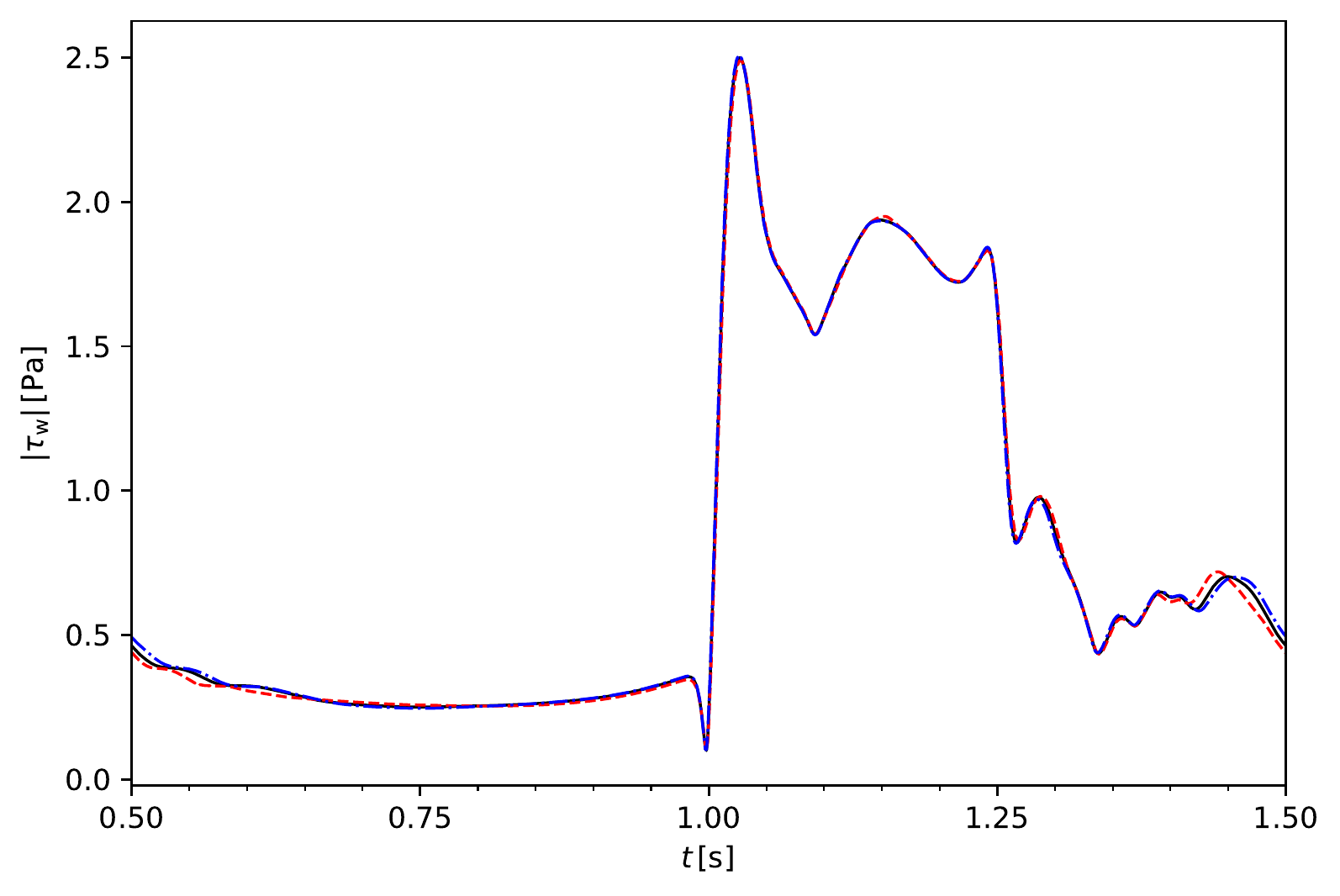}
        \includegraphics[width = 0.49 \textwidth]{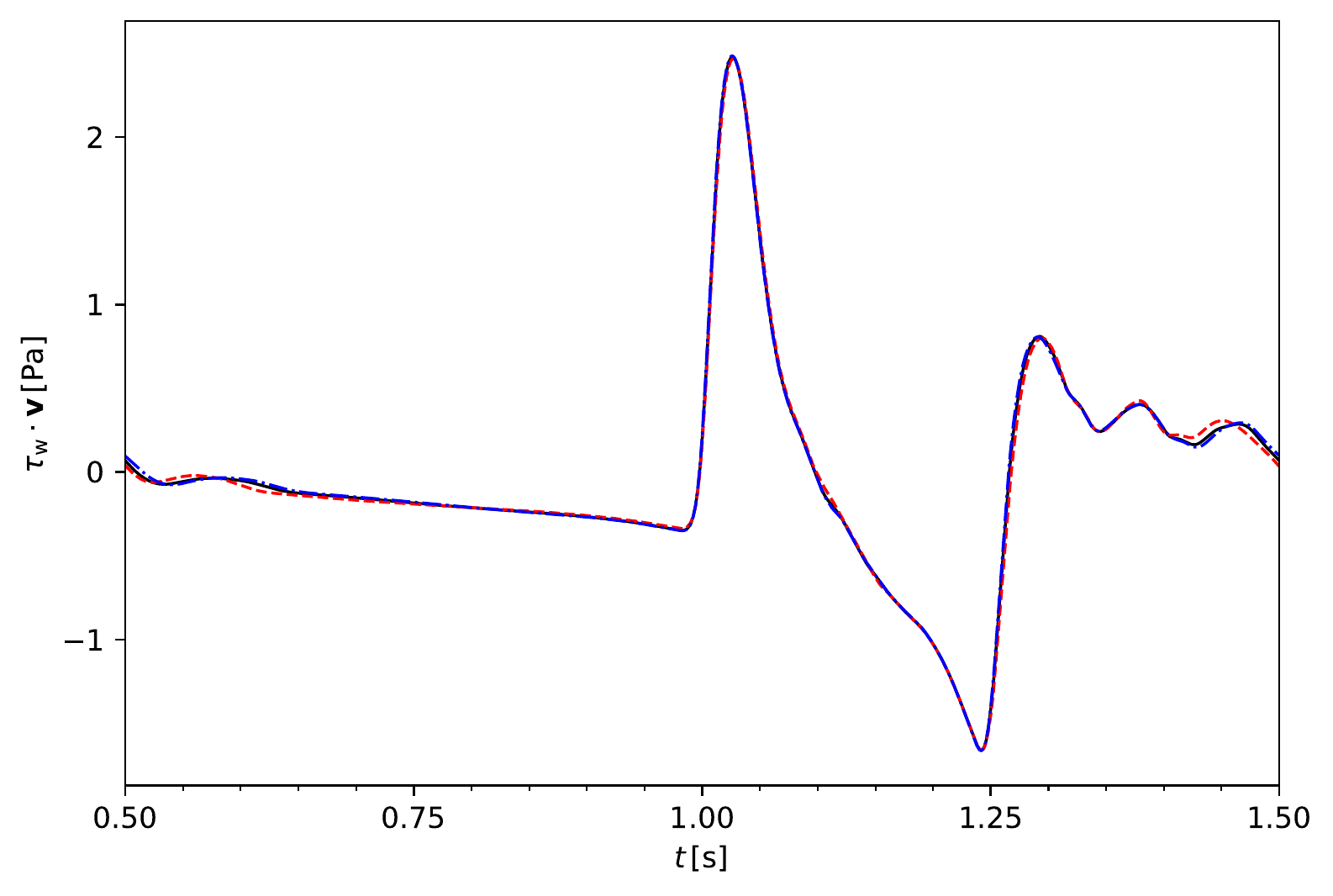}
        \caption[\impsvr wall shear stress]
        {
          Impact of the variation of SVR. Left: average wall shear stress magnitude over reference patch (see Figure \ref{fig:qoi-planes}, right).
          Right: average forward wall shear stress (shear stress along the main flow direction) over reference patch.
          Simulations with Smagorinsky model, $C_\mathrm{Sma} = 0.01$,
          $\RSV [\unitfrac{MPa \cdot s}{m^3}] = 160$ (dash-dot blue line), 115 (solid black line), 70 (dashed red line).
        }
        \label{fig:svr_wss}
    \end{center}
\end{figure}

Figure~\ref{fig:svr_wss} presents the magnitude of the wall shear stress and of
its ``forward'' component, i.e., the component in the main direction of flow,
averaged over the reference patch depicted in Figure~\ref{fig:qoi-planes}
(right).
Also in this case, only minor differences are visible.

Finally, Table~\ref{tab:svr_wss} provides information on the time-averaged WSS
magnitude and the OSI values over one pulse period. Both WSS and OSI increase
for increasing $\RSV$. However, in the considered SVR range, the differences are
less than 1\%.

\subsection{Impact of turbulence model selection} \label{ssec:turb_impact}

This section starts by providing an overall comparison of the flow field
obtained with some of the considered turbulence models. Next, the results with
respect to the considered quantities of interest will be presented in detail.
The time-averaged quantities were computed for all turbulence models for one pulse period,
concretely in the time interval $[0.5,1.5]~\unit{s}$. In addition, for selected models, long-term
computations were performed over a longer time interval of 
$31$ periods,
in order to investigate the differences from period to period. In
this case, results are shown in terms of long time-averages over the
interval $[1, 31]~\unit{s}$ ($30$ periods), discarding the first time interval used for a smooth start.

In light of the relatively small influence of the systemic vascular resistance,
the simulations presented in this section were all performed using
$\RSV = \unitfrac[115]{MPa \cdot s}{m^3}$.

\subsubsection{Flow field}

For the sake of brevity, this section focuses on the RB-VMS model
used in combination with $\mathrm P_1 / \mathrm P_1$ elements, since this is the only approach that uses
first order elements for the velocity.

\begin{figure}[t!]
    \begin{center}
        \includegraphics[width = 0.8 \textwidth]{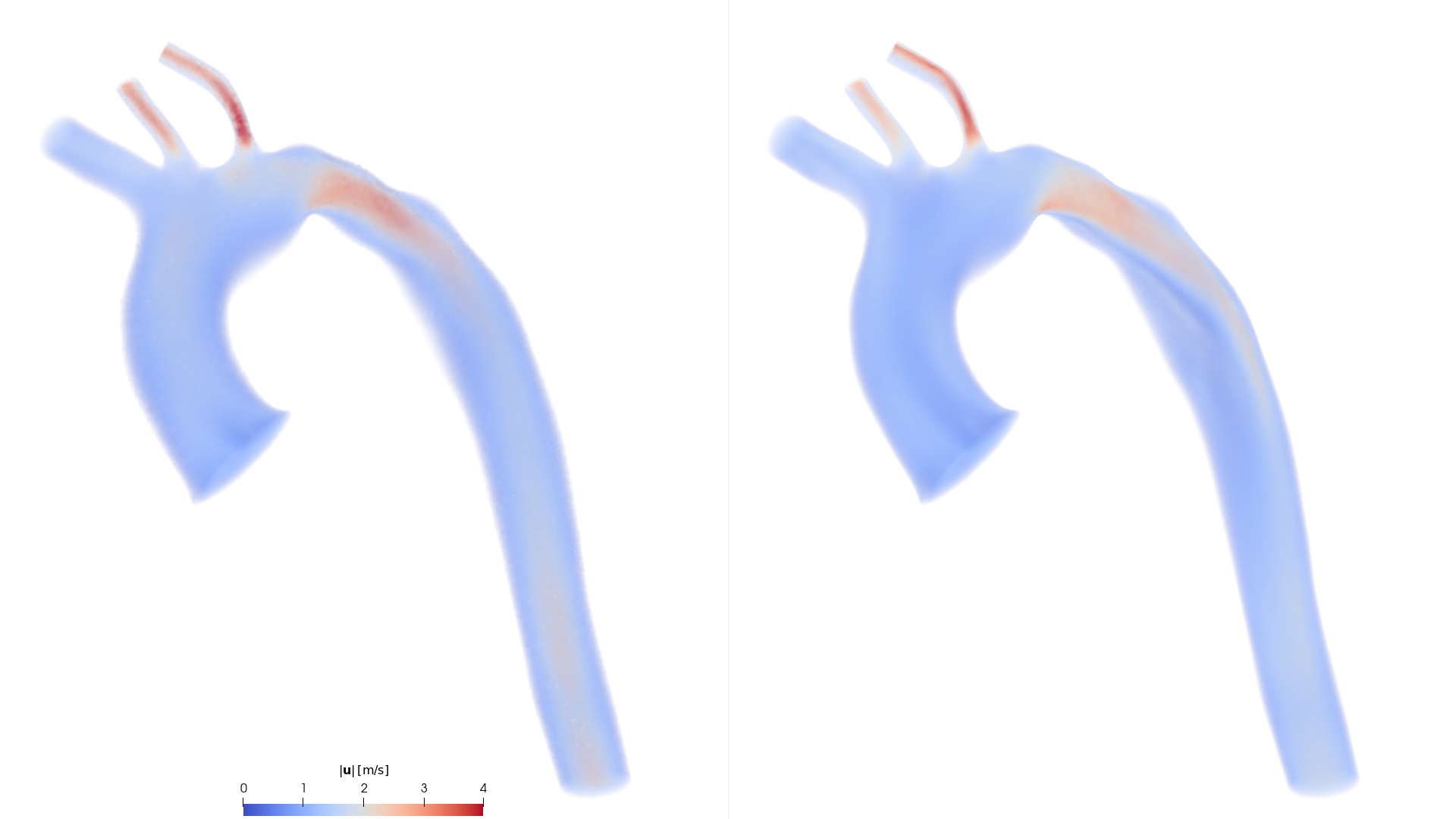}
        \caption[\impturb Velocity field comparison at peak inflow, RB-VMS, coarse vs. fine mesh]
        {
          Magnitude of the velocity field (volume plot) $t = \unit[1.125]{s}$ (peak inflow) for the
          RB-VMS model, $\mathrm P_1 / \mathrm P_1$ elements.
          Left: Solution on the coarse mesh. Right: Solution on the fine mesh.
        }
        \label{fig:vol_1125_rbvms_p1_p1_coarse_fine}
    \end{center}
\end{figure}

\begin{figure}[t!]
    \begin{center}
        \includegraphics[width = 0.8 \textwidth]{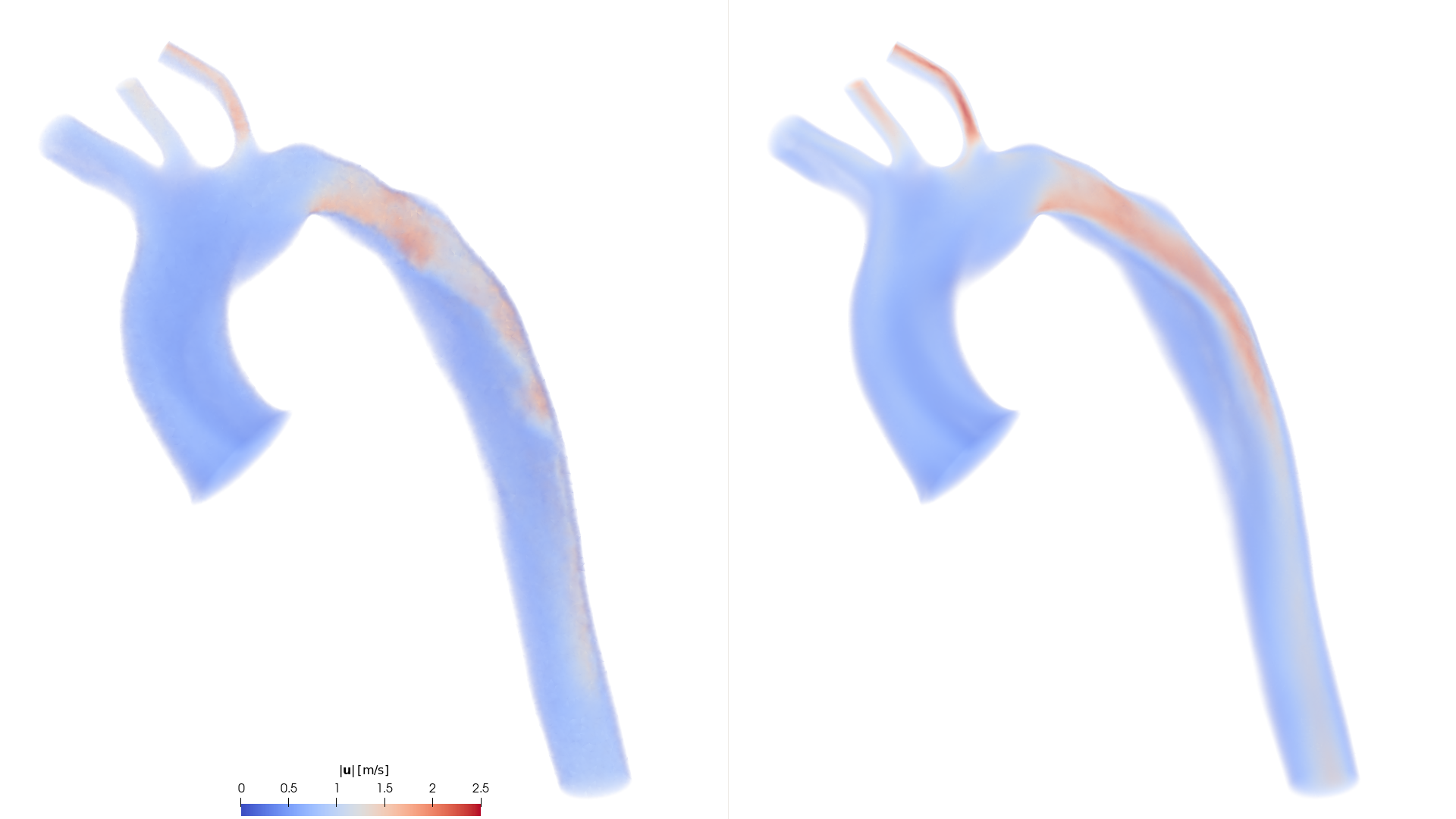}
        \caption[\impturb Velocity field comparison during deceleration, $\bsig$-model vs. refined RB-VMS]
        {
          Magnitude of the velocity field (volume plot) at $t = \unit[1.2]{s}$ (decelerating phase).
          Left: $\bsig$-model, $C_\sigma = 1.35$.
          Right: RB-VMS, $\mathrm P_1 / \mathrm P_1$ elements, fine mesh.
        }
        \label{fig:vol_1200_sigma_1.35_rbvms_p1_p1}
    \end{center}
\end{figure}

Figures~\ref{fig:vol_1125_rbvms_p1_p1_coarse_fine}
and~\ref{fig:vol_1200_sigma_1.35_rbvms_p1_p1} present the flow fields computed with
the RB-VMS model ($\mathrm P_1 / \mathrm P_1$) 
at one time instant
at peak flow and another time instant in the decreasing phase, where in the
latter figure also the corresponding picture for the $\bsig$-model is shown. It
can be observed that the flow fields for the RB-VMS model ($\mathrm P_1 / \mathrm P_1$)
are rather smooth, in particular on the
coarse mesh. Using a low order velocity space introduces therefore a comparatively large
amount of numerical diffusion.

\begin{figure}[t!]
    \begin{center}
        \includegraphics[width = 0.49 \textwidth]{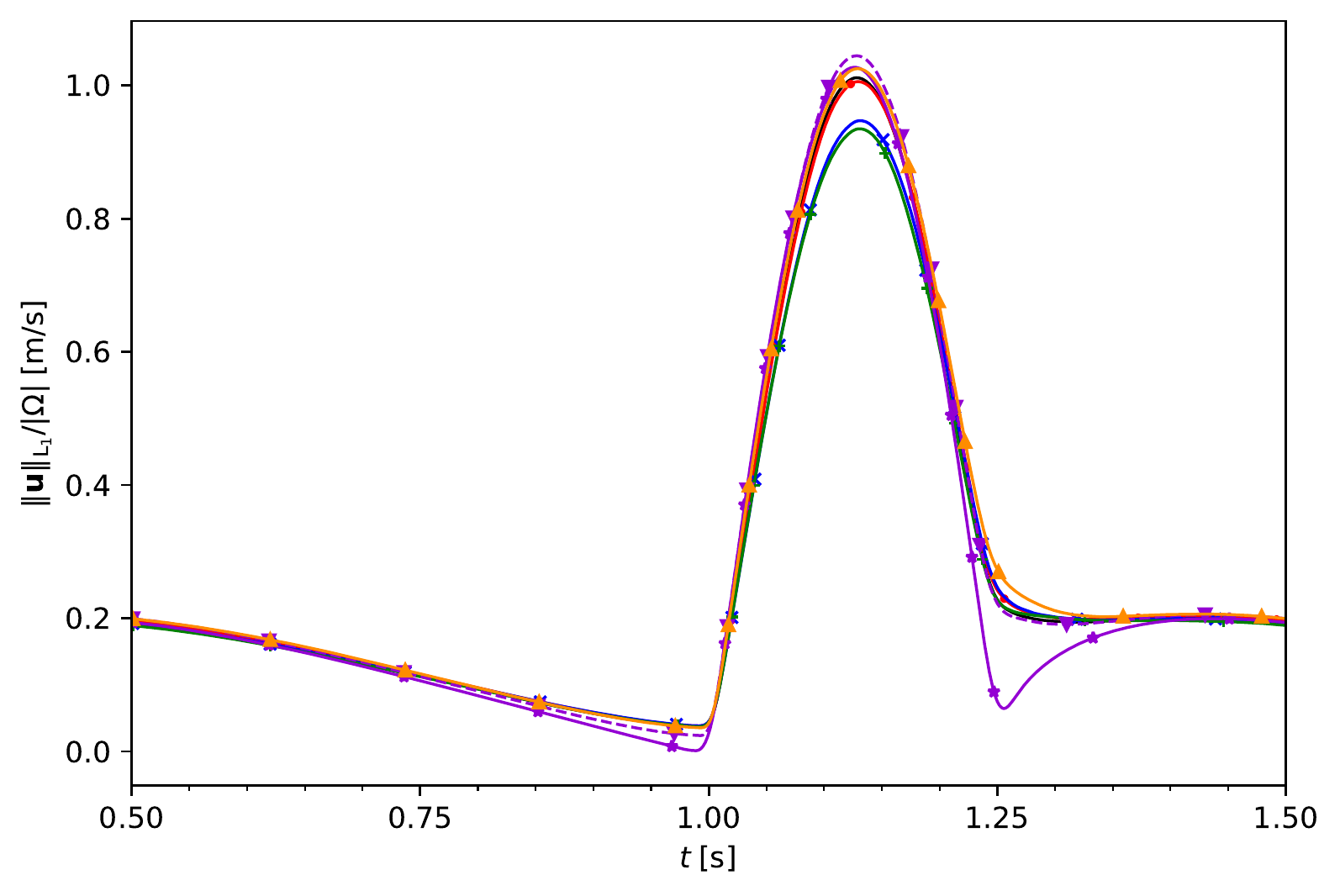}
        \includegraphics[width = 0.49 \textwidth]{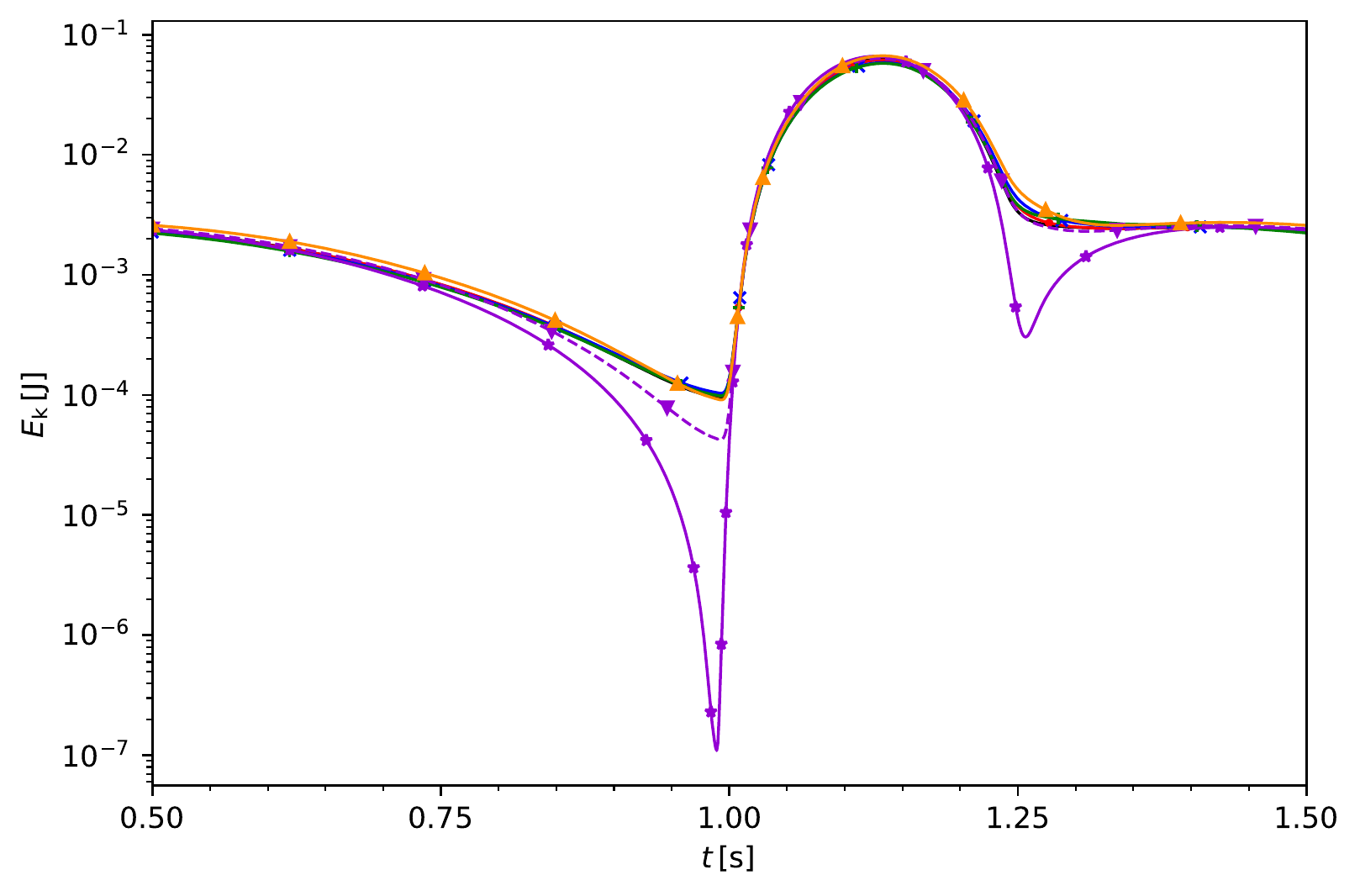}
        \caption[\impturb average velocity magnitude and kinetic energy]
        {
          Left: spatially averaged velocity magnitude over time. Right: kinetic energy over time.
          \turbleg
        }
        \label{fig:turb_energy}
    \end{center}
\end{figure}

This observation is confirmed by the quantitative comparisons in
Figure~\ref{fig:turb_energy}, showing that the velocity field computed with the
RB-VMS model using equal-order linear elements decays more quickly with
decreasing inflow, especially on the coarse mesh. The simulations with second
order velocity retain finer features, resulting in slower dissipation of energy
carried by small eddies.

\begin{figure}[t!]
    \begin{center}
        \includegraphics[width = 0.33 \textwidth]{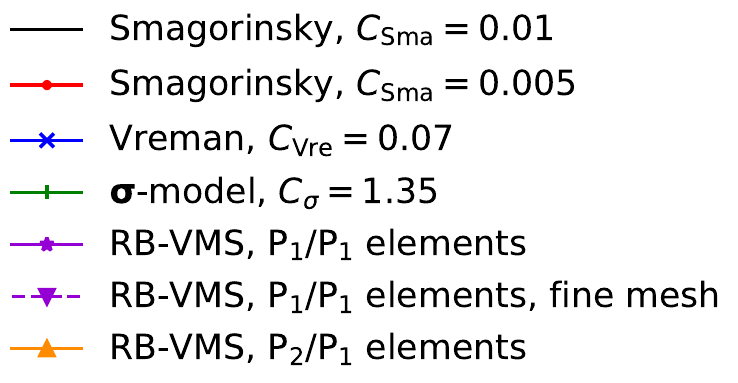}
        \caption[\impturb common legend]
        {
          Common legend for the figures describing the comparison among turbulence models.
        }
        \label{fig:turb_legend}
    \end{center}
\end{figure}

\subsubsection{Pressure difference} \label{sssec:turb_p}

Figure~\ref{fig:turb_legend} shows the legend convention used for the detailed
comparison of turbulence models below.
\begin{figure}[t!]
    \begin{center}
        \includegraphics[width = 0.49 \textwidth]{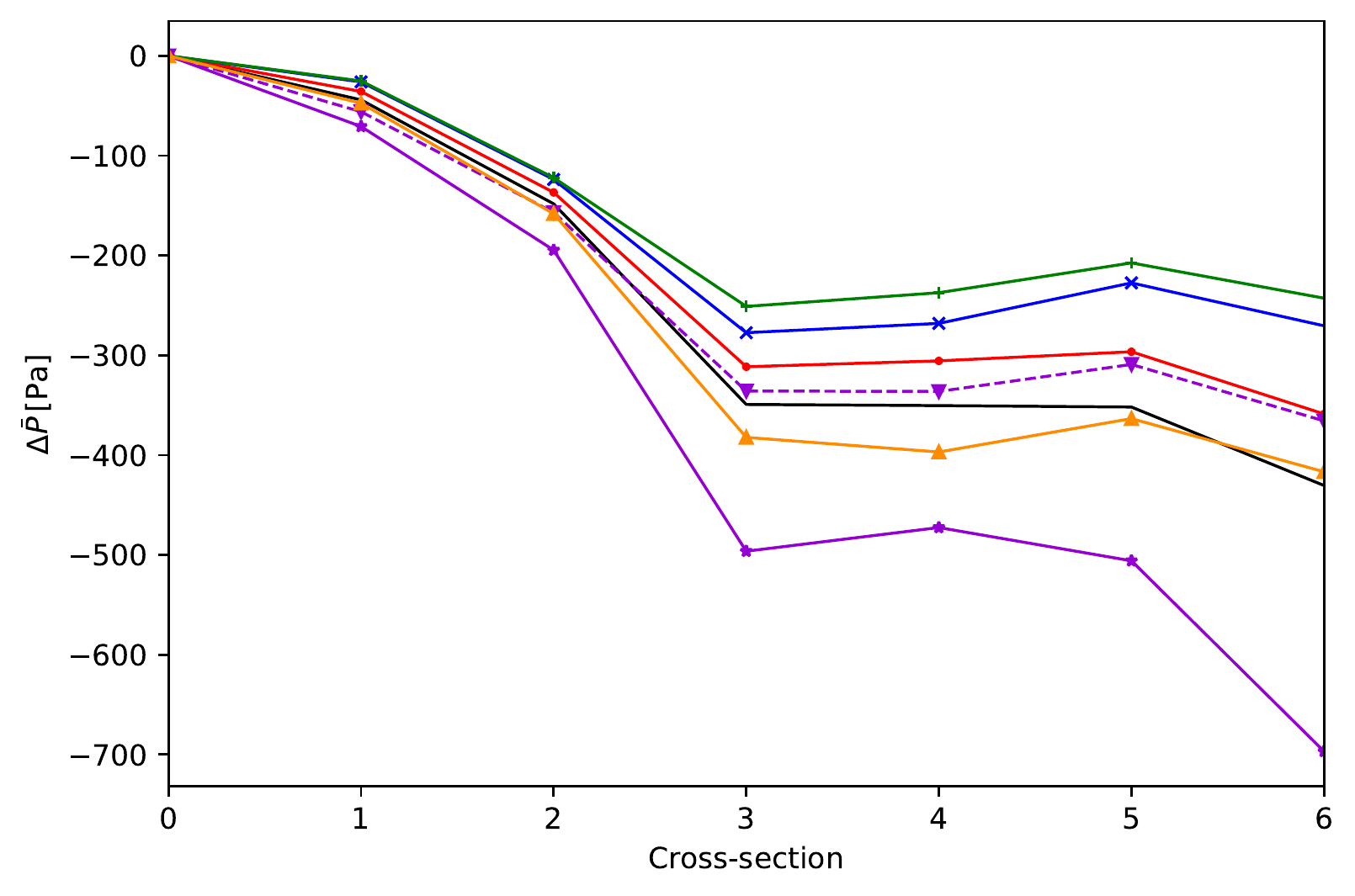}
        \includegraphics[width = 0.49 \textwidth]{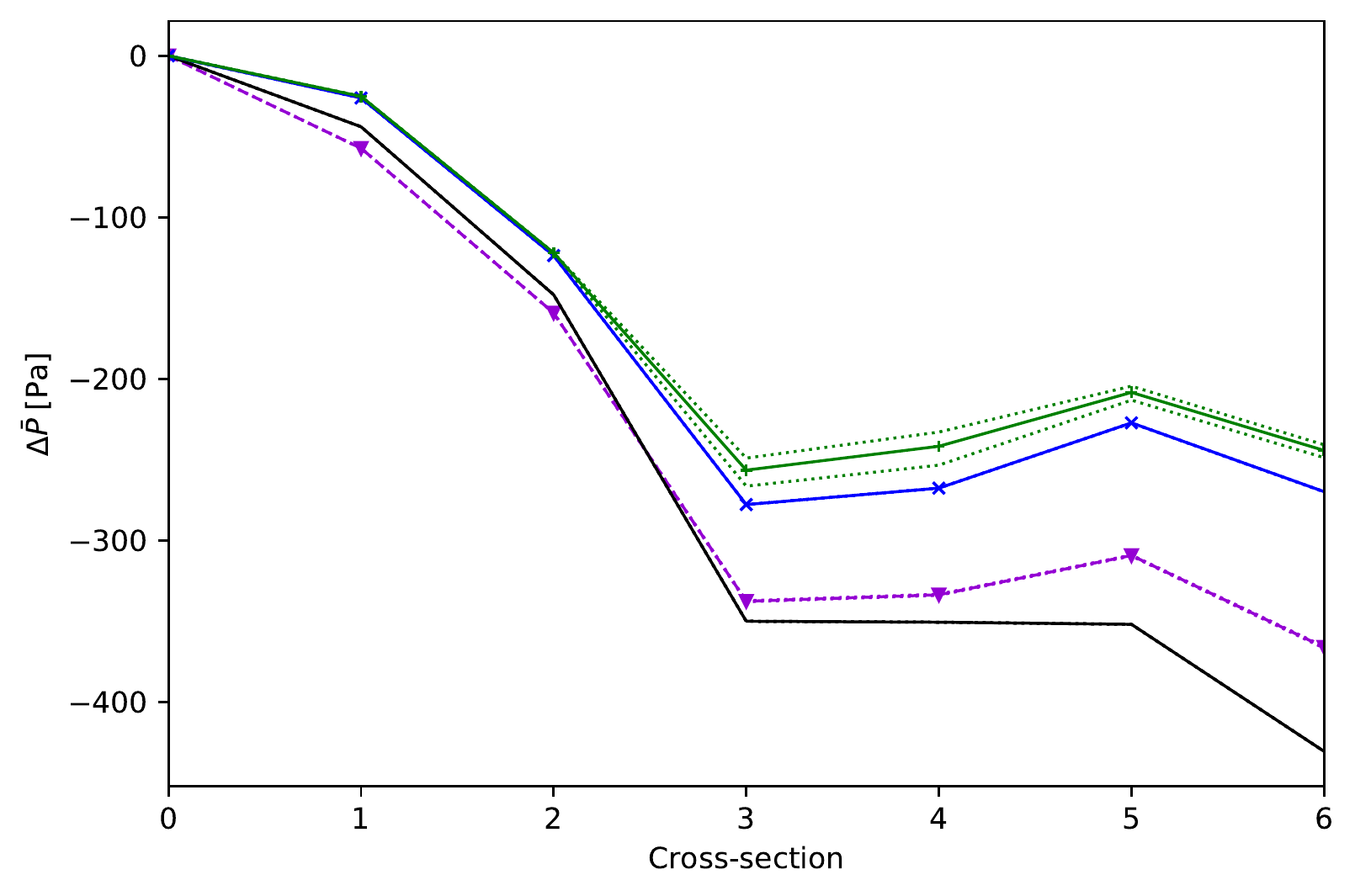}
        \includegraphics[width = 0.49 \textwidth]{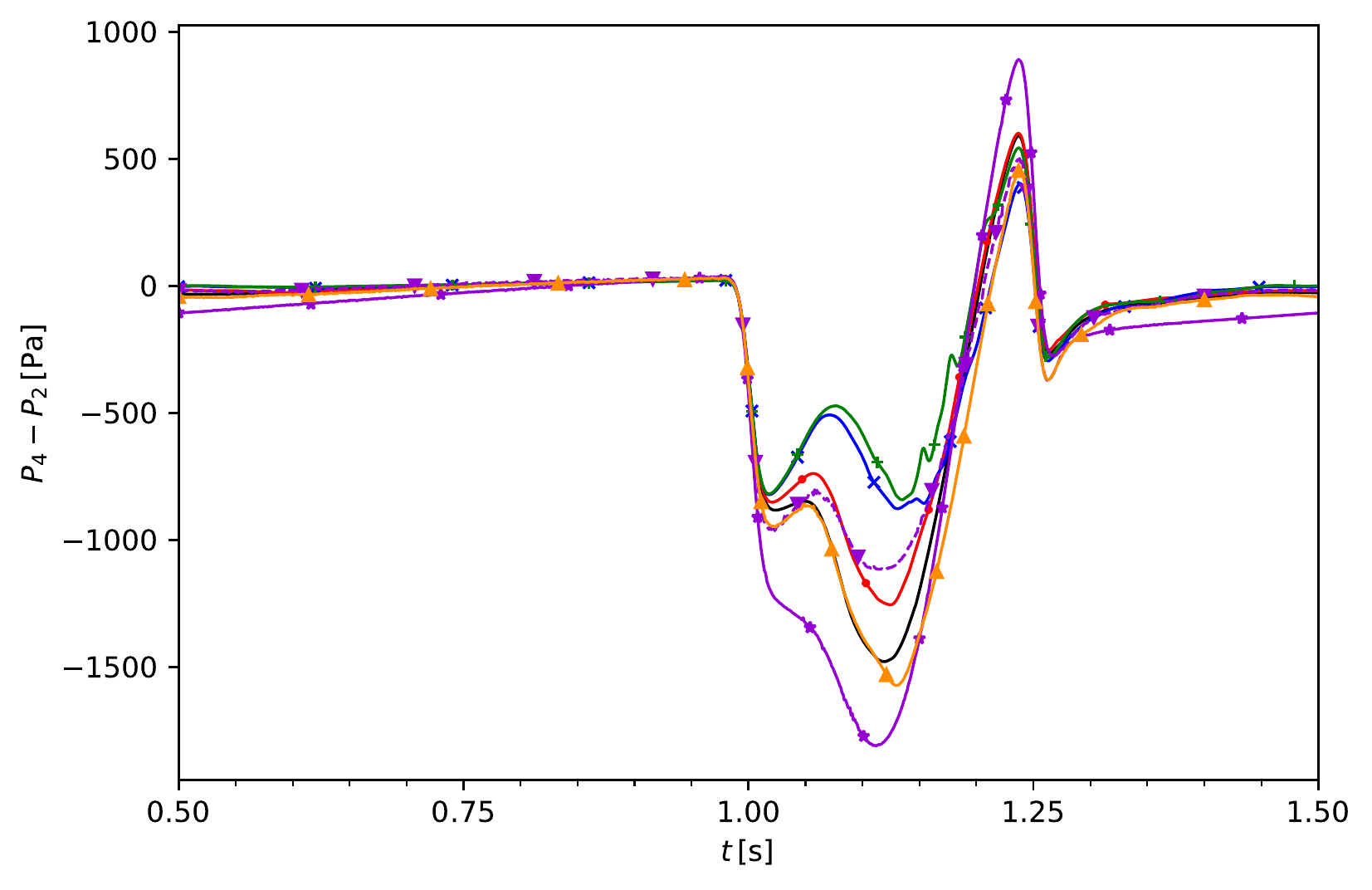}
        \includegraphics[width = 0.49 \textwidth]{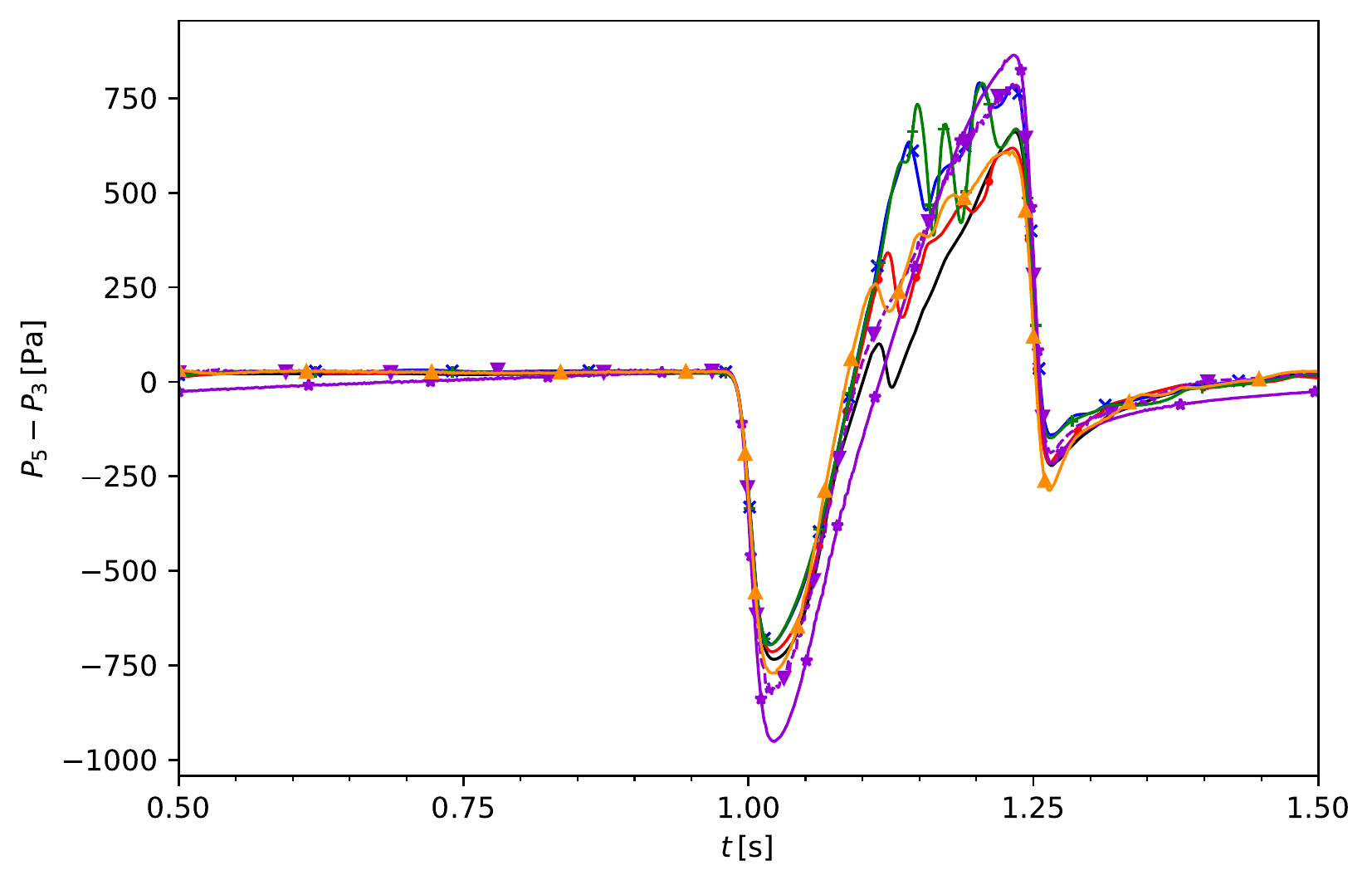}
        \caption[\impturb pressure differences]
        {
          Upper left: time-averaged pressure difference per cross-section.
          Upper right: long time-averaged pressure difference per cross-section.
          Lower left: pressure difference between cross-sections~4 and 2 over time.
          Lower right: pressure difference between cross-sections~5 and 3 over time.
          \turbleg
        }
        \label{fig:turb_p}
    \end{center}
\end{figure}
\begin{figure}[t!]
  \begin{center}
    \includegraphics[width = 0.49 \textwidth]{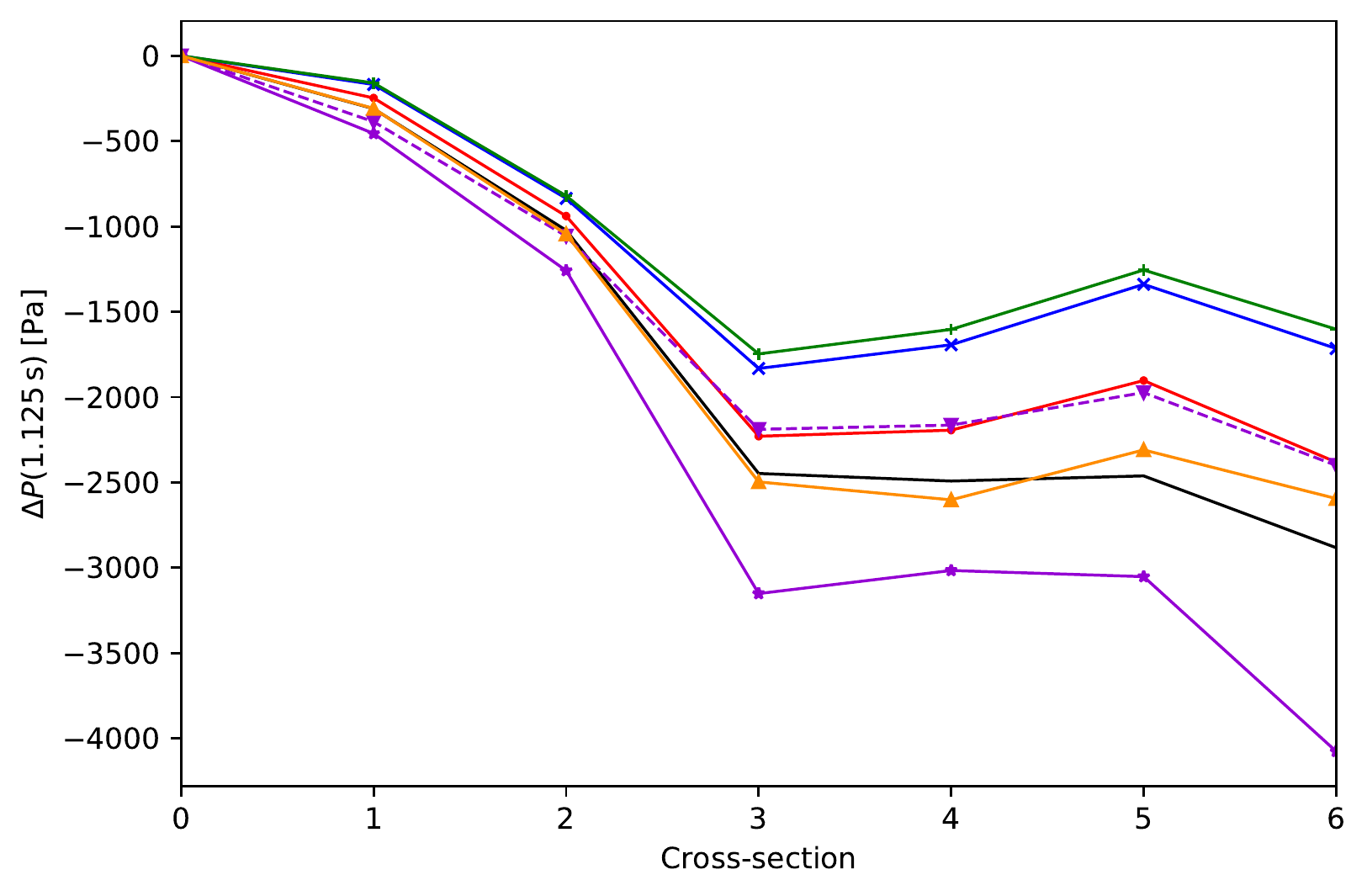}
    \includegraphics[width = 0.49 \textwidth]{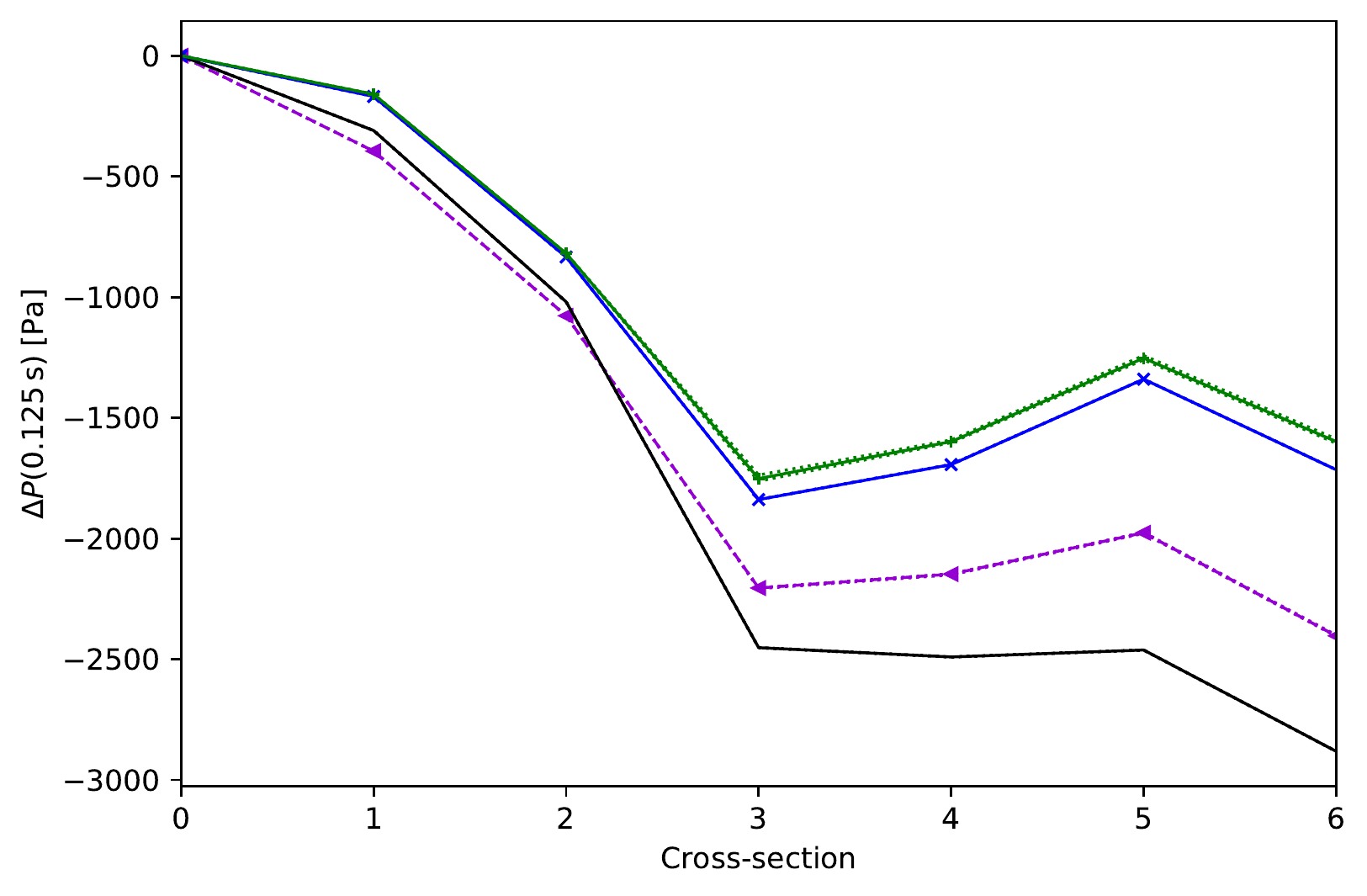}
    \includegraphics[width = 0.49 \textwidth]{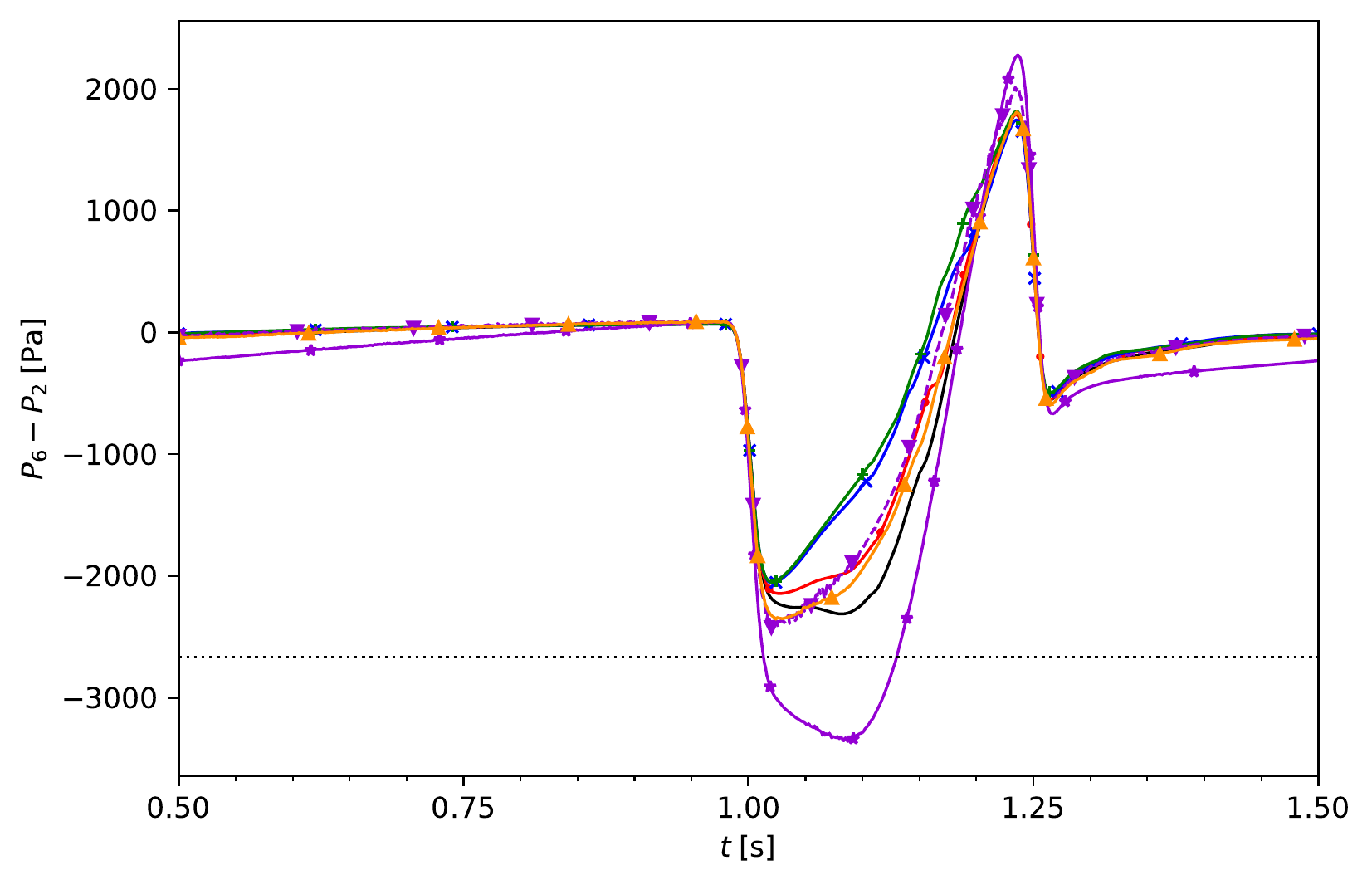}
    \includegraphics[width = 0.49 \textwidth]{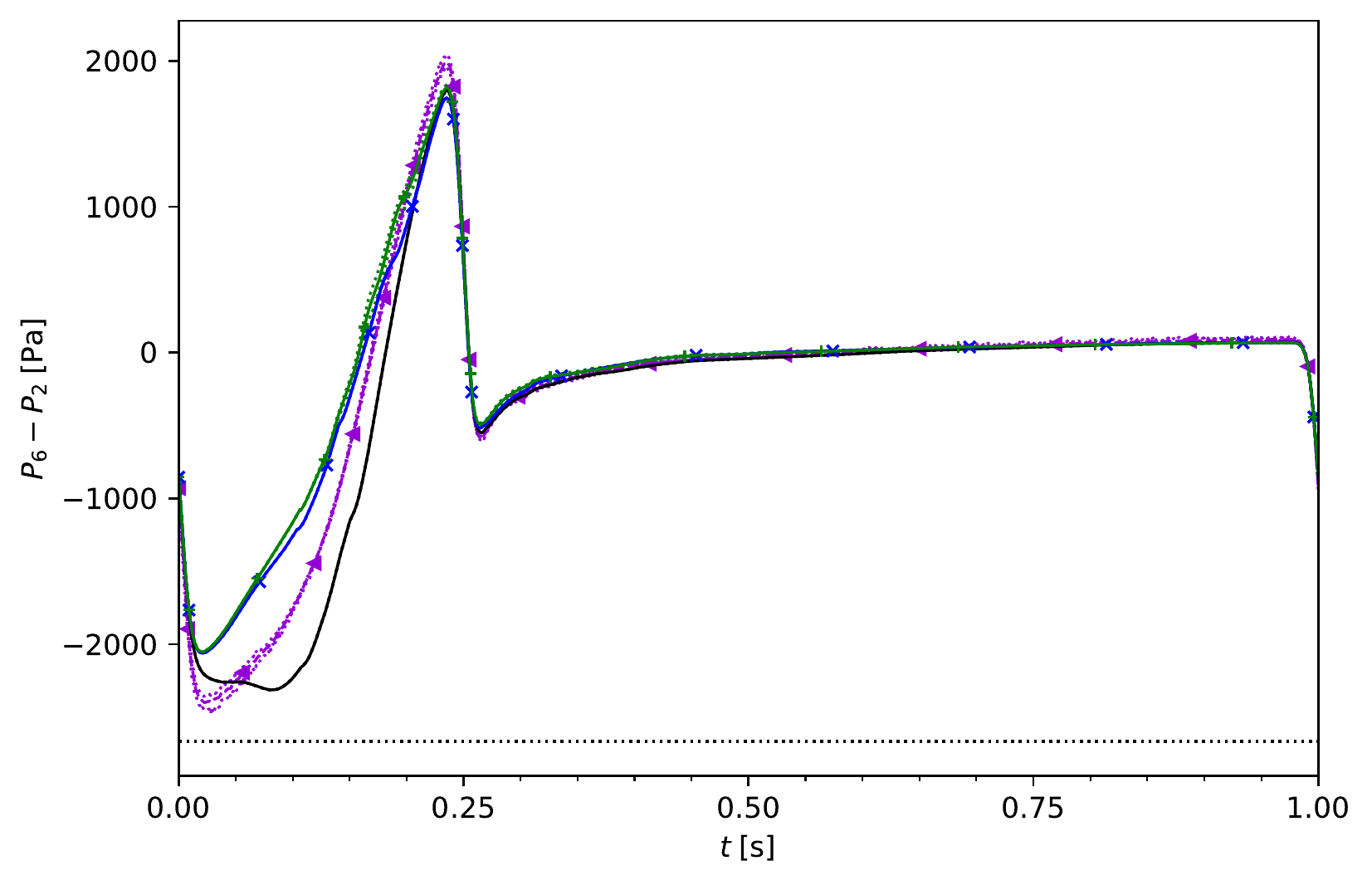}
    \caption[\impturb peak pressure differences]
    {
      Upper left: pressure difference per cross-section at peak flow.
      Upper right: pressure difference per cross-section at peak flow averaged over all simulation periods.
      Lower left: pressure difference between cross-sections~6 and 2 over time.
      Lower right: pressure difference between cross-sections~6 and 2 over time
        over a heartbeat, averaging the results for the corresponding time
        instant over the simulated beats. The $\unit[20]{mmHg}$ threshold, which
        is considered an indication of severe coarctation, is marked by a dotted
        line.
      \turbleg
    }
    \label{fig:turb_max_p}
  \end{center}
\end{figure}
Figures~\ref{fig:turb_p} and \ref{fig:turb_max_p} present the results for the
cross-sectional pressures.

The plots in the upper row of Figure~\ref{fig:turb_p} depict the difference with
respect to the first cross section. Mean pressures are averaged over a single
period (upper-left) and over the longer interval of 30 periods (upper-right).
The lower pictures show the pressure difference between two selected pairs of
cross-sections over one pulse period.

Figure~\ref{fig:turb_max_p} instead shows, in the upper row, the instantaneous
pressure differences  at peak flow, for one period (left) and averaged over $30$
periods (top right).

Although the qualitative behavior is similar, the models differ widely in scale.
The RB-VMS model with $\mathrm{P}_1 / \mathrm{P}_1$ elements on the coarse mesh
in particular shows a pronounced overestimation of pressure differences compared
to the other models, which is likely an artifact of the excessive numerical
dissipation discussed above. Refining the mesh results in values closer to the other
models. The average pressure differences computed with the
$\bsig$-model and the Vreman model are very similar for all cross-sections; the
results computed using the Smagorinsky model with $C_{\mathrm{Sma}} = 0.005$ is
closer to these than those computed using $C_{\mathrm{Sma}} = 0.01$. The latter
results, in turn, are close to those obtained with the RB-VMS model with
$\mathrm{P}_2 / \mathrm{P}_1$ elements.

Discarding the RB-VMS model with $\mathrm{P}_1 / \mathrm{P}_1$ elements on the
coarse mesh, the largest time-averaged pressure difference is still about twice
the smallest. For instance, the difference between cross-sections~4 and 2 ranges
from $\unit[115.2]{Pa}$ ($\bsig$-model, $C_\sigma = 1.35$) to $\unit[238.7]{Pa}$
(RB-VMS model with $\mathrm P_2 / \mathrm P_1$ elements).

Whereas the average pressure increases slightly from cross-section~3 to
cross-section~5 in most models, by up to $\unit[49.9]{Pa}$ for the Vreman model
with $C_\mathrm{Vre} = 0.07$, the Smagorinsky model with $C_\mathrm{Sma} = 0.01$
exhibits a decrease by $\unit[2.6]{Pa}$. The RB-VMS model with
$\mathrm{P}_1 / \mathrm{P}_1$ elements on the coarse mesh is the only other
model not to show an increase here.

The pressures at peak flow behave similarly. Peak pressure
generally occurs during acceleration, slightly before peak flow.

Looking at the pressure's behavior over time gives some insight into these
differences. Every model exhibits an inversion of the pressure difference
between consecutive cross-sections as the flow decelerates towards diastole, as
one would expect. However, as a prominent jet forms at the narrowed exit of the
aortic arch and begins to shed vortices beneath it, an inverted pressure
gradient emerges from the jet's deceleration as it dissipates into the wider
descending aorta, and the pressure waves associated with the shed vortices
manifest as oscillations in the pressure plots in Figure~\ref{fig:turb_p}. These
effects emerge earlier and more clearly in the results given by less diffusive
models, particularly the $\bsig$-model. In other models, increased numerical
diffusion results in a much cleaner jet that remains coherent further down the
descending aorta (compare Figures~\ref{fig:vol_1125_rbvms_p1_p1_coarse_fine}
and \ref{fig:vol_1200_sigma_1.35_rbvms_p1_p1}), resulting in the larger
pressure difference between cross-sections~2 and 4 (compare
Figure~\ref{fig:turb_p}, lower left).

In Figure~\ref{fig:turb_max_p} (bottom row) an additional dotted line shows the
pressure difference of $\unit[20]{mmHg}$. Peak systolic pressure exceeding this
value is indicated in recent guidelines as a marker of a severe
coarctation\cite{circulation-2019-guidelines}\cite{esc_aorta_2014}.
One can see that only the RB-VMS model with
$\mathrm{P}_1 / \mathrm{P}_1$ on the coarse mesh exceeds the $\unit[20]{mmHg}$
threshold. While the finer models vary in the shape and timing of the pressure
peak during acceleration, this peak remains in all cases slightly below the critical threshold.

The long term averages show very little distinction from period to period in all
simulations except those using the $\bsig$-model. Here one can observe that
moderate period-wise differences increase after the coarctation, as the eddying
around the jet is not identical each time. The overall effect of this behavior
on the pressure is smaller further down the aorta, where the downward flow
begins to relaminarize somewhat as the smallest eddies dissipate.

\subsubsection{Maximum velocity} \label{sssec:turb_max_u}

\begin{figure}[t!]
  \begin{center}
    \includegraphics[width = 0.49 \textwidth]{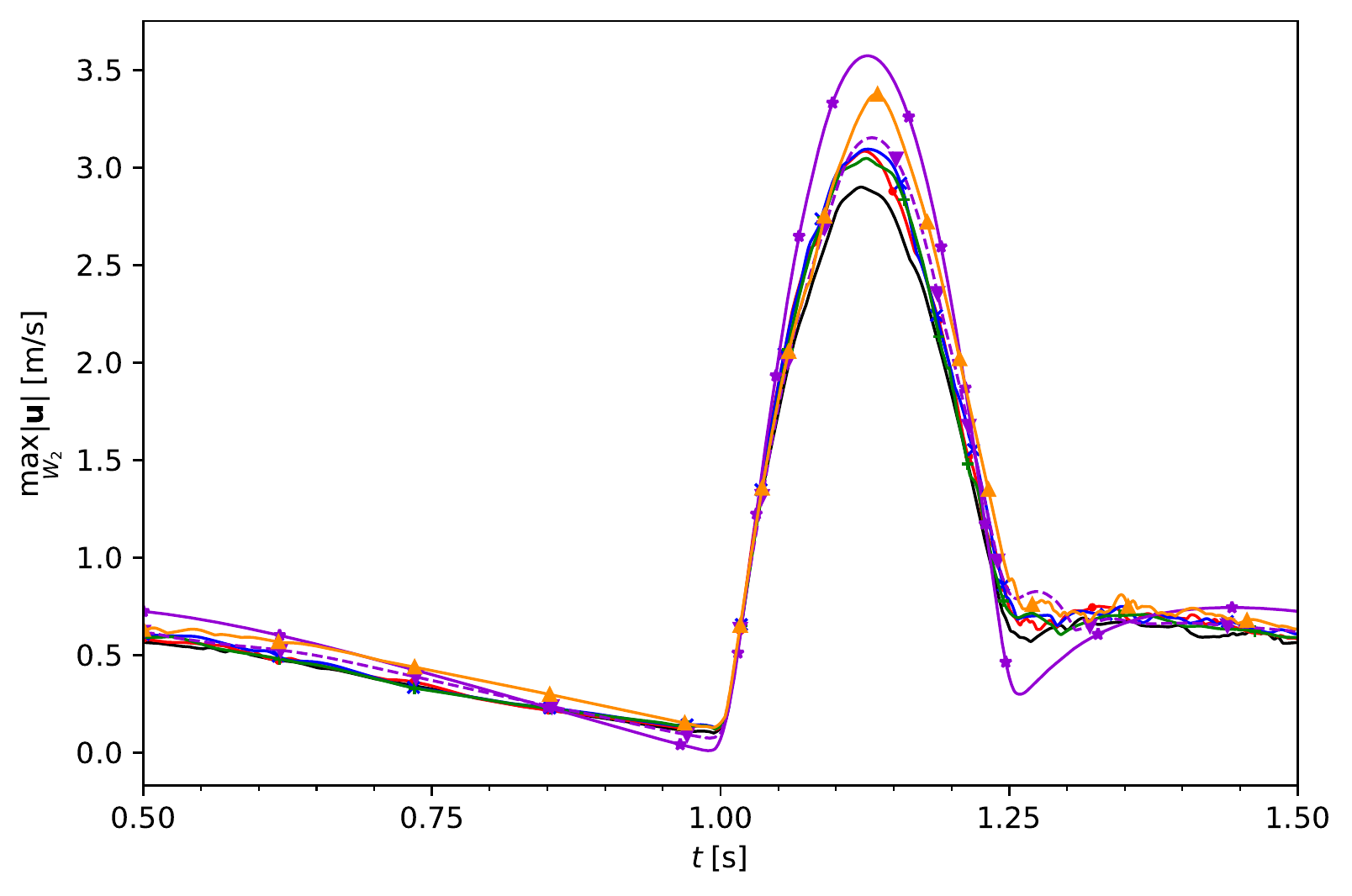}
    \includegraphics[width = 0.49 \textwidth]{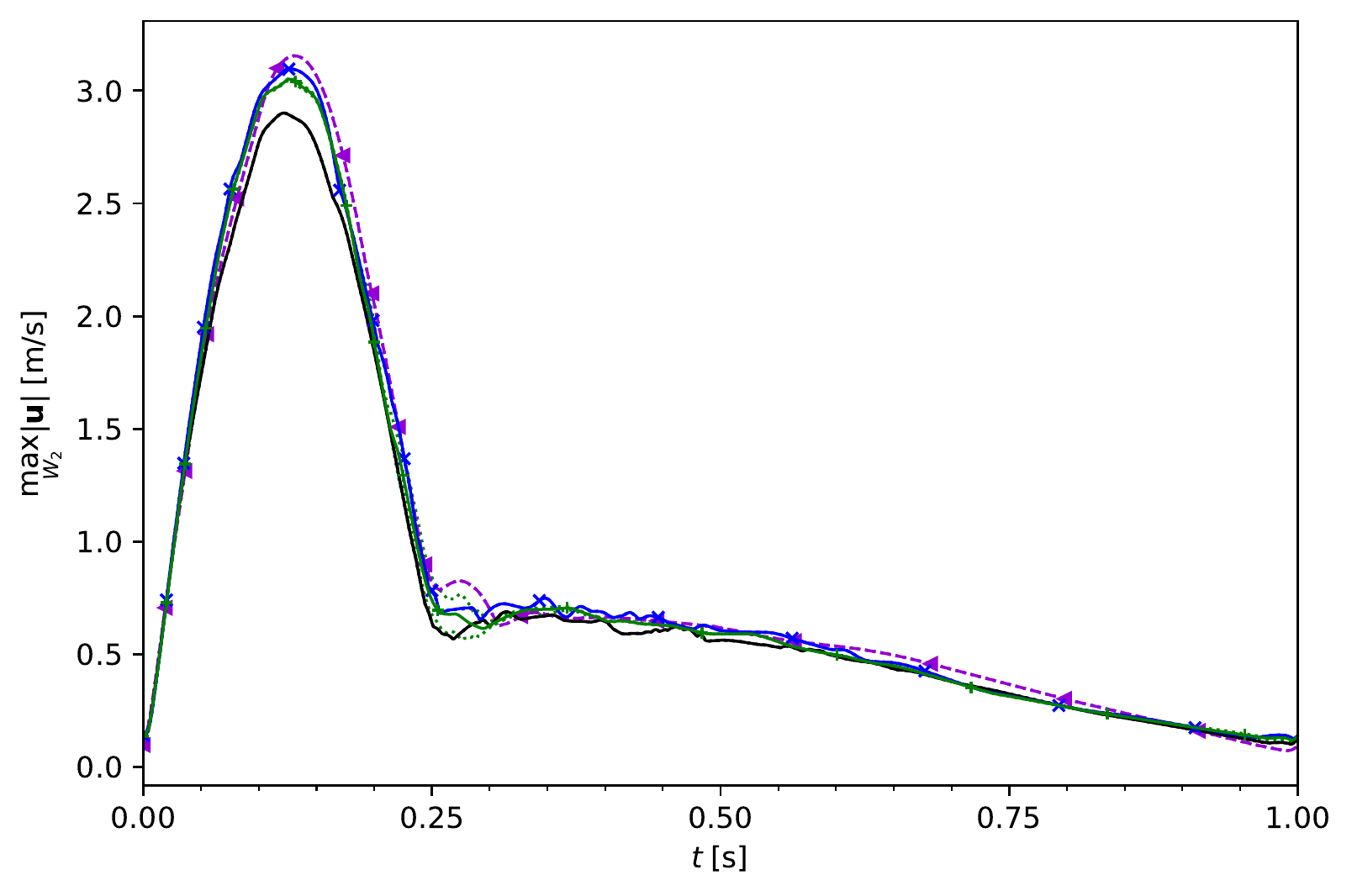}
    \includegraphics[width = 0.49 \textwidth]{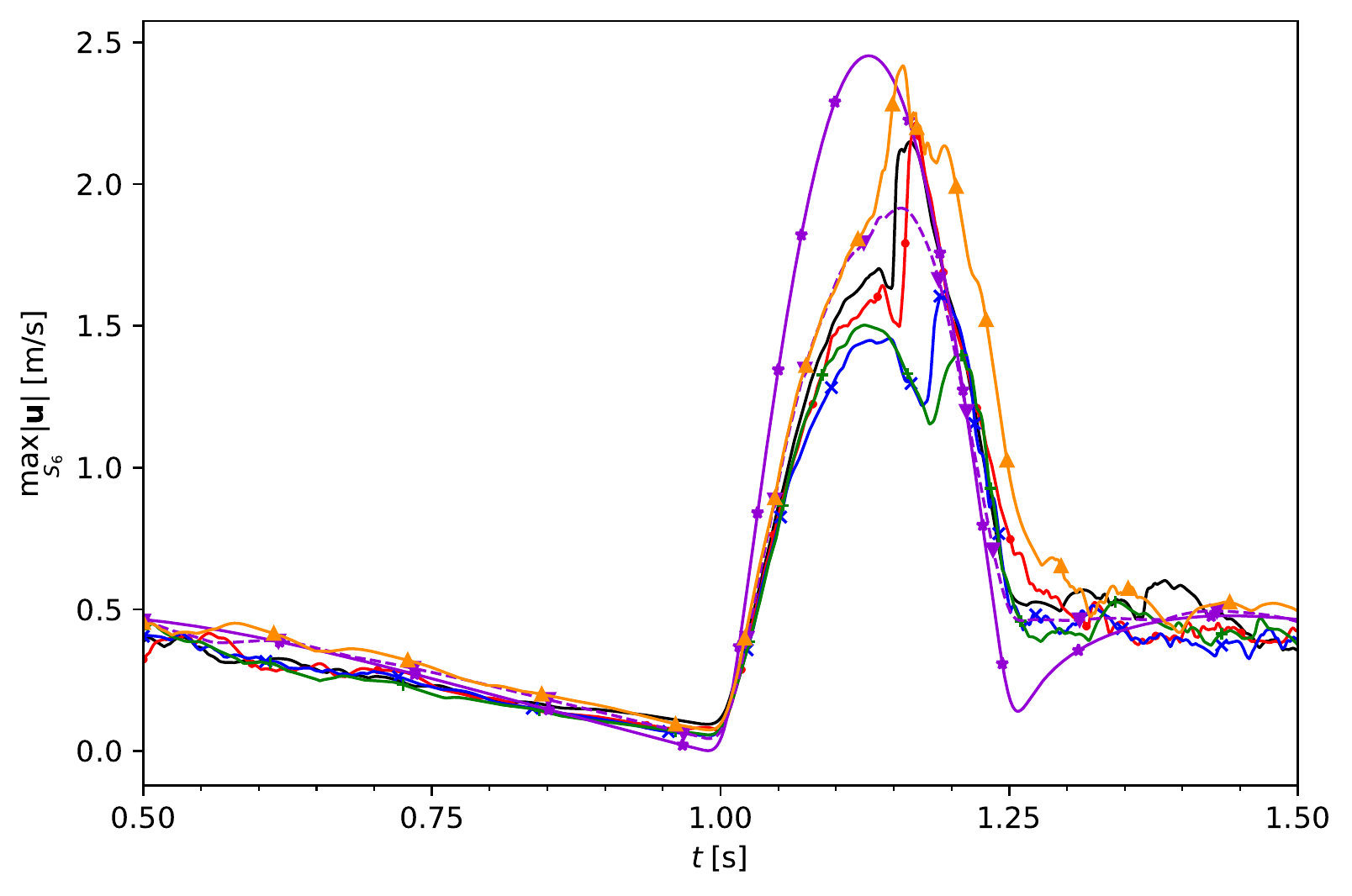}
    \includegraphics[width = 0.49 \textwidth]{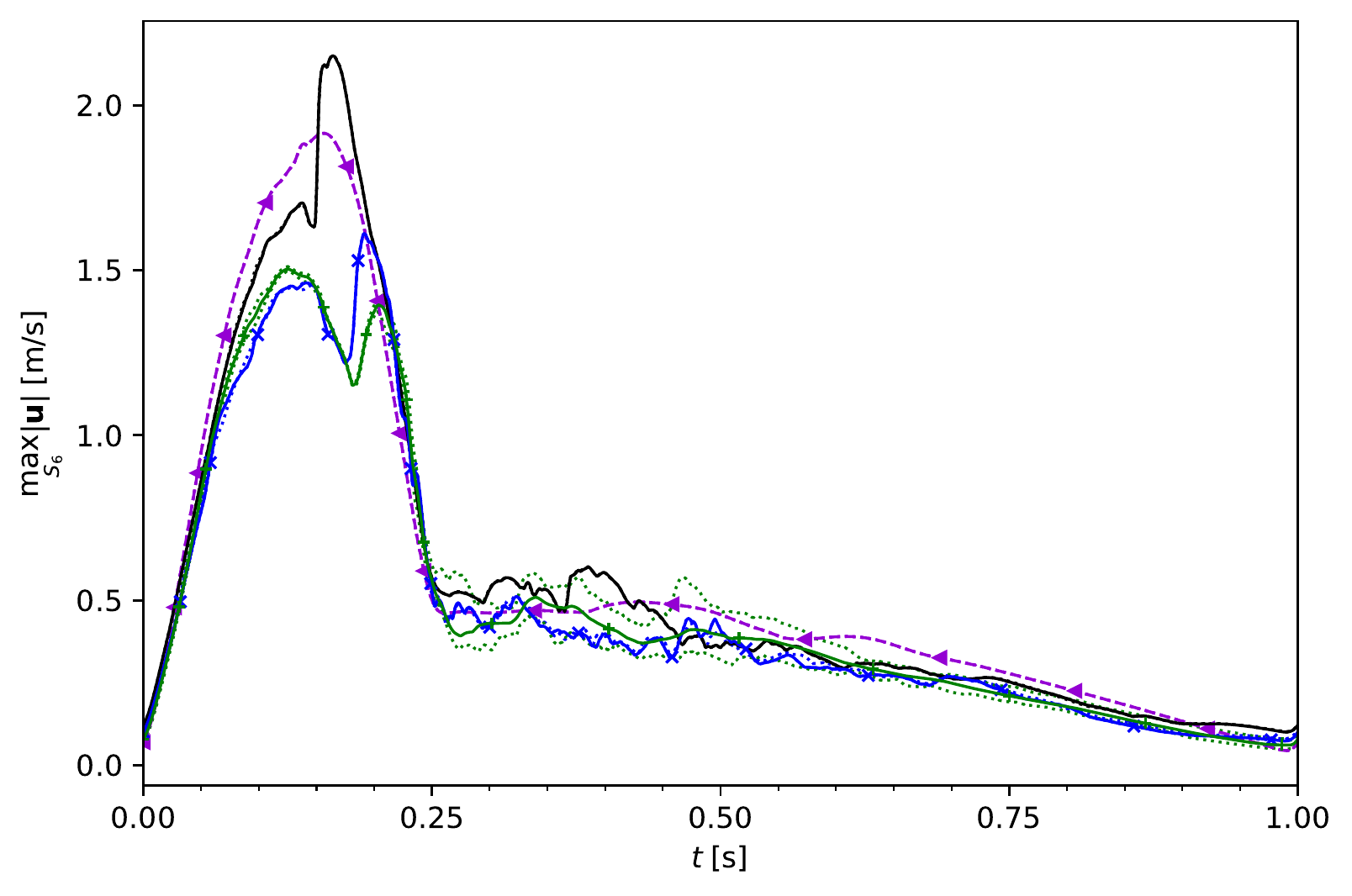}
    \caption[\impturb maximum velocity]
    {
      Upper left: maximum velocity through the wedge between cross-sections~2
        and 3 over time.
      Upper right: maximum velocity through the wedge between cross-sections~2
        and 3 over time over a heartbeat, averaging the results for the
        corresponding time instant over the simulated beats.
      Lower left: maximum velocity through cross-section~6 over time.
      Lower right: maximum velocity through cross-section~6 over time over a
        heartbeat, averaging the results for the corresponding time instant over
        the simulated beats.
      \turbleg
    }
    \label{fig:turb_max_u}
  \end{center}
\end{figure}

Figure~\ref{fig:turb_max_u} displays the maximum velocity magnitude over the
wedge between cross-sections~2 and 3 (top), straddling the coarctation, and
across cross-section~6 (bottom), further down the descending aorta.

The maximum stenotic velocity (the peak in Figure~\ref{fig:turb_max_u}, top row) shows a
clustering of most models around $\unitfrac[3.1]{m}{s}$.
 The RB-VMS models on
the coarse mesh notably exceed this value, whilst  the Smagorinsky model with
$C_\mathrm{Sma} = 0.01$ yields a slightly lower peak velocity ($\unitfrac[2.9]{m}{s}$).
In image-based clinical assessment, this range of values might indicate the
presence of a mild stenosis, but it is well below the critical value of
$\unitfrac[5]{m}{s}$ considered as a marker of a severe condition.

The behavior of the velocity in the descending aorta
(Figure~\ref{fig:turb_max_u}, bottom), while less clinically relevant, is
helpful in distinguishing the models' behaviors. The eddy viscosity models each
match the jet's development by showing a distinct dip followed by a secondary
peak, but the height and timing vary considerably. The RB-VMS model with
$\mathrm{P}_1 / \mathrm{P}_1$ elements -- notably on both meshes -- behaves much
more smoothly and indistinctly. The decaying vortices shed by the jet above are
also much less visible in these models' plots during diastole.

Only the $\bsig$ model shows minor variability from period to period,
particularly during diastole.

\subsubsection{Secondary flow degree} \label{sssec:turb_sfd}

\begin{figure}[t!]
    \begin{center}
        \includegraphics[width = 0.49 \textwidth]{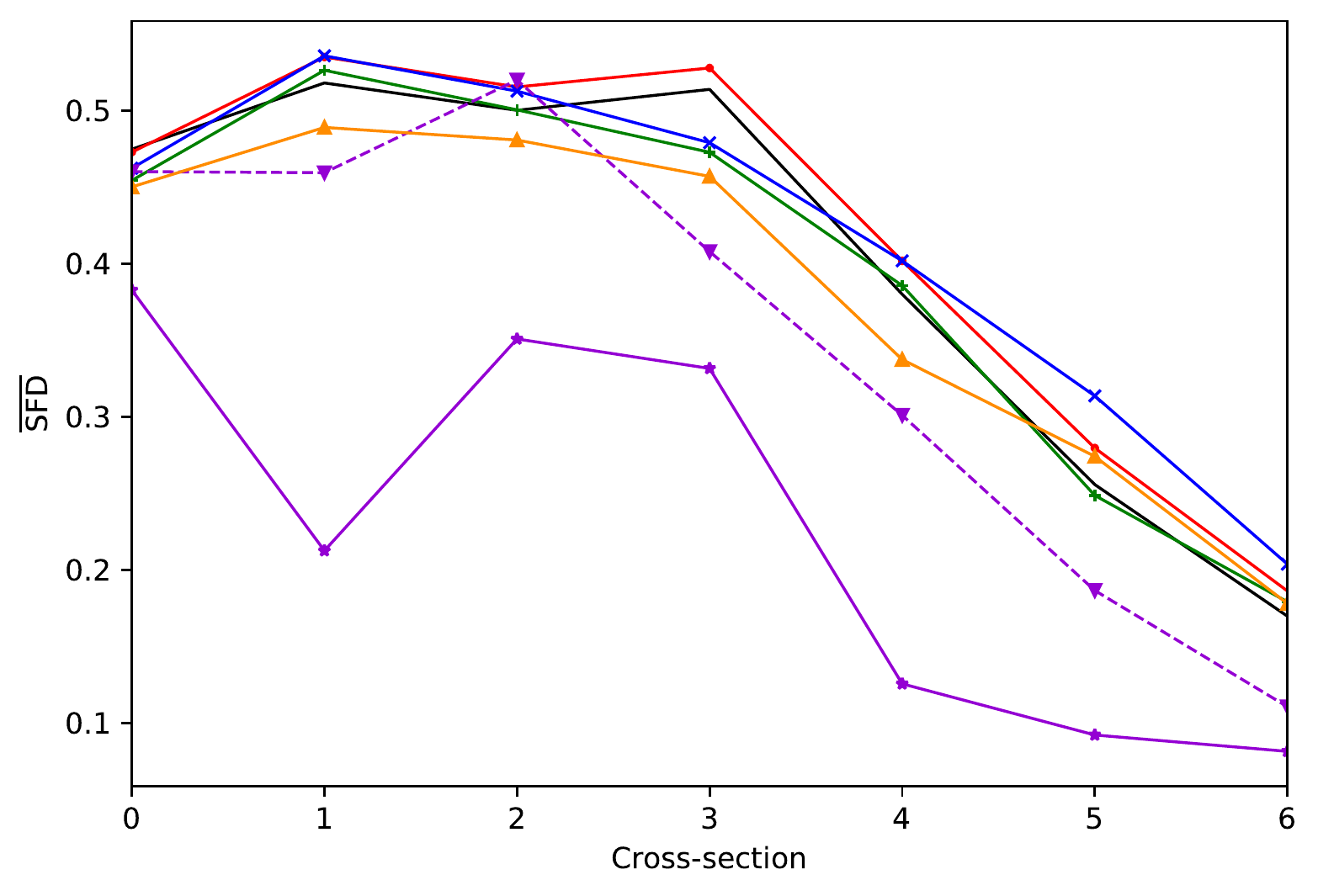}
        \includegraphics[width = 0.49 \textwidth]{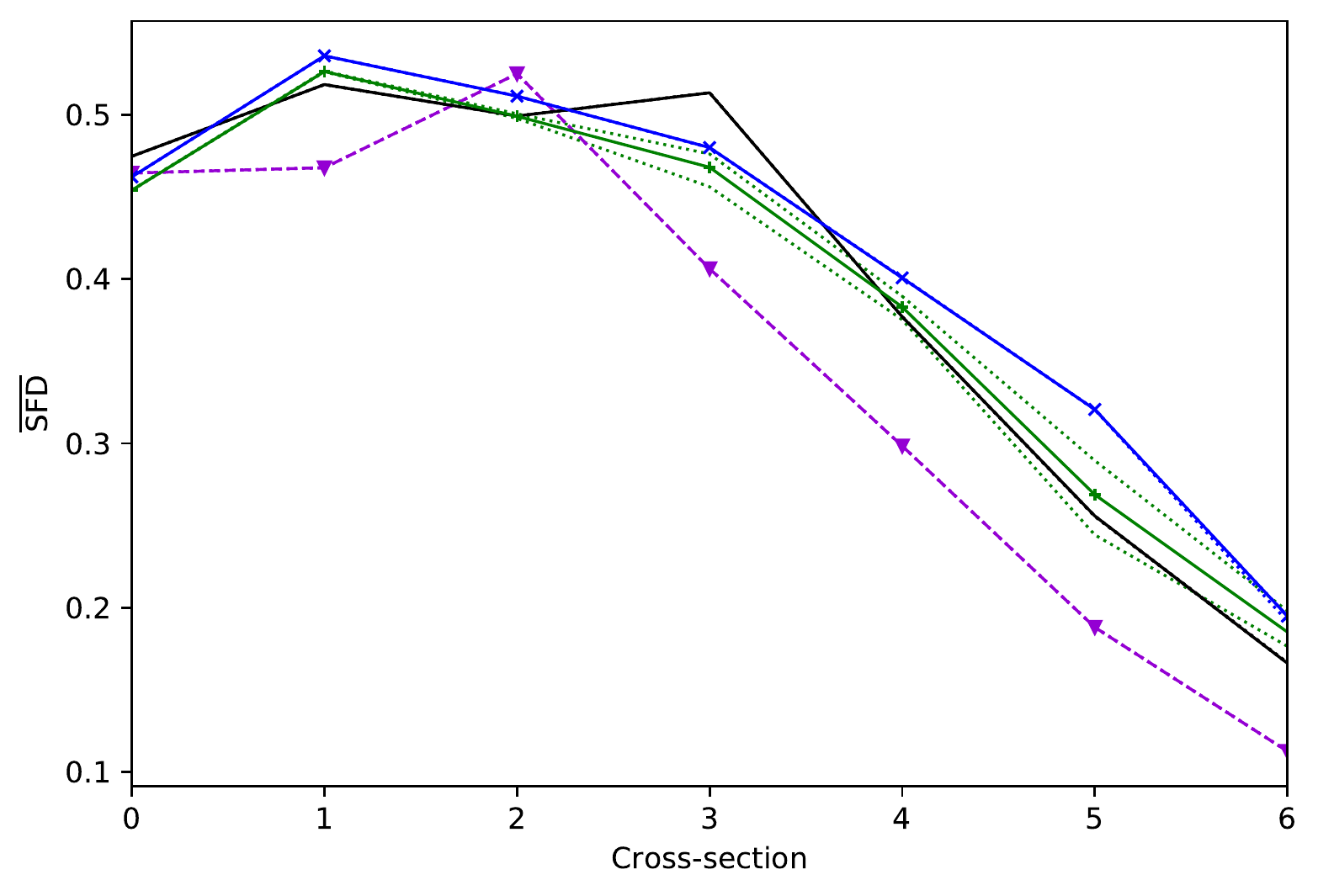}
        \includegraphics[width = 0.49 \textwidth]{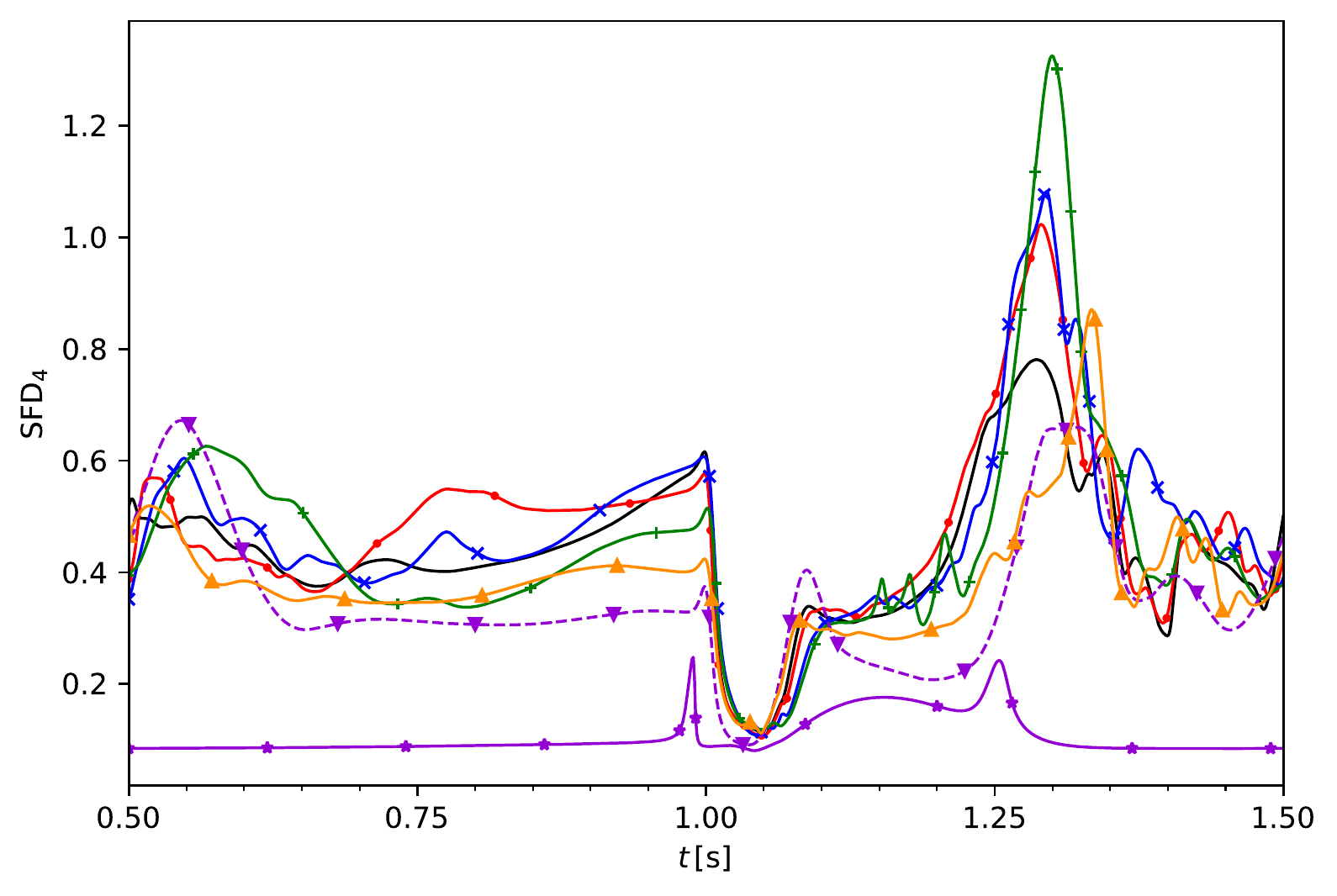}
        \includegraphics[width = 0.49 \textwidth]{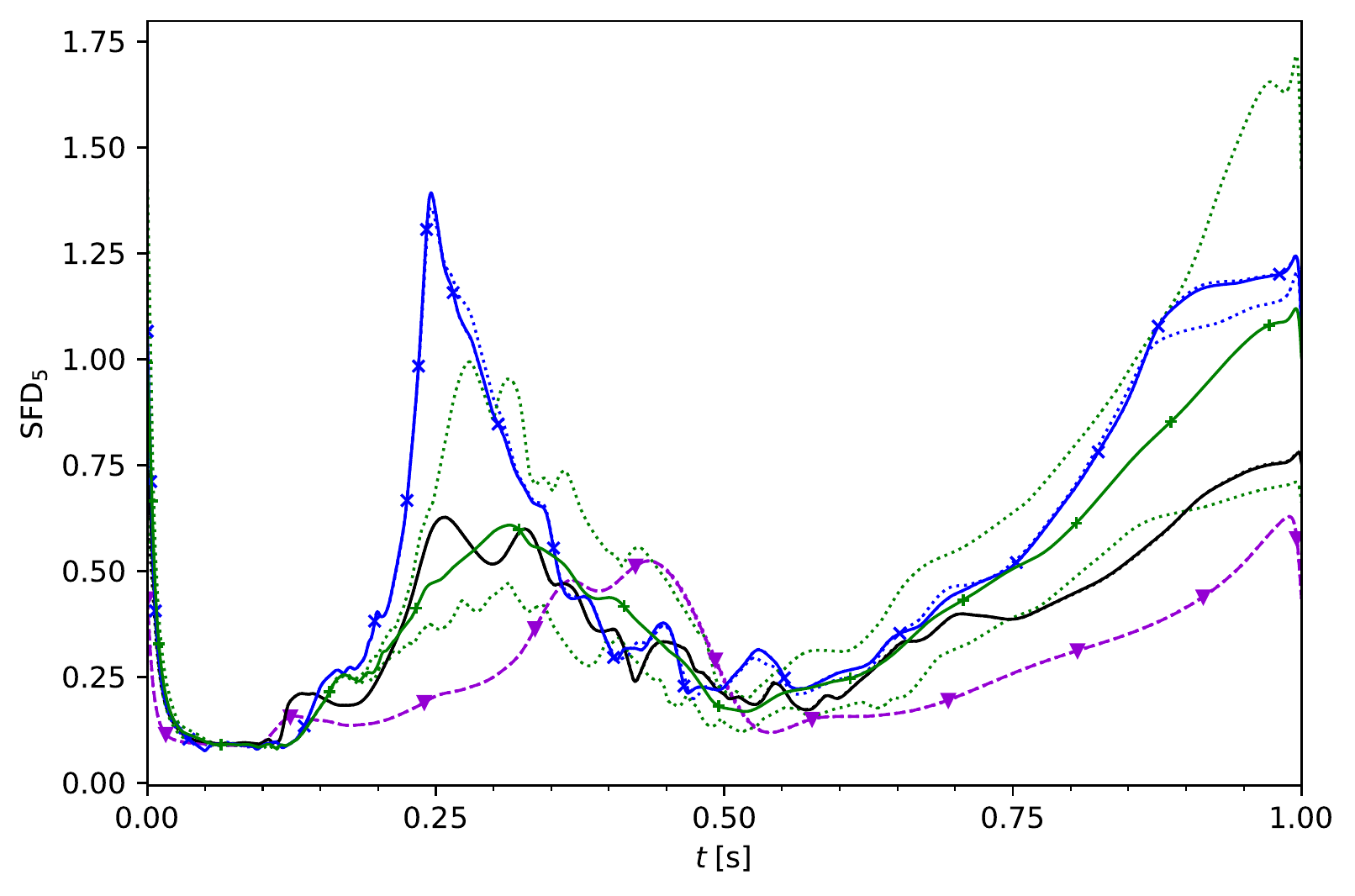}
        \caption[\impturb secondary flow degree]
        {
          Upper left: time-averaged secondary flow degree per cross-section.
          Upper right: long time-averaged secondary flow degree per cross-section.
          Lower left: secondary flow degree across cross-section 4 over time.
          Lower right: secondary flow degree across cross-section 5 over time
            during a heartbeat, averaging the results for the corresponding time
            instant over the simulated beats.
          \turbleg
        }
        \label{fig:turb_sfd}
    \end{center}
\end{figure}

The secondary flow degree averaged over time is shown in Figure~\ref{fig:turb_sfd}  (top).
Excluding the RB-VMS model with $\mathrm{P}_1 / \mathrm{P}_1$ elements on the
coarse grid, which delivers numerical results which are clearly different from
all other models,
the time-averages are clustered rather closely together. The largest difference
between models is $0.127$ (on cross-section~5) when the refined
$\mathrm{P}_1 / \mathrm{P}_1$ model is included; the largest difference between
the $\mathrm{P}_2 / \mathrm{P}_1$ models is $0.071$ (on cross-section~3).

The plots of the SFD across two cross-sections over time
(Figure~\ref{fig:turb_sfd}, bottom) show a more irregular behavior. However, the
curves exhibit peaks and valleys at roughly the same times for most models: SFD
increases as the velocity field decays away from forward flow towards the end of
diastole, then decreases rapidly as the inflow begins to accelerate (from
$t = \unit[1]{s}$ to $t = \unit[1.05]{s}$). The vortex shedding in the
descending aorta (see also the pronounced jet visible in
Figure~\ref{fig:vol_1200_sigma_1.35_rbvms_p1_p1}) increases well before
peak inflow time ($t = \unit[1.125]{s}$), and the SFD peaks again just before
the secondary inflow increase visible in Figure~\ref{fig:inflow_amplitude}.

Apart from small peaks around the inflow minima, the SFD predicted with the
$\mathrm{P}_1 / \mathrm{P}_1$ RB-VMS model on the coarse grid remains almost
constant, close to its minimum, indicating effectively laminar flow not
precisely normal to the cross-section. Using a second order velocity in the
RB-VMS model or refining the mesh gives results that are much more similar to
the other models.

In the long term simulation the $\bsig$-model is the only one to show a large
variation in the SFD from period to period. Here, the differences in position
and dissipation of the vortices formed below the jet at the exit of the aortic
arch lead to substantial SFD variance in the upper descending aorta.

It is clear that the results obtained with the $\mathrm{P}_1 / \mathrm{P}_1$
RB-VMS model on the coarse grid are very inaccurate for the quantities of
interest discussed so far. This model will no longer be considered in detail
below, although the corresponding results will still be displayed in the
figures.

\subsubsection{Normalized flow displacement} \label{sssec:turb_nfd}

\begin{figure}[t!]
    \begin{center}
        \includegraphics[width = 0.49 \textwidth]{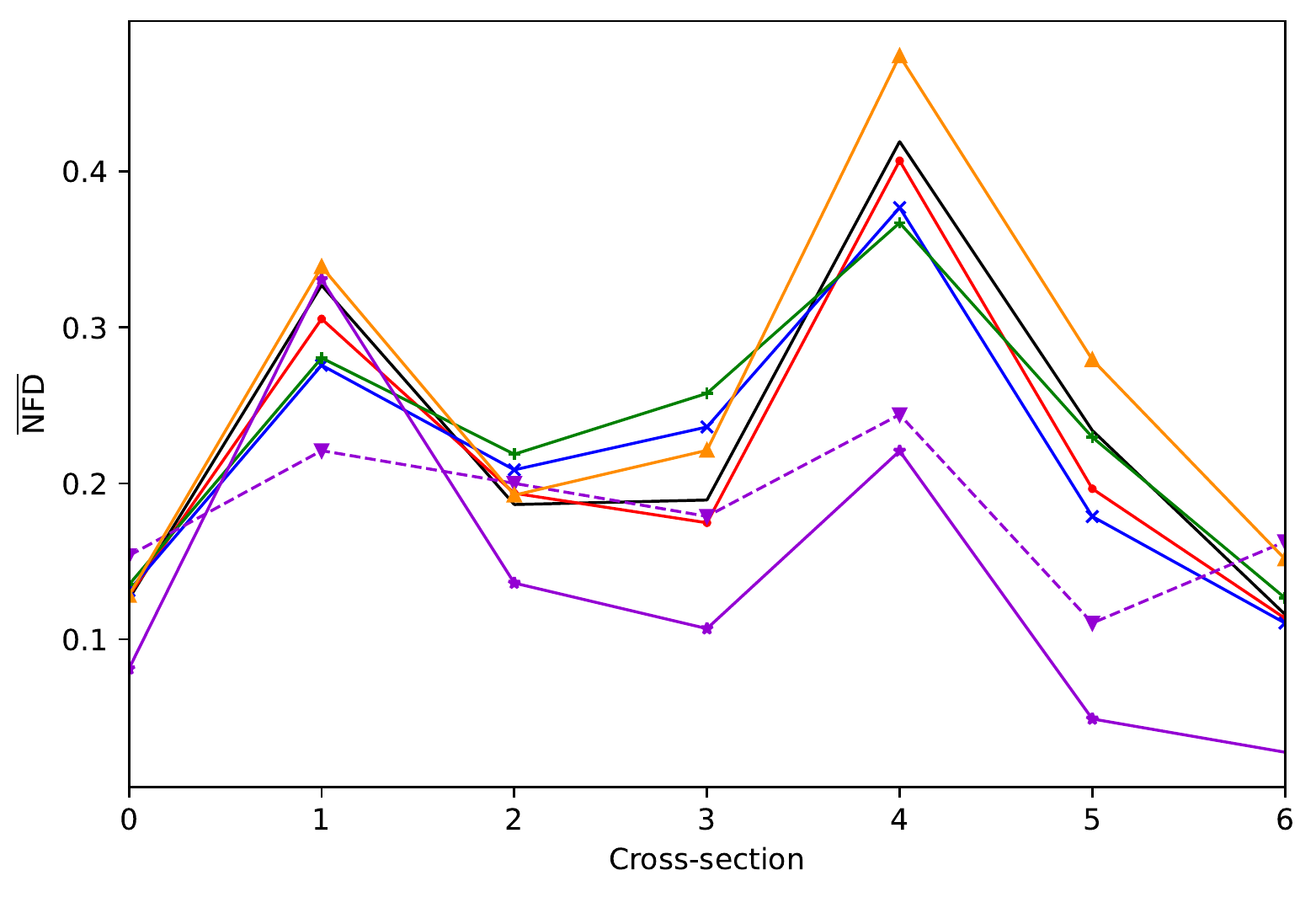}
        \includegraphics[width = 0.49 \textwidth]{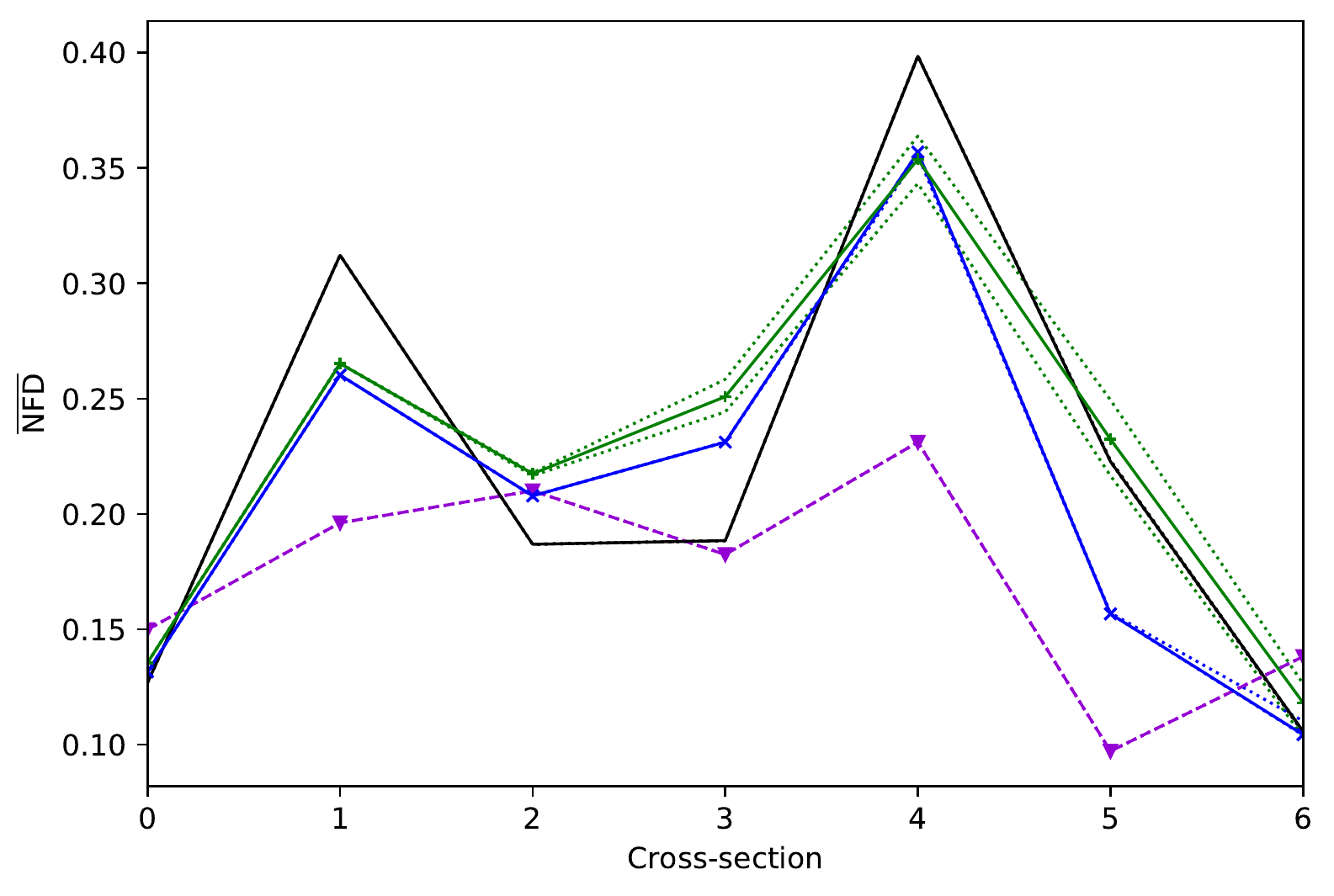}
        \includegraphics[width = 0.49 \textwidth]{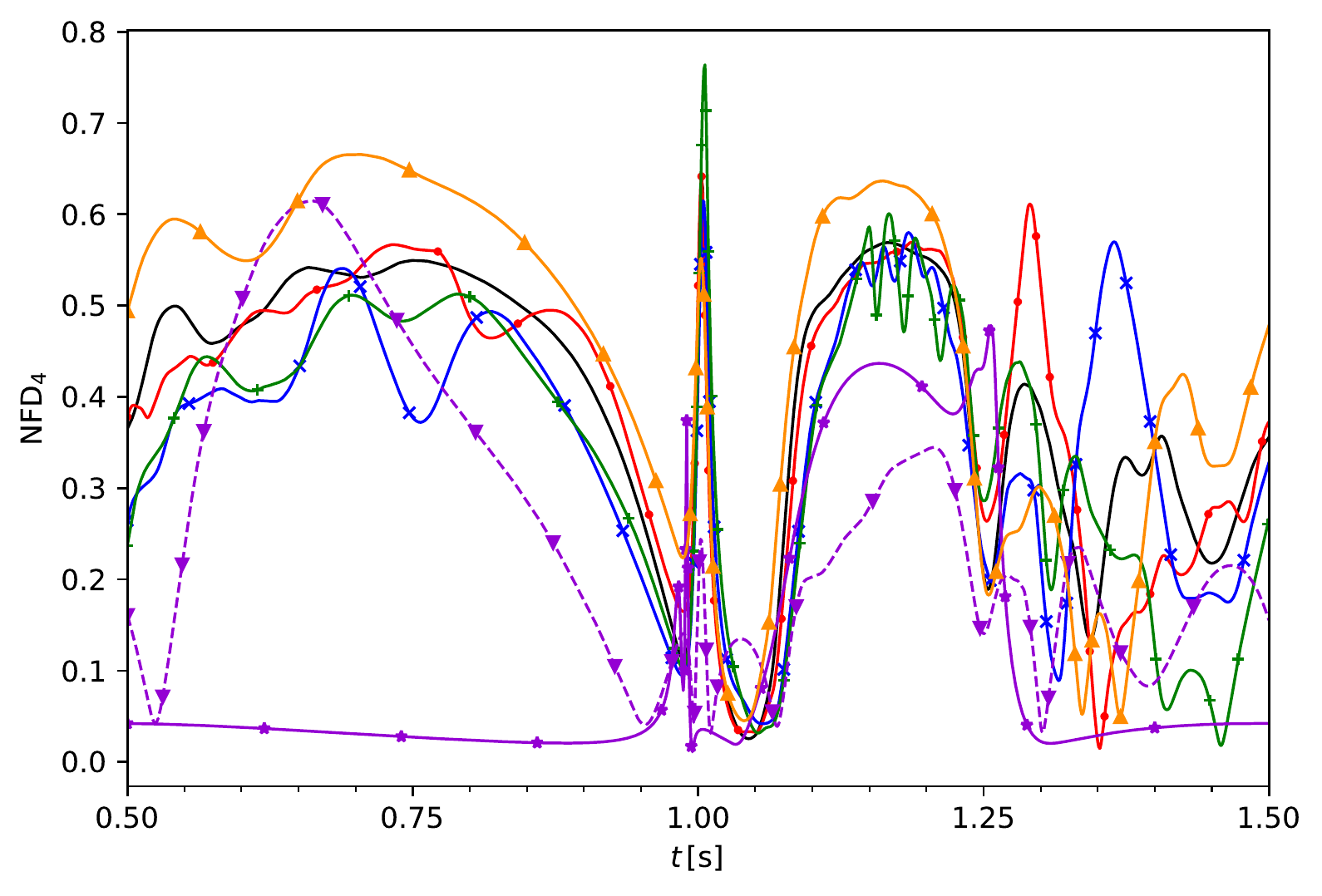}
        \includegraphics[width = 0.49 \textwidth]{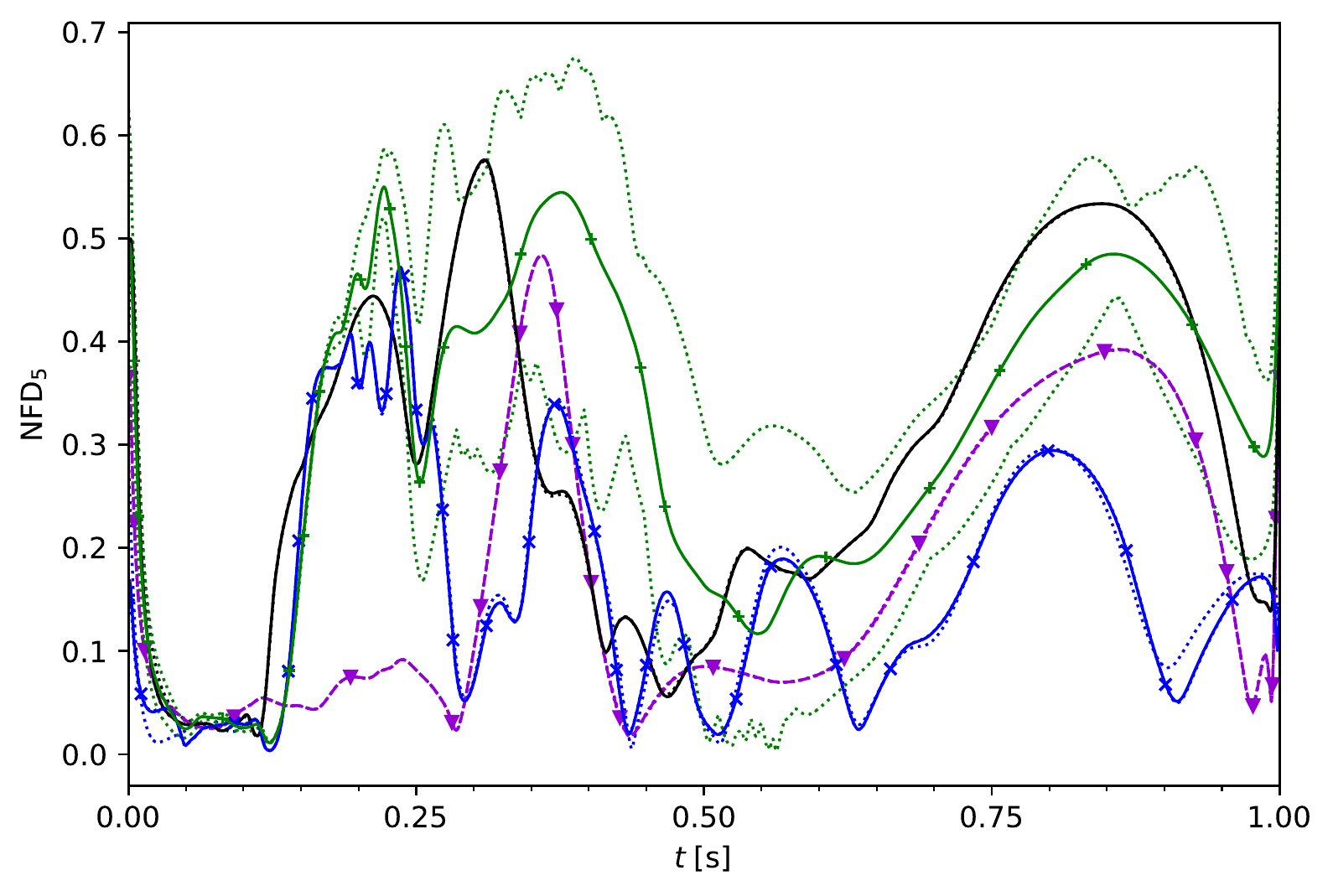}
        \caption[\impturb normalized flow displacement]
        {
          Upper left: time-averaged normalized flow displacement per cross-section.
          Upper right: long time-averaged normalized flow displacement per cross-section.
          Lower left: normalized flow displacement across cross-section 4 over time.
          Lower right: Normalized flow displacement across cross-section 5 over
            time during a heartbeat, averaging the results for the corresponding
            time instant over the simulated beats.
          \turbleg
        }
        \label{fig:turb_nfd}
    \end{center}
\end{figure}

Results for the normalized flow displacement are presented in
Figure~\ref{fig:turb_nfd}.
It is noteworthy that the results for the RB-VMS model with
$\mathrm{P}_1 / \mathrm{P}_1$ elements on the finer grid differ considerably
from the others. However, using second order velocity on the coarse grid leads
to qualitatively the same behavior as predicted by the other methods.

All curves show substantial differences on at least one cross-section.
The results obtained with the Vreman model and the $\bsig$-model
differ significantly only on cross-section~5. This cross-section is also the only
one on which the two Smagorinsky models do not predict a similar
average NFD.

Looking at the NFD behavior over time (Figure~\ref{fig:turb_nfd}, bottom), one
sees a more irregular dynamics than those of the pressure differences or of the
SFD, especially during the phases with lower velocity. This observation is not
surprising, as a largely undirected flow dominated by decaying fluctuations
should be expected not to have a strongly defined center. Nevertheless, during
systole, the eddy viscosity models lead to more similar results. It is perhaps
notable that the Vreman and $\bsig$-models, which are conceptually the most
concerned with avoiding unnecessary artificial dissipation, exhibit oscillations
as the flow decelerates.

As with pressure differences and the SFD, the $\bsig$-model shows by far the
most variation from period to period in the long time simulation. This effect is
largely due to the fact that, as the consistent forward flow disappears during
the decelerating phase, the normal components of slowly dissipating eddies begin
to dominate. As previously observed, the period-wise variation of these eddies
is much less prominent in the other models.

\begin{figure}[t!]
    \begin{center}
        \includegraphics[width = 0.9 \textwidth]{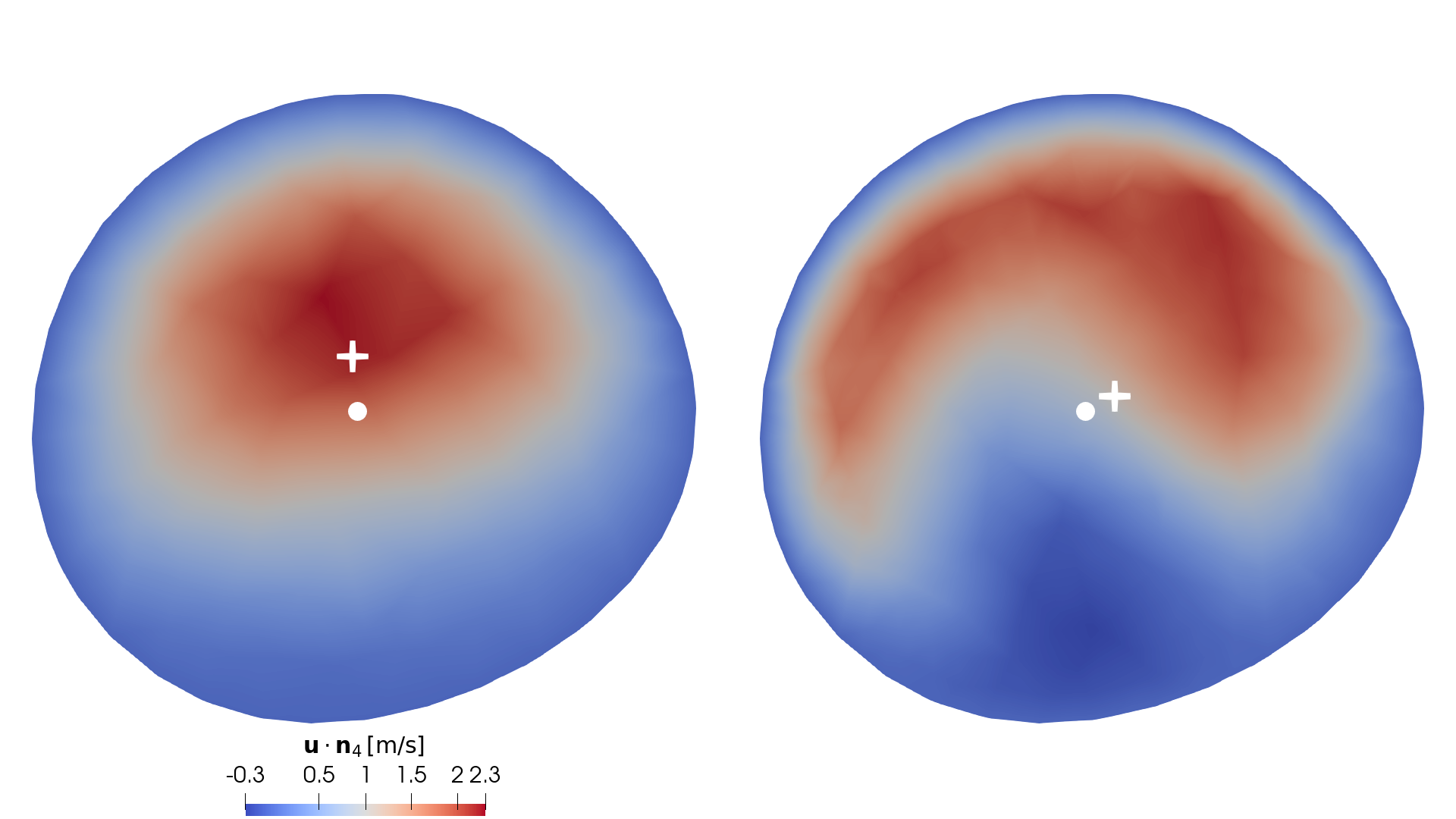}
        \includegraphics[width = 0.9 \textwidth]{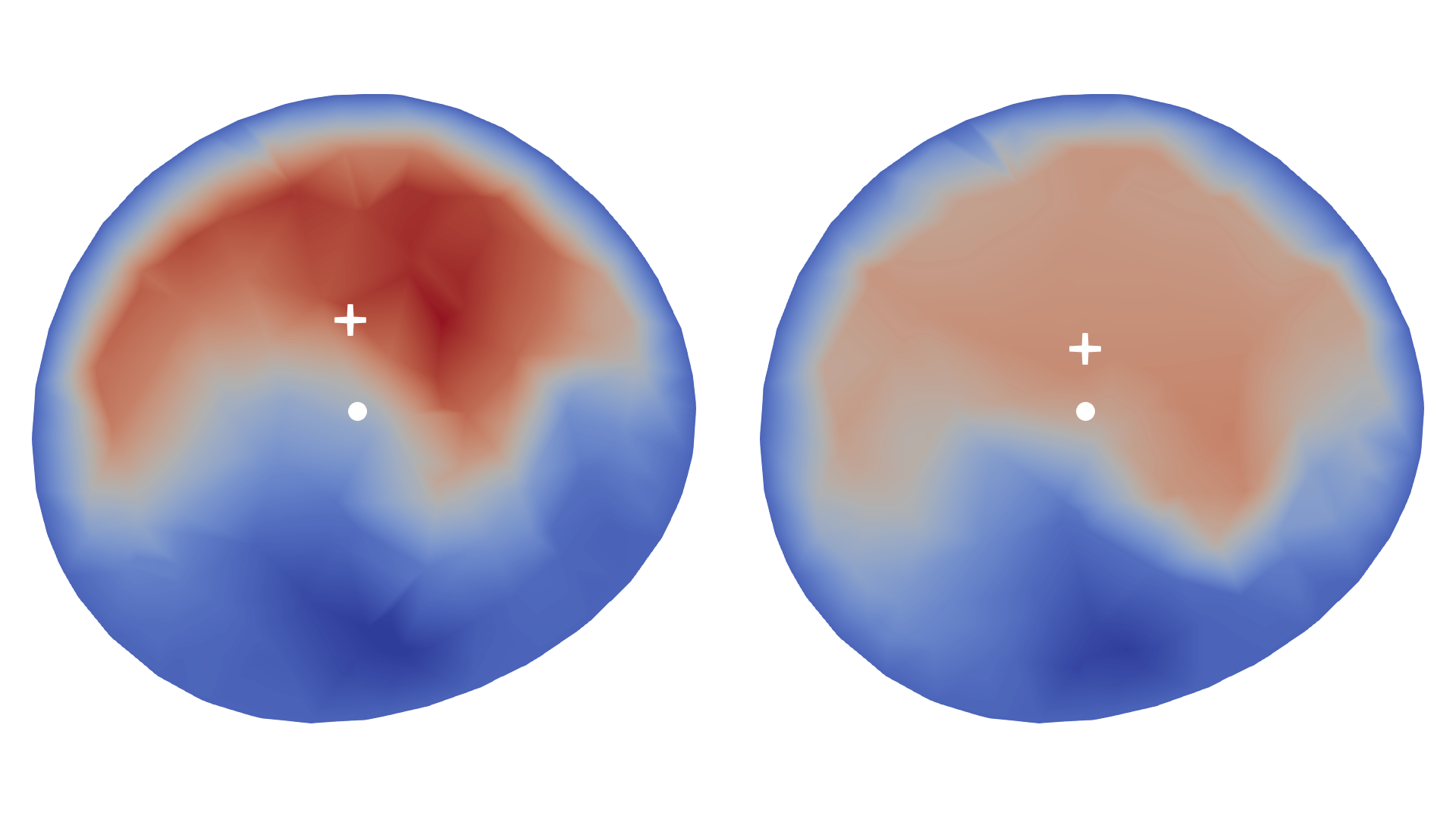}
        \caption[\impturb normal flow moment comparison]
        {
          Normal velocity across cross-section 4 at time $t = \unit[1.107]{s}$ (late accelerating phase).
          The geometric center of mass of the cross section is marked with a circle, whilst the normal flow moment is marked with a cross.
          Upper left: RB-VMS model with $\mathrm{P}_1 / \mathrm{P}_1$ elements.
          Upper right: RB-VMS model with $\mathrm{P}_1 / \mathrm{P}_1$ elements, fine mesh.
          Lower left: RB-VMS model with $\mathrm{P}_2 / \mathrm{P}_1$ elements.
          Lower right: $\bsig$-model, $C_\sigma = 1.35$.
        }
        \label{fig:turb_nfd_slice_4}
    \end{center}
\end{figure}

Figure~\ref{fig:turb_nfd_slice_4} shows the normal component of the velocity
across cross-section~4 at time instant $t = \unit[1.107]{s}$ computed using the
three RB-VMS models and using the $\bsig$-model.
In this case, one can observe large differences of the NFD across these models.
Investigating the slices suggests some weakness in the NFD's ability to
characterize ring-like flow structures. In fact,
the RB-VMS model with $\mathrm{P}_1 / \mathrm{P}_1$ elements on the refined mesh
(upper right) has the smallest NFD of these four examples, despite the clearly
visible concentration of the forward flow near the cross-section's boundaries.

\subsubsection{Wall shear stress} \label{sssec:turb_wss}

\begin{figure}[t!]
    \begin{center}
        \includegraphics[width = 0.49 \textwidth]{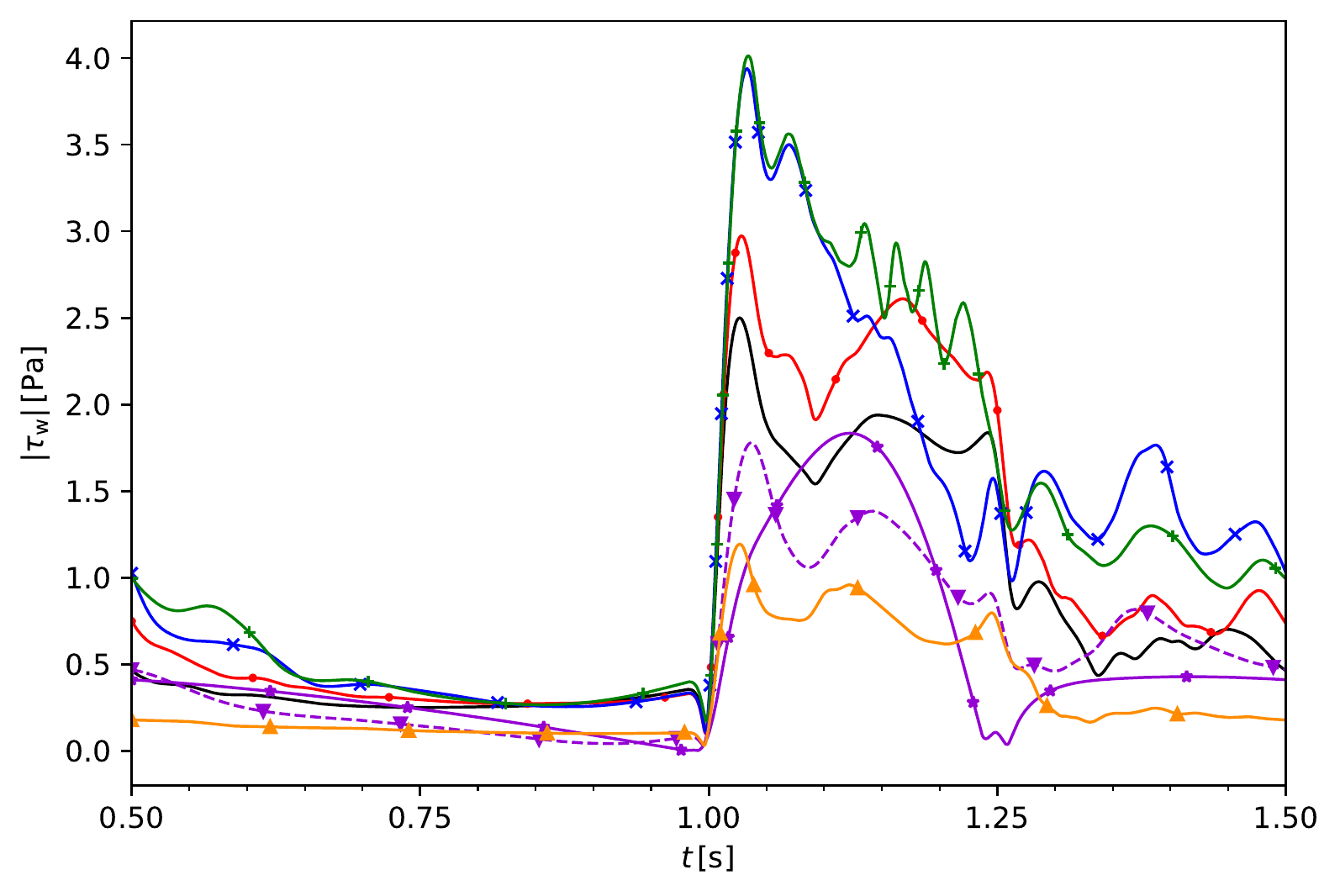}
        \includegraphics[width = 0.49 \textwidth]{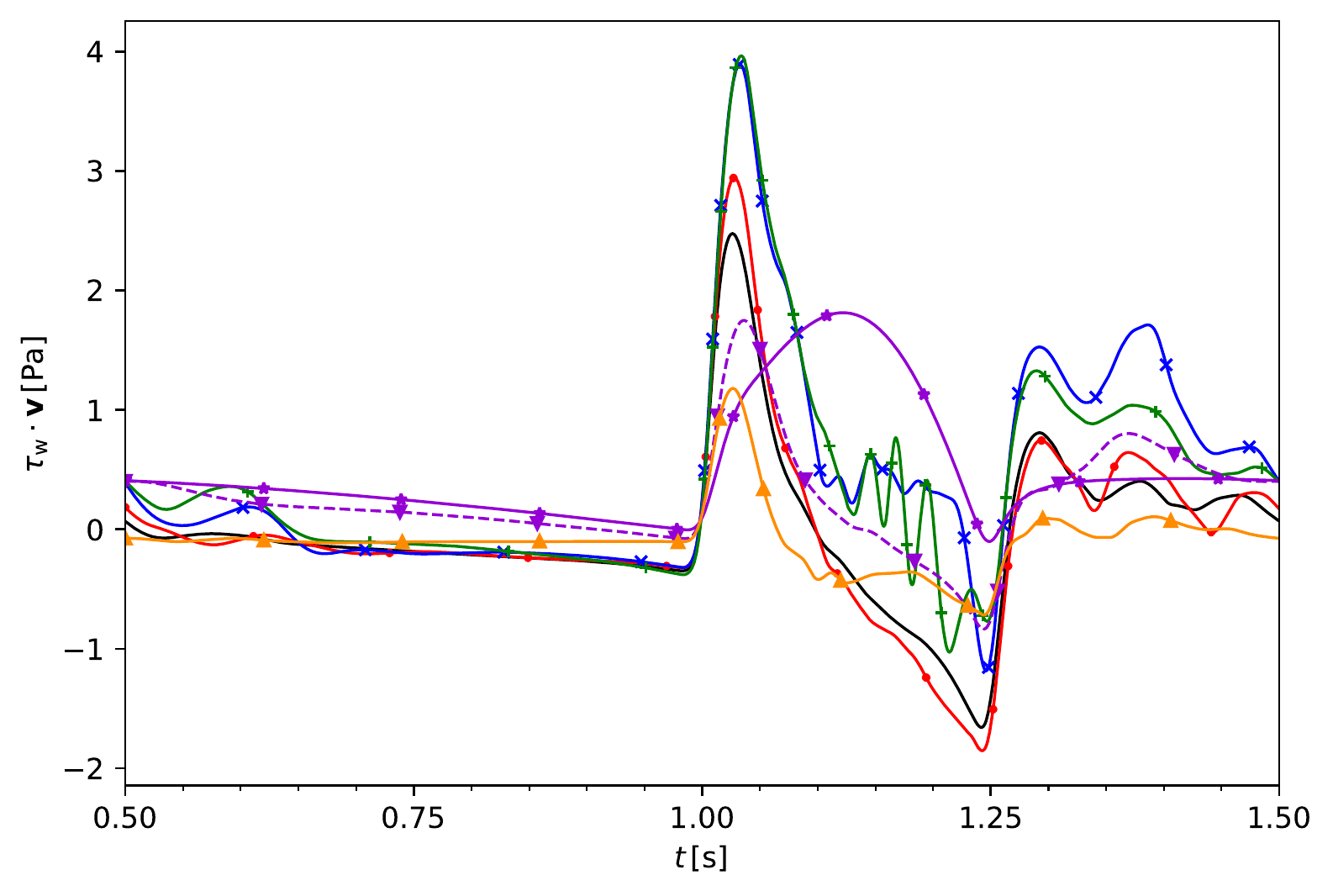}
        \includegraphics[width = 0.49 \textwidth]{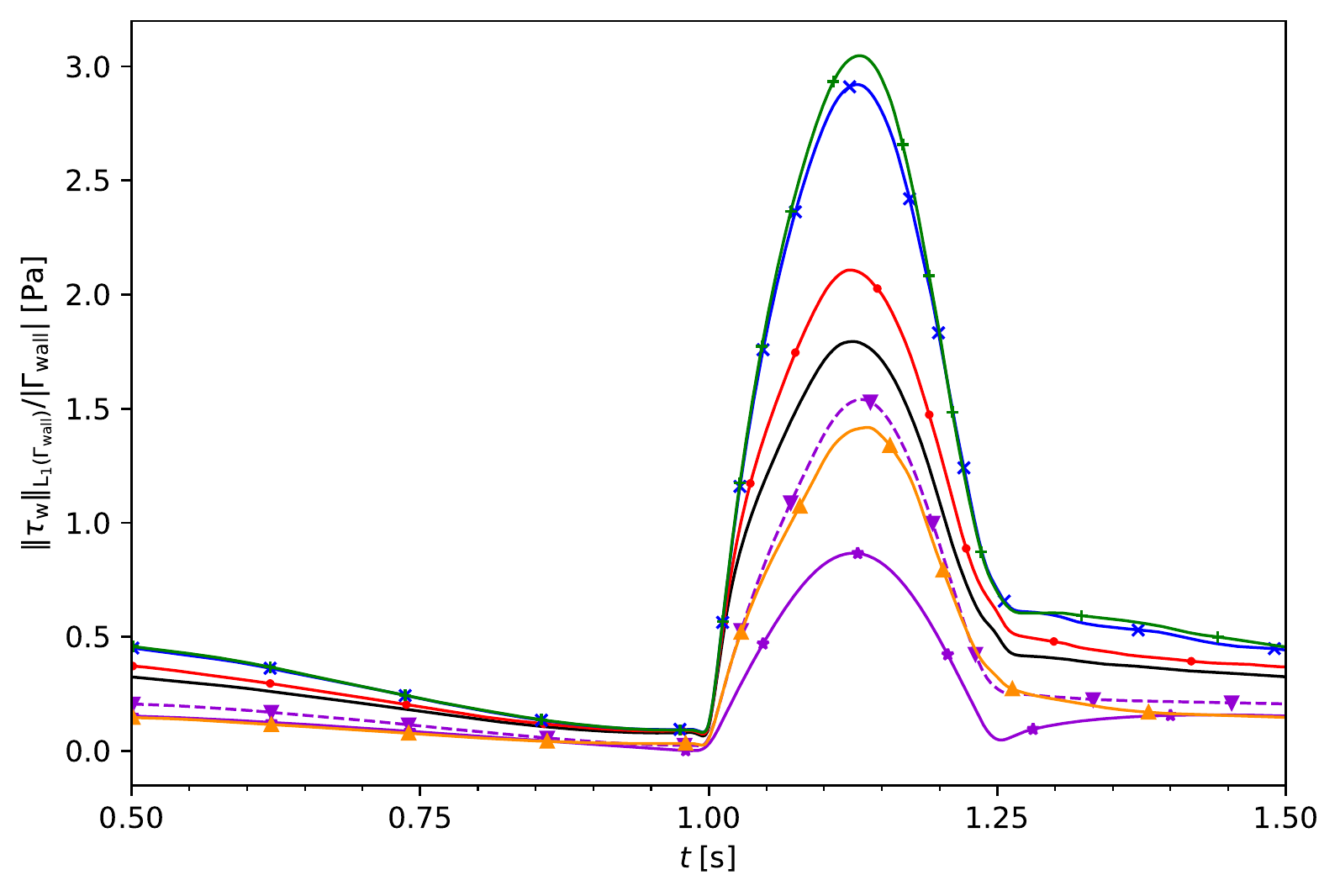}
        \includegraphics[width = 0.49 \textwidth]{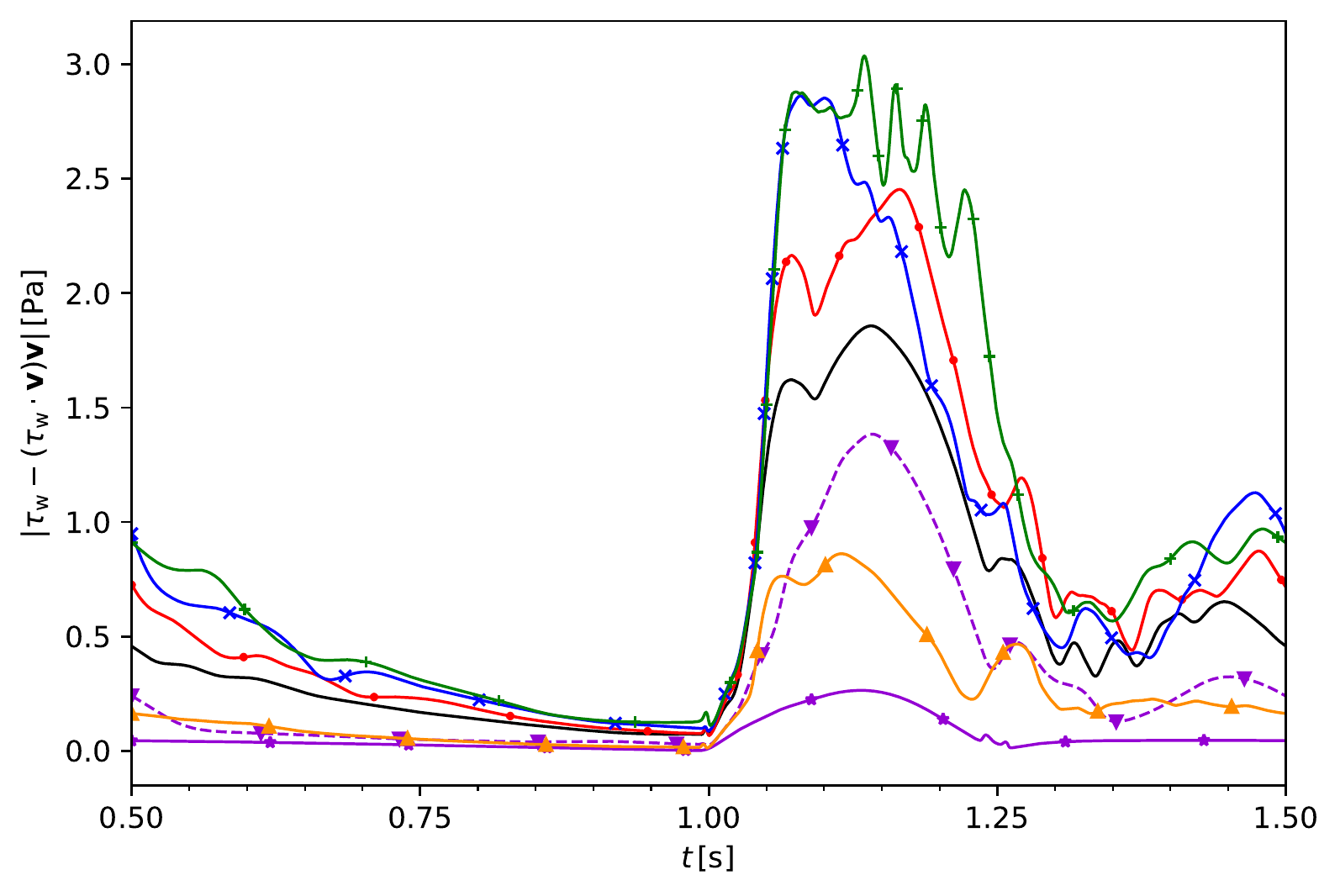}
        \caption[\impturb wall shear stress]
        {
          Upper left: spatially-averaged wall shear stress magnitude over the reference patch.
          Upper right: spatially-averaged forward wall shear stress over the reference patch.
          Lower left: spatially-average wall shear stress magnitude over $\Gamma_\mathrm{wall}$.
          Lower right: spatially-average lateral wall shear stress over the reference patch.
          \turbleg
        }
        \label{fig:turb_wss}
    \end{center}
\end{figure}

\begin{figure}[t!]
    \begin{center}
        \includegraphics[width = 0.49 \textwidth]{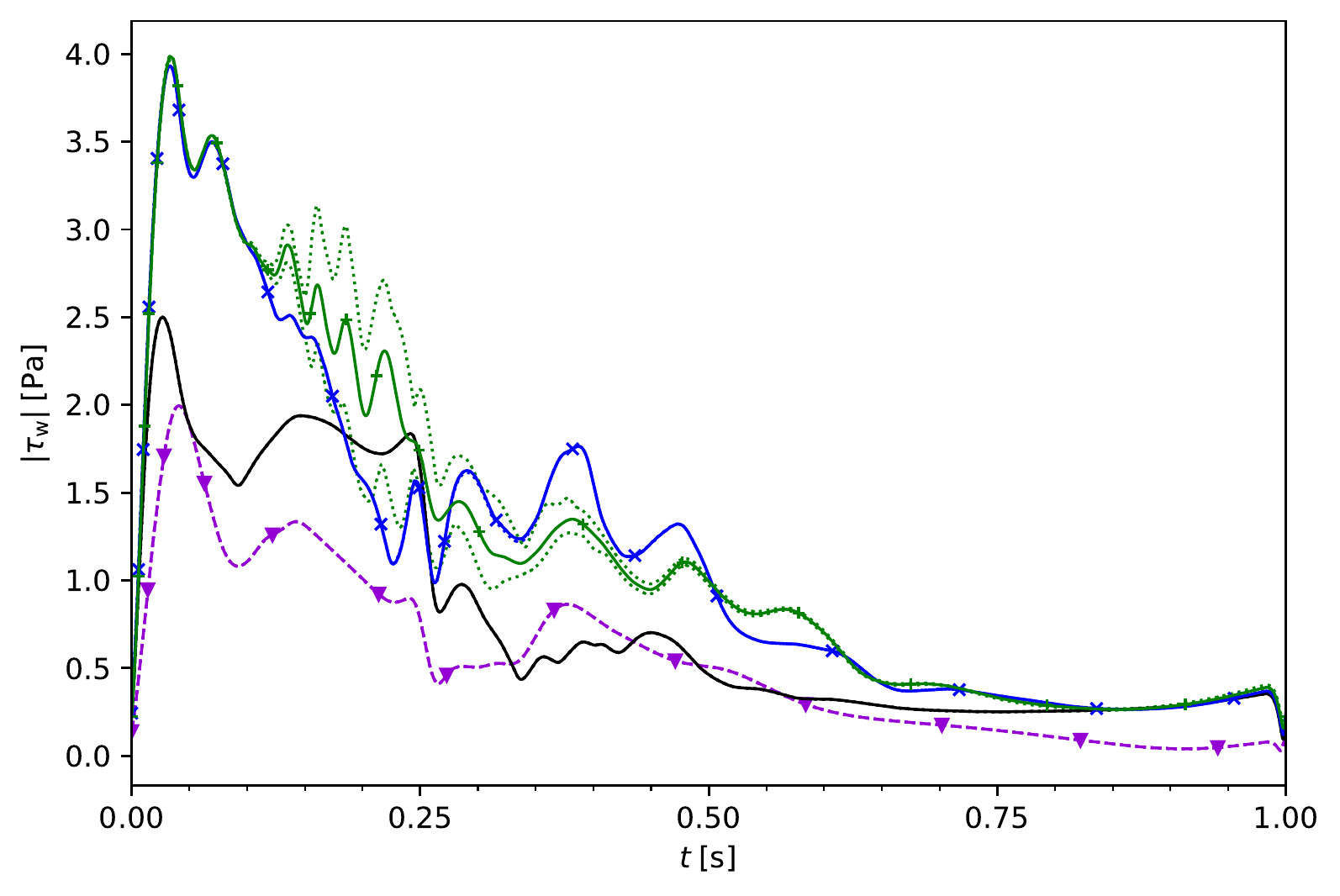}
        \includegraphics[width = 0.49 \textwidth]{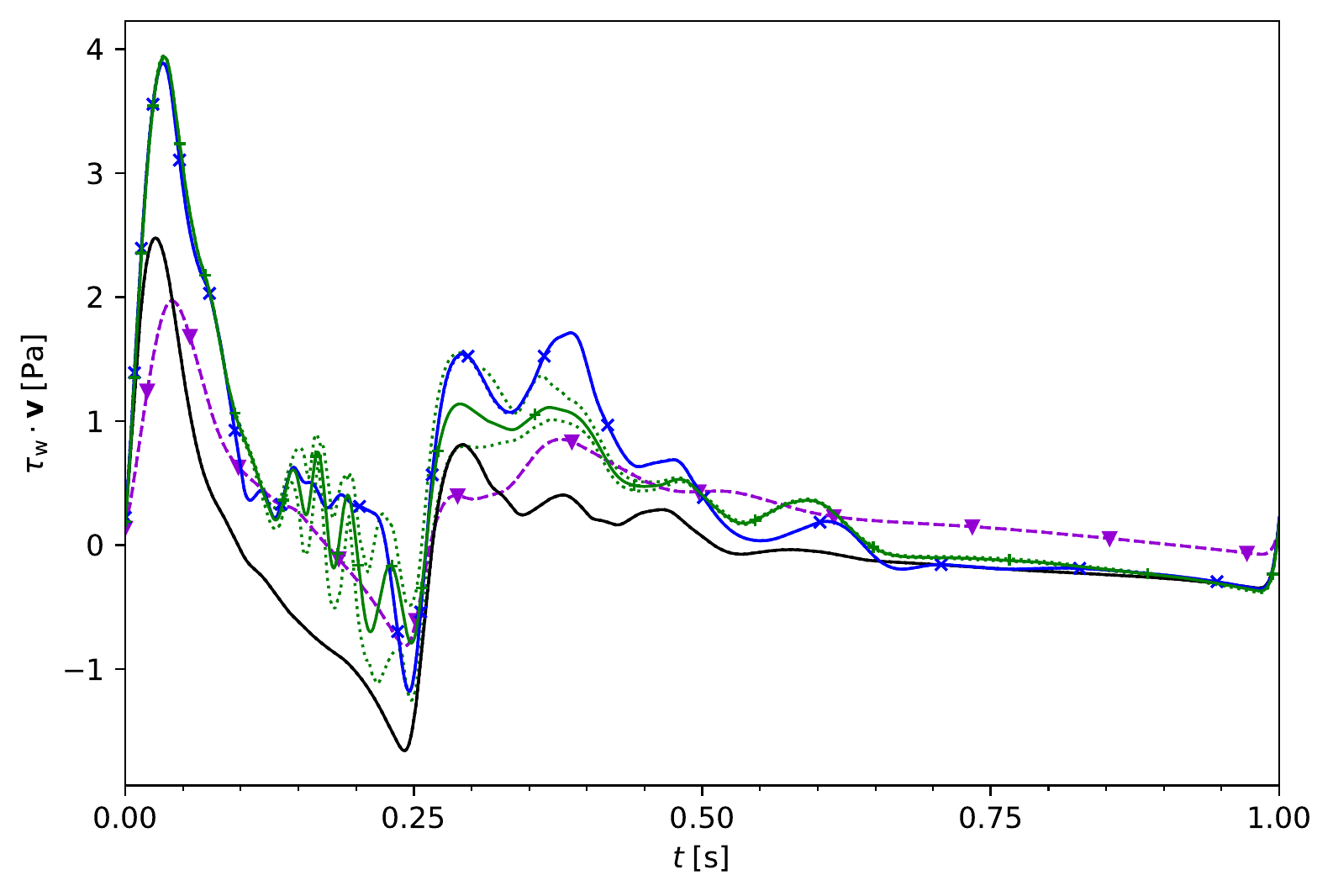}
        \caption[\impturb wall shear stress, long-time simulations]
        {
          Left: spatially-averaged wall shear stress magnitude over the
            reference patch during a heartbeat, averaging the results at the
            corresponding time instant over the simulated beats.
          Right: spatially-averaged forward wall shear stress over the reference
            patch during a heartbeat, averaging the results at the corresponding
            time instant over the simulated beats.
          \turbleg
        }
        \label{fig:turb_long_wss}
    \end{center}
\end{figure}

\begin{figure}[t!]
    \begin{center}
        \includegraphics[width = 0.49 \textwidth]{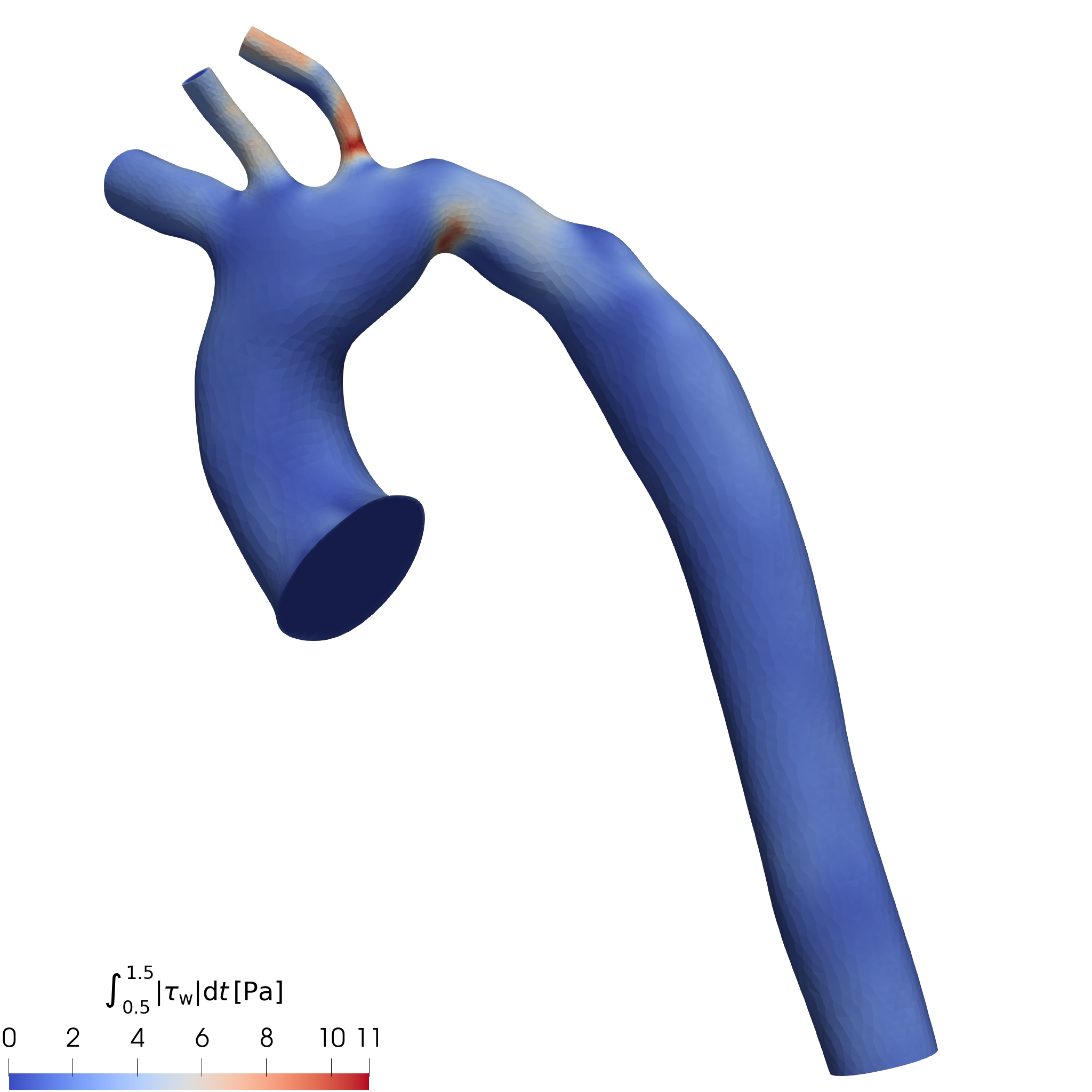}
        \includegraphics[width = 0.49 \textwidth]{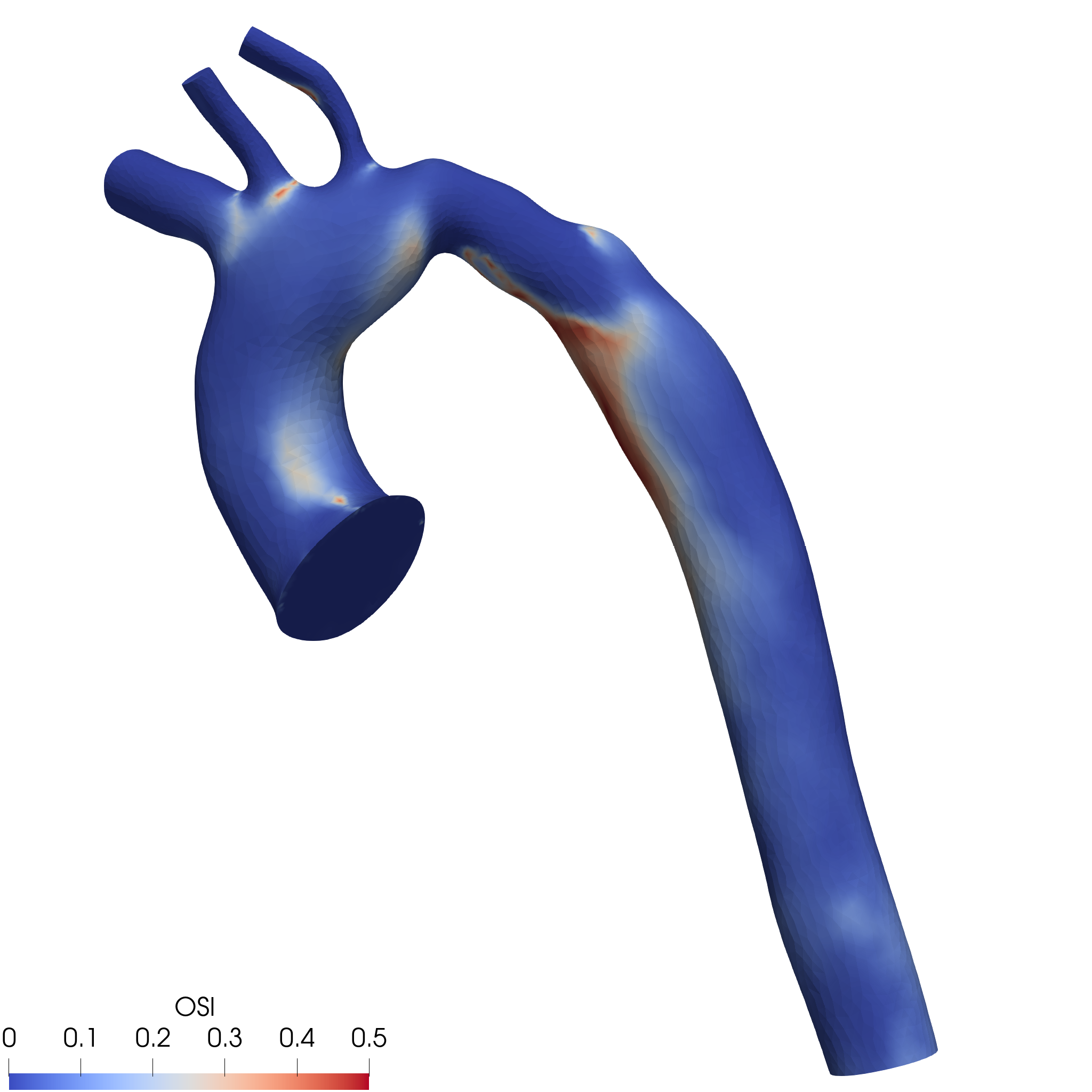}
        \caption[Wall shear stress and OSI surface plot, $\bsig$-model]
        {
          Left: time-averaged wall shear stress over the arterial wall for $\bsig$-model, $C_\sigma = 1.35$.
          Right: oscillatory shear index over the arterial wall for $\bsig$-model, $C_\sigma = 1.35$.
        }
        \label{fig:turb_osi}
    \end{center}
\end{figure}

Figure~\ref{fig:turb_wss} presents the space-averaged magnitude (upper left),
the forward component (upper right), and the lateral magnitude (lower right) of the wall shear
stress over the reference patch shown in
Figure~\ref{fig:qoi-planes} (right), as well as the WSS magnitude averaged over
the entirety of $\Gamma_\mathrm{wall}$ (lower left). Table~\ref{tab:turb_wss}
lists the time-averaged WSS magnitude and the OSI over the reference patch, as
in Section~\ref{sssec:svr_wss}. Here, the models differ widely in scale.
Unsurprisingly, the Vreman model and the $\bsig$-model, which aim to avoid
excessive artificial dissipation near walls, give larger WSS values particularly
during accelerating flow. As the flow reaches its peak and decelerates towards
the end of systole, only the two Smagorinsky model simulations exhibit
significant backward stress, matching the higher OSI values seen in
Table~\ref{tab:turb_wss}.
Due to the smaller constant in the model, the curves for the Smagorinsky model
with $C_{\mathrm{Sma}}=0.005$ in Figure~\ref{fig:turb_wss} are usually closer to
the Vreman and $\bsig$-model than the curves for $C_{\mathrm{Sma}}=0.01$. The
latter results have some similarity with those computed with the RB-VMS model
with second order velocity and the RB-VMS model with first order velocity on the
fine grid.

Figure~\ref{fig:turb_long_wss} demonstrates that, as with the cross-sectional
quantities of interest, the wall shear stress on the reference patch does not
change much from period to period, again with the exception of the
$\bsig$-model. For the $\bsig$-model, the amplitude and timing of the oscillations
associated with the vortices just above the reference patch during deceleration
vary along with the vortices themselves.

Exemplarily, Figure~\ref{fig:turb_osi} depicts the pointwise time-averaged WSS
magnitude and OSI for the $\bsig$-model.

\subsubsection{Computational costs} \label{sssec:turb_cpu}

\begin{figure}[t!]
    \begin{center}
        \includegraphics[width = 0.49 \textwidth]{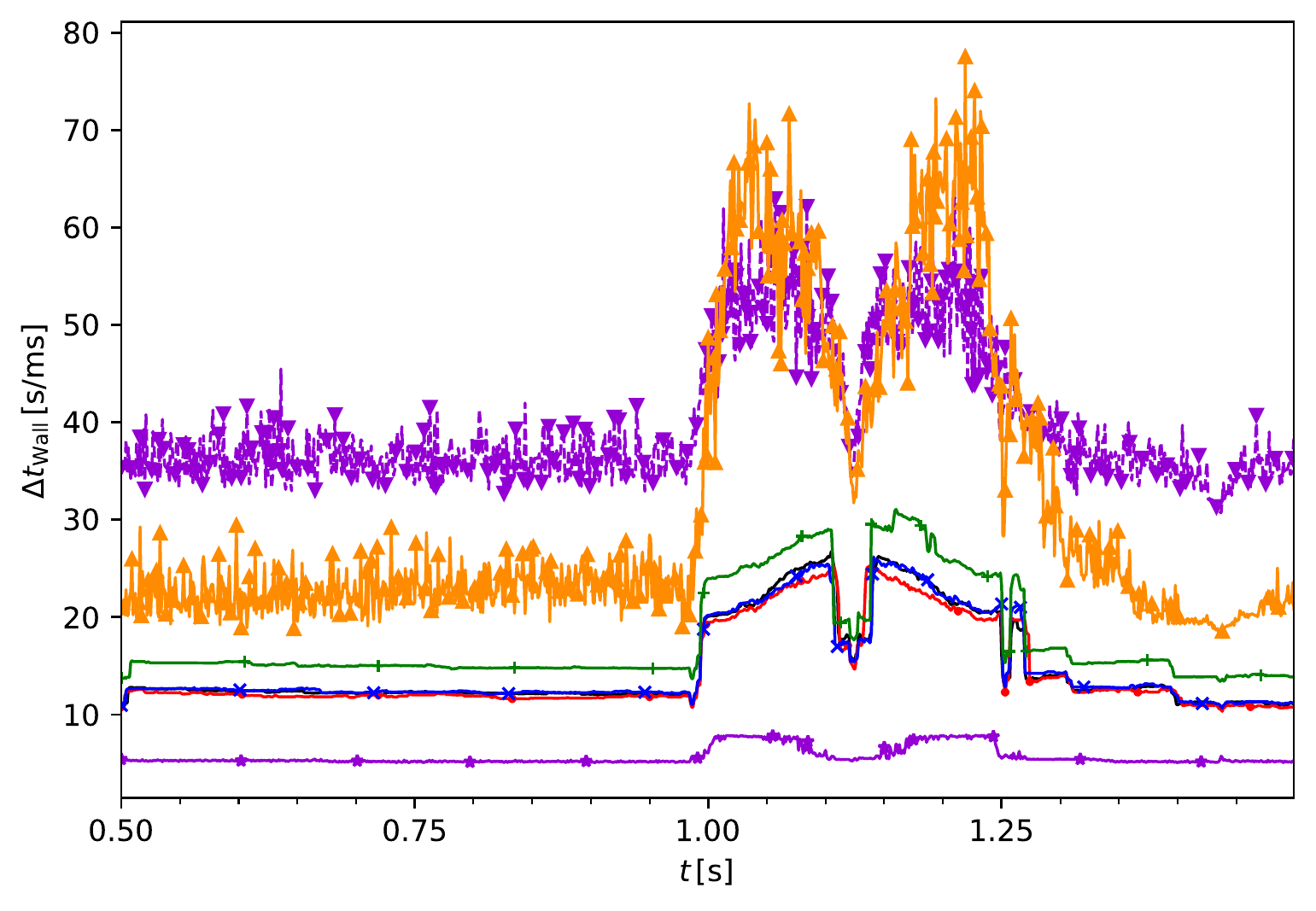}
        \caption[Computation time]
        {
          Computation time per output step.
          \turbleg
        }
        \label{fig:turb_cpu}
    \end{center}
\end{figure}

The final comparison among the considered methods concerns the computational costs.
The CPU times for each turbulence model are summarized in
Table~\ref{tab:turb_cpu}.
Among the turbulence models with $\mathrm{P}_2 / \mathrm{P}_1$ elements, there
are only minor differences with respect to the computing time for both
Smagorinsky models and the Vreman model. The $\bsig$-model was somewhat less
efficient, which is likely due to the computation of the singular values. The
RB-VMS model with $\mathrm{P}_2 / \mathrm{P}_1$ elements needed considerably
more time. There the reason is the lack of a good solver for the arising linear
problems, as explained in Section~\ref{sec:sim_setup}. Finally, the most
time-consuming simulations were those run on the fine grid, with the RB-VMS model and using
$\mathrm{P}_1 / \mathrm{P}_1$ elements.

Figure~\ref{fig:turb_cpu} breaks down the CPU time depending on the time step.
This graph clearly shows the increased computational cost of computing the numerical solution
in presence of rapid changes in the flow (during
acceleration and deceleration at systole). Interestingly, the computational cost
of using the RB-VMS model with $\mathrm{P}_2 / \mathrm{P}_1$ elements on the
coarse mesh increases above those of $\mathrm{P}_1 / \mathrm{P}_1$ elements on
the fine mesh, despite the smaller number of degrees of freedom.

\section{Conclusions}\label{sec:conclusions}

This paper presented a study on the impact of the turbulence model that is used
in numerical simulations of an aortic blood flow on clinically significant
quantities of interest.

The computational model is based on a patient-specific mesh, and the numerical
simulations have been tuned in order to match available inflow (velocity
profile) and outflow (flow rates) measurements.
Due to the lack of clinical data, the outflow boundary conditions have been
restricted to purely resistive lumped parameter models (one parameter per
outlet).
This choice possesses an additional unknown parameter, the systemic vascular
resistance. However, our numerical studies showed that the impact of this
parameter on all quantities of interest is very small.
As an alternative, more general models could be considered (e.g., 3-elements
Windkessel models). These models might affect the time-dependent behavior of the
quantities of interest, but it may be expected that the impact on averaged
quantities would remain small.
Furthermore, the fluid model assumed rigid vessel walls, neglecting
fluid-structure interaction effects, which are out of the scope of the current
work. Although this aspect could have a quantitative influence on the wall shear
stresses, one can expect that it does not affect the qualitative differences
between the turbulence models, which is the main focus of this study.
The numerical simulations are based on the assumption of a Newtonian flow.
Further studies taking into account non-Newtonian models are currently in
preparation.

The main outcome of the presented numerical study is that the impact of the
turbulent model choice is non-negligible, and in selected cases also rather
strong, both for averaged quantities of interest and for their temporal
evolution.

The effect of the order of the finite element velocity space was investigated,
exemplarily for the RB-VMS model. On the same grid, results with second order
velocity were by far more accurate than results obtained with first order
velocity. Using a piecewise linear velocity on a refined grid led often to a
considerable increase of the accuracy, but not always, as could be seen for the
normalized flow displacement in Figure~\ref{fig:turb_nfd}. Consequently, in our
opinion, from the point of view of accuracy using second order velocity is the
better choice.

Inspecting the results obtained with second order velocity simulations, one can
divide the considered turbulence models into two groups. On the one hand, the
Smagorinsky and the RB-VMS models and on the other hand, the Vreman and the
$\bsig$-models. The results given by the models of each group were often, though
not always, qualitatively similar. This division corresponds to the
amount of numerical diffusion that is introduced by the models, particularly
near walls and in transitional regions. Whereas the models from the first group
computed rather smooth solutions, due to their comparatively large numerical
diffusion, the flow fields predicted with the methods from the second group
possess much more small eddies. In our opinion, the results with the less
diffusive turbulence models are the more trustworthy ones.
The $\bsig$-model was less dissipative in our simulations, but its computational
costs were higher by a factor of around $1.2$ compared to the Vreman model. The
numerical results presented in this paper do not show a clear advantage in preferring one of these two models
to the other.

\section*{Funding}
The work of S.~Katz has been supported by the Deutsche Forschungsgemeinschaft (DFG) within the RTG 2433
\emph{Differential Equation- and Data-driven Models in Life Sciences and Fluid Dynamics (DAEDALUS).}

\bibliographystyle{WileyNJD-AMA}
\bibliography{aorta}

\listoffigures

\FloatBarrier

\section{Tables}

\begin{table}[!h]
  \begin{center}
    \caption{Mesh statistics: number of tetrahedra and vertices, maximum and average boundary layer height, maximum and average cell volume.}
    \label{tab:grid_stat}
    \begin{tabular}{l|c|c|c|c|c|c}
      Mesh                & Tetrahedra &  Vertices &  $y_\mathrm{max}$ & $\bar y$           & $V_\mathrm{max}$    & $\bar V$ \\ [0.2em]
      \hline
      $\mathcal T$        &  $106,983$ &  $21,495$ &  $\unit[3.8]{mm}$ &  $\unit[1.06]{mm}$ & $\unit[43.2]{mm^3}$ & $\unit[4.72]{mm^3}$ \\ [0.2em]
      \hline
      $\mathcal T^\prime$ &  $855,864$ & $158,335$ & $\unit[2.18]{mm}$ & $\unit[0.537]{mm}$ & $\unit[5.39]{mm^3}$ & $\unit[0.59]{mm^3}$ \\ [0.2em]
    \end{tabular}
  \end{center}
\end{table}

\begin{table}[!h]
  \begin{center}
    \caption{Estimated flow rates and corresponding fraction of the inlet flow (absolute value) for each outlet
      (see also Figure \ref{fig:domain_sketch}, left) used in the simulations.}
    \label{tab:flowsplit}
    \begin{tabular}{l|c|c}
      Boundary & Flow $Q^*_i$ [m$^3$/s] & Flow fraction $\mathrm{out}_i$\\[0.2em]
      \hline
      $\Gamma_{\mathrm{out},1}$ (brachiocephalic artery)        & $7.43 \cdot 10^{-5}$ &  16.81 \% \\[0.2em]
      $\Gamma_{\mathrm{out},2}$ (left common carotid artery)    & $3.80 \cdot 10^{-5}$ &   8.60 \% \\[0.2em]
      $\Gamma_{\mathrm{out},3}$ (left common subclavian artery) & $3.63 \cdot 10^{-5}$ &   8.21 \% \\[0.2em]
      $\Gamma_{\mathrm{out},4}$ (descending aorta)              & $2.93 \cdot 10^{-4}$ &  66.38 \% \\[0.2em]
      \hline
      Inlet                           & $4.42 \cdot 10^{-4}$ & 100.00 \%
    \end{tabular}
  \end{center}
\end{table}

\begin{table}[!h]
  \begin{center}
    \caption{Mesh size and velocity/pressure space dimensions. The pair $\mathrm P_2^3 \times \mathrm P_1$ was not
      used on $\mathcal T^\prime$ since it results in about $3,75$ million degrees of freedom.}
    \label{tab:grid_dof}
    \begin{tabular}{l|c|c|c|c|c}
      Mesh & Tetrahedra & $\dim(\mathrm P_1)$ & $\dim(\mathrm P_2)$ & $\dim(\mathrm P_1^3 \times \mathrm P_1)$ & $\dim(\mathrm P_2^3 \times \mathrm P_1)$ \\ [0.2em]
      \hline
      $\mathcal T$       & $106,983$ &  $21,495$ &   $158,335$ &  $85,980$ &   $496,500$ \\ [0.2em]
      \hline
      $\mathcal T^\prime$ & $855,864$ & $158,335$ & $-$ 
      & $633,340$ & $-$ 
      \\ [0.2em]
    \end{tabular}
  \end{center}
\end{table}

\begin{table}[!h]
  \begin{center}
    \caption{Number of points used for the calculation of cross-sectional QOIs
      and cross-section statistics.}
    \begin{tabular}{l|c|c|c|c|c|c|c}
      Plane                          &     0 &     1 &     2 &     3 &     4 &     5 &     6 \\ \hline
      \# points                      & $464$ & $341$ & $225$ & $249$ & $329$ & $313$ & $258$ \\
      Area $[\unit{mm^2}]$           & $462$ & $344$ & $225$ & $249$ & $328$ & $312$ & $257$ \\
      Perimeter $[\unit{mm}]$        &  $76$ &  $66$ &  $53$ &  $56$ &  $64$ &  $62$ &  $57$ \\
      Hydraulic radius $[\unit{mm}]$ & $6.1$ & $5.2$ & $4.2$ & $4.5$ & $5.1$ & $5.0$ & $4.5$
    \end{tabular}
    \label{tab:points}
  \end{center}
\end{table}

\begin{table}[!h]
  \begin{center}
    \caption{Resistances $[\unitfrac{MPa \cdot s}{m^3}]$ at each outlet for the considered turbulence models and for
      different values of the SVR.} \label{tab:resistances_coarse}
    \begin{tabular}{l|c|c|c|c|c}
      Turbulence model                                          & $\RSV$ &    $R_1$ &    $R_2$ &    $R_3$ &    $R_4$ \\[0.2em]
      \hline
      &  $160$ & $992.99$ & $1855.9$ & $1829.9$ & $240.52$ \\[0.2em]
      Smagorinsky, $C_\mathrm{Sma} = 0.01$                      &  $115$ & $725.77$ & $1333.4$ & $1282.9$ & $172.76$ \\[0.2em]
      &   $70$ & $458.87$ & $811.55$ & $736.64$ & $105.08$ \\[0.2em]
      \hline
      &  $160$ & $986.97$ & $1858.0$ & $1863.7$ & $240.27$ \\[0.2em]
      Smagorinsky, $C_\mathrm{Sma} = 0.005$                     &  $115$ & $719.67$ & $1335.3$ & $1316.6$ & $172.48$ \\[0.2em]
      &   $70$ & $452.54$ & $813.04$ & $769.84$ & $104.74$ \\[0.2em]
      \hline
      &  $160$ & $978.57$ & $1860.4$ & $1902.0$ & $240.11$ \\[0.2em]
      Vreman, $C_\mathrm{Vre} = 0.07$                           &  $115$ & $711.19$ & $1337.6$ & $1354.7$ & $172.30$ \\[0.2em]
      &   $70$ & $443.89$ & $814.95$ & $807.61$ & $104.52$ \\[0.2em]
      \hline
      &  $160$ & $976.45$ & $1858.7$ & $1905.4$ & $240.21$ \\[0.2em]
      $\bsig$-model, $C_\sigma = 1.35$                          &  $115$ & $709.05$ & $1335.9$ & $1358.1$ & $172.40$ \\[0.2em]
      &   $70$ & $441.73$ & $813.18$ & $810.92$ & $104.61$ \\[0.2em]
      \hline
      &  $160$ & $1004.5$ & $1861.0$ & $1791.6$ & $240.44$ \\[0.2em]
      RB-VMS, $\mathrm{P}_1 / \mathrm{P}_1$ elements            &  $115$ & $737.50$ & $1338.8$ & $1245.0$ & $172.72$ \\[0.2em]
      &   $70$ & $471.02$ & $817.77$ & $699.56$ & $105.15$ \\[0.2em]
      \hline
      &  $160$ & $984.37$ & $1856.0$ & $1873.5$ & $240.30$ \\[0.2em]
      RB-VMS, $\mathrm{P}_1 / \mathrm{P}_1$ elements, fine mesh &  $115$ & $717.03$ & $1333.3$ & $1326.3$ & $172.50$ \\[0.2em]
      &   $70$ & $449.85$ & $810.87$ & $779.39$ & $104.75$ \\[0.2em]
      \hline
      &  $160$ & $974.81$ & $1851.1$ & $1880.8$ & $240.83$ \\[0.2em]
      RB-VMS, $\mathrm{P}_2 / \mathrm{P}_1$ elements            &  $115$ & $707.43$ & $1328.3$ & $1333.5$ & $173.03$ \\[0.2em]
      &   $70$ & $440.15$ & $805.72$ & $786.44$ & $105.25$ \\[0.2em]
    \end{tabular}
  \end{center}
\end{table}

\begin{table}[!h]
  \begin{center}
    \caption{Impact of the variation of SVR. Time-averaged WSS magnitude and OSI over reference patch.
      Simulation with the Smagorinsky model, $C_\mathrm{Sma} = 0.01$.}
    \label{tab:svr_wss}
    \begin{tabular}{c|c|c}
      $\RSV \; [\unitfrac{MPa \cdot s}{m^3}]$ & $|\tau_\mathrm{w}| \; [\unit{Pa}]$ & OSI \\ \hline
      $70$ & $0.76756$ & $0.45781$ \\
      $115$ & $0.77128$ & $0.45779$ \\
      $160$ & $0.77321$ & $0.45792$
    \end{tabular}
  \end{center}
\end{table}

\begin{table}[!h]
  \begin{center}
    \caption{Time-averaged wall shear stress magnitude and OSI over reference patch.}
    \label{tab:turb_wss}
    \begin{tabular}{l|c|c}
      Turbulence model                                          & $|\tau_\mathrm{w}| \; [\unit{Pa}]$ & OSI \\ \hline
      Smagorinsky, $C_\mathrm{Sma} = 0.01$                      & $0.77128$                          & $0.45779$ \\
      Smagorinsky, $C_\mathrm{Sma} = 0.005$                     & $0.96752$                          & $0.44082$ \\
      Vreman, $C_\mathrm{Vre} = 0.07$                           & $1.15169$                          & $0.29806$ \\
      $\bsig$-model, $C_\sigma = 1.35$                          & $1.22330$                          & $0.32456$ \\
      RB-VMS, $\mathrm{P}_1 / \mathrm{P}_1$ elements            & $0.50879$                          & $0.00662$ \\
      RB-VMS, $\mathrm{P}_1 / \mathrm{P}_1$ elements, fine mesh & $0.52881$                          & $0.25779$ \\
      RB-VMS, $\mathrm{P}_2 / \mathrm{P}_1$ elements            & $0.32271$                          & $0.36761$
    \end{tabular}
  \end{center}
\end{table}

\begin{table}[!h]
  \begin{center}
    \caption
    {
      CPU statistics for each model. Step time in units of seconds wall time
      per millisecond simulated time.
    }
    \label{tab:turb_cpu}
    \begin{tabular}{l|c|c}
      Turbulence model                                          & Total wall time [\unit{h}:\unit{min}:\unit{s}] & Average step time [\unitfrac{s\,}{ms}] \\ \hline
      Smagorinsky, $C_\mathrm{Sma} = 0.01$                      &  6:35:42.4                                     & 15.8 \\
      Smagorinsky, $C_\mathrm{Sma} = 0.005$                     &  6:25:23.9                                     & 15.4 \\
      Vreman, $C_\mathrm{Vre} = 0.07$                           &  6:34:45.0                                     & 15.8 \\
      $\bsig$-model, $C_\sigma = 1.35$                          &  7:48:53.3                                     & 18.7 \\
      RB-VMS, $\mathrm{P}_1 / \mathrm{P}_1$ elements            &  2:26:41.2                                     & 5.9 \\
      RB-VMS, $\mathrm{P}_1 / \mathrm{P}_1$ elements, fine mesh & 17:17:12.5                                     & 41.5 \\
      RB-VMS, $\mathrm{P}_2 / \mathrm{P}_1$ elements            & 14:17:21.9                                     & 34.3
    \end{tabular}
  \end{center}
\end{table}
\clearpage

\end{document}